\documentclass[10pt,twocolumn,aip,jcp]{revtex4-1}
\usepackage[T1]{fontenc}
\usepackage[latin9]{inputenc}
\usepackage{bm}
\usepackage{amsmath}
\usepackage{amssymb}
\usepackage{graphicx}

\usepackage{amstext,amsbsy}
\usepackage{times}
\usepackage{units}

\usepackage{braket}
\usepackage{color}
\begin{document}
\title{Reduced Rovibrational Coupling Cartesian Dynamics for Semiclassical
Calculations: Application to the Spectrum of the Zundel Cation}
\author{G. Bertaina}
\email{gianluca.bertaina@unimi.it}
\affiliation{Dipartimento di Chimica, Universit\`{a} degli Studi di Milano, via 
C. Golgi 19, 20133 Milano, Italy}

\author{G. Di Liberto}
\altaffiliation{Current address: Dipartimento di Scienza dei Materiali, 
Universit\`{a} di Milano-Bicocca, via R. Cozzi 55, 20125 Milano, Italy}
\affiliation{Dipartimento di Chimica, Universit\`{a} degli Studi di Milano, via 
C. Golgi 19, 20133 Milano, Italy}

\author{M. Ceotto}
\email{michele.ceotto@unimi.it}
\affiliation{Dipartimento di Chimica, Universit\`{a} degli Studi di Milano, via 
C. Golgi 19, 20133 Milano, Italy}

\begin{abstract}
We study the vibrational spectrum of the protonated water dimer, by
means of a divide-and-conquer semiclassical initial value representation
of the quantum propagator, as a first step in the study of larger protonated water clusters. We use the potential energy surface from
{[}Huang et al., J. Chem. Phys. $\bm{122}$, 044308 (2005){]}. To
tackle such an anharmonic and floppy molecule, we employ fully Cartesian
dynamics and carefully reduce the coupling to global rotations in
the definition of normal modes. We apply the time-averaging filter
and obtain clean power spectra relative to suitable reference states,
that highlight the spectral peaks corresponding to the fundamental
excitations of the system. Our trajectory-based approach allows us for physical interpretation of the very challenging proton
transfer modes. We find that it is important, for such a floppy molecule, to selectively avoid to initially excite lower energy modes, in order to obtain cleaner spectra. The estimated vibrational energies display a mean
absolute error (MAE) of $\sim29\text{cm}^{-1}$ with respect to available
Multi-Configuration time-dependent Hartree calculations and $\text{MAE}\sim14\text{cm}^{-1}$
when compared to the optically active experimental excitations of
the Ne-tagged Zundel cation. The reasonable scaling in the number of trajectories for Monte Carlo convergence is promising for applications to higher dimensional protonated cluster systems.
\end{abstract}
\maketitle

\section{Introduction}

Floppy molecules are one of the major vibrational spectroscopic challenges
for ab initio simulations.\cite{Bowman_Meyer_Polyatomic_2008} The
strong couplings between vibrations and global and internal hindered
rotations present in these moieties generate a high density of strongly
anharmonic energy levels. A theoretical accurate method able to calculate
these levels and, at the same time, to assign them, is very much desired.
Besides grid approaches,\cite{Bowman_Meyer_Polyatomic_2008,Thomas_Carrington_SevenAtoms_2015}
which rely on pre-computed and fitted potential energy surfaces (PES)
and suitable basis set representations, or imaginary-time correlation
function calculations from path-integral methods,\cite{bertaina_statistical_2017}
classical trajectories are a direct and ab initio dynamics way to
calculate vibrational density of states via Fourier transform of correlation
functions. In particular, semiclassical molecular dynamics,\cite{miller2001semiclassical,Miller_PNAScomplexsystems_2005}
which relies on classical trajectories, allows for the calculation
of quantum wave-packet correlation functions, together with their
Fourier transform, the quantum power spectrum. This spectrum reproduces
quantum features, such as zero point energy (ZPE) values, tunneling,
delocalization effects, overtones, quantum resonances, etc. These phenomena are particularly relevant
in systems featuring hydrogen bonds and containing water molecules.\cite{Yang_Deconstructingwaterdiffuse_2019,Gabas_Protonatedglycinesupramolecular_2018}
Instead, a classical Fourier transform of the velocity correlation
function can only provide the effect of classical PES anharmonicity
on the frequencies of vibration. Another advantage of the semiclassical approach is that a suitable partitioning of phase space sampling,\cite{DiLiberto_Divideandconquersemiclassicalmolecular_2018} or the use of single trajectories,\cite{Gabas_Protonatedglycinesupramolecular_2018} allows for a favorable scaling with the dimensionality of the considered molecules.

Given the relevance of protonated water clusters, both from the point
of view of experimental accuracy and theoretical challenge, the protonated
water dimer $\text{H}_{5}\text{O}{}_{2}^{+}$ (also known as the Zundel
cation),\cite{Zundel_EnergiebandertunnelndenUberschussProtonen_2011}
is a good test case for our purposes. This molecule features two bands
of high-frequency O-H stretching modes, that can be recovered, in
a semiclassical treatment, only with high-energy trajectories; it
displays strongly anharmonic dynamics for the shared proton, that
manifests itself in a distinctive proton transfer doublet. This feature
involves both proton transfer, wagging of the two water moieties,
and stretching of the two oxygens. The low-frequency barriers between
equivalent global minima,\cite{Wales_Rearrangementstunnelingsplittings_1999}
accessible via the wagging modes and internal torsion, render classical
trajectories particularly unstable, a property that presents a clear
challenge for theoretical methods in general and for semiclassical
ones in particular, because they rely on the evaluation of the stability
matrix to include quantum corrections. Moreover, the resulting enlarged
symmetry group of the molecule requires careful consideration in trajectory
sampling.

The Zundel cation is the most representative member of the family
of protonated water clusters, towards which many computational efforts
are being devoted, mainly motivated by a flourishing of experimental
results,\cite{miyazaki_mikami_largeprotonatewaterclusters_2004,Shin_InfraredSignatureStructures_2004,douberly_duncan_expsmallprotwaterclusters_2010,douberly_duncan_expsmallprotwaterclusters_2010,Thamer_Tokmakoff_2DIRZundel_2015,Wolke_Spectroscopicsnapshotsprotontransfer_2016,Fagiani_Gasphasevibrational_2016}
and the request for higher accuracy.\cite{singh_kim_magicacid_2006,agostini_ciccotti_smallprotonatedwaterclus_2011,Yu_Bowman_protonatedwater_2017,Yu_ClassicalThermostatedRing_2019,Egan_AssessingManyBodyEffects_2019}In
this respect, this molecule is a prototypical example that has
been tackled by various approaches, given the great biological relevance
of the charge transport mechanism in aqueous solutions.\cite{deGrotthuss_decompositioneaucorps_1806,Agmon_Grotthussmechanism_1995,Tuckerman_Parrinello_Abinitiotransport_1995,mohammed_Nibbering_waterbrideges_2005,marx_abinitio_200years_2006,Berkelbach_Tuckerman_concertedhydrogenbond_2009}
On the experimental side, the vibrational spectrum of the Zundel cation
has been investigated by infrared multiphoton photodissociation spectroscopy
\cite{Asmis_GasPhaseInfraredSpectrum_2003,fridgen2_maitre_expIRMPDzundel_2004}
and noble gases predissociation spectroscopy, in particular Argon
and Neon. \cite{Yeh_Vibrationalspectroscopyhydrated_1989,headrick_johnson_expartaggedzundel_2004,hammer_carter_expzundelspectrum_2005}

The theoretical literature about the vibrational spectrum of the Zundel
cation is quite vast, since its strong anharmonicity provides the
ideal test-bed for theoretical methods. The PES computed at the level
of coupled cluster theory and devised in Ref.~\onlinecite{huang_Bowman_ZundelPES_2005},
has been employed by a plethora of methods for vibrational calculations,
such as vibrational configuration interaction (VCI),\cite{Dai_Bowman_Multimode_Zundel_2003} diffusion Monte
Carlo,\cite{mccoy_bowman_VCIzundel_2005,hammer_carter_expzundelspectrum_2005}
classical molecular dynamics,\cite{Kaledin_VibrationalAnalysisH5O2_2006,kaledin_jordan_CPMDzundel_2009}
ring polymer molecular dynamics,\cite{huang_bowman_RPMDzundel_2008,rossi_manolopoulos_TRPDM_2014,rossi_ceriotti_TRPMDLangevin_2017}
and semiclassical methods.\cite{DiLiberto_Ceotto_Jacobiano_2018}
In a series of papers, the static and dynamical properties of the
Zundel cation have been studied with the Multi-Configuration Time-Dependent
Hartree (MCTDH) method, elucidating in particular the nature of the
proton-transfer doublet.\cite{vendrell_Meyer_ZundelHamiltonian_2007,vendrell_Meyer_Zundeldynamics_2007,vendrell_Meyer_Zundeldynamics_2007,vendrell_Meyer_zundelspectra_2007,vendrell_meyer_zundelquantumdynamics_2008,vendrell_Meyer_Jacobianparametriz_2009,vendrell_Meyer_isotopeeffects_2009,vendrell_Meyer_isotopeffects2_2009}
Ab initio molecular dynamics has been used to investigate the role
of tagging atoms in messenger spectroscopy.\cite{baer_mathias_AIMDzundel_2010}
Ref.~\onlinecite{Pitsevich_MP4StudyAnharmonic_2017} shows results from
perturbative theory, together with an extensive review of the literature.
Recently, effort is being devoted to studying static properties of
the protonated water dimer by new methods, employing on-the-fly coupled
cluster electronic structure, \cite{Spura_Ontheflycoupledcluster_2015,Spura_CorrectionOntheflycoupled_2015}
neural network potentials, \cite{Schran_Highdimensionalneuralnetwork_2017,Schran_ConvergedColoredNoise_2018}
and variational Monte Carlo.\cite{Dagrada_QuantumMonteCarlo_2014,Mouhat_FullyQuantumDescription_2017}

This paper describes a reduced rovibrational coupling Cartesian dynamics
approach for semiclassical calculations that we apply to the vibrational
spectrum of the Zundel cation, as a first step towards the study of bigger protonated water clusters. 
The standard way to perform semiclassical (SC) molecular dynamics
is using a normal mode coordinate framework determined by diagonalizing
the Hessian matrix at the optimized equilibrium geometry. However,
the numerical procedure is not free of ro-vibrational couplings and,
even if this approximation leads commonly to satisfactory outcomes,
it may be too drastic for small floppy molecules, where rovibrational
coupling is strong. In this paper, we use full Cartesian dynamics
and we analytically remove rovibrational coupling from initial Cartesian
conditions and from the normal modes used in the evaluation of 
wave-packet overlaps and stability matrix, finding this to be quite beneficial for the spectrum quality.
In particular, we are able to drop some of the approximations employed
in previous semiclassical calculations,\cite{DiLiberto_Ceotto_Jacobiano_2018}
by carefully reducing numerical noise which arises from the use of
normal-mode dynamics and non optimal rovibrational decoupling. Moreover, we determine
that a careful choice of the initial conditions, where no kinetic energy is given to the floppiest modes,\cite{DiLiberto_Divideandconquersemiclassicalmolecular_2018}
is necessary for an accurate determination of the frequencies of the higher energy modes in water systems. At variance with the standard phase-space sampling, which would prevent the convergence of the results, this approach is promising for larger water clusters as well. In addition, we show how our semiclassical approach can provide useful physical insight into the dynamics of the proton-transfer modes, when it is reduced to a single classical trajectory picture.

In Section \ref{sec:methods}, we describe in some detail 
the methodology used in this work, to ease the reproducibility of our results. More specifically, in Section \ref{subsec:propagator},
we introduce the semiclassical propagator in the Cartesian coherent
states set, we then define normal modes in Section \ref{subsec:cleannormal},
focusing on the analytical determination of infinitesimal translations
and rotations, and, in Section \ref{subsec:Semiclassical-normal-mode},
we introduce the time-averaging filter and the divide-and-conquer
semiclassical approach. In
Section \ref{subsec:referencestates}, we characterize the types of
reference states whose survival amplitude is to be Fourier transformed
for power spectrum evaluation. In Section \ref{subsec:sampling},
we explain the phase-space sampling of the initial conditions for
the classical trajectories. In Section \ref{sec:results}, we report
the results, regarding the stretching (Sec.~\ref{subsec:stretching}),
bending (Sec.~\ref{subsec:bending}), proton transfer (Sec.~\ref{subsec:transfer}),
proton perpendicular (Sec.~\ref{subsec:perp}), and O-O stretching
modes (Sec.~\ref{subsec:other}). In Section \ref{subsec:single},
we qualitatively analyze proton transfer by means of suitable trajectories.
In Section \ref{sec:conclusions}, we draw future perspectives. Appendix
\ref{app:coherent} recaps known results on coherent states and Appendix
\ref{app:cnorm} shows some details in the derivation of normal modes.

\section{Methods}

\label{sec:methods} 
In this work,
to simulate the Zundel cation, we employ the accurate PES by Huang
et al.,\cite{huang_Bowman_ZundelPES_2005} which was fitted to coupled
cluster level calculations. The kinetic nuclear energy of the $N$
atoms is evaluated in Cartesian coordinates, employing the bare nuclear
masses $m_{i}$, and the resulting Hamiltonian is $H(\bm{P},\bm{X})=\frac{1}{2}\sum_{i}P_{i}^{2}+V(\bm{X})$.
The Cartesian coordinates $\bm{x}=(x_{1x},x_{1y},x_{1z},\dots,x_{Nx},x_{Ny},x_{Nz})$
are mass scaled $X_{k}=x_{k}\sqrt{m_{k}}/\hbar$, where the index
$k$ indicates both the atom and the Cartesian axis. Correspondingly,
Cartesian momenta $P_{k}$ imply a factor $1/\sqrt{m_{k}}$. Moreover,
it is understood that time includes a factor $1/\hbar$, so that we
have energies and frequencies interchangeably.

We calculate the quantum vibrational spectral density of a molecular
system, described by the Hamilton operator $\hat{H}$, as the Fourier
transform of the survival amplitude $\braket{\chi|e^{-i\hat{H}t}|\chi}$
of a suitable reference state $\ket{\chi}$: 
\begin{equation}
I_{\chi}(E)\equiv\int_{-\infty}^{+\infty}\frac{dt}{2\pi}\braket{\chi|e^{-i\hat{H}t}|\chi}e^{iEt}\;.\label{eq:spectral_density}
\end{equation}
In Eq.~\eqref{eq:spectral_density} the spectral peak intensities $I_{\chi}(E)$
strongly depend on the reference state choice (to the point that they may be zero if the reference state is orthogonal to the eigenstate of interest), while their positions
are invariant.

\subsection{Semiclassical Cartesian propagator}

\label{subsec:propagator}To reduce the amount of rovibrational couplings
in our spectra calculations, we choose to perform symplectic classical
dynamics in Cartesian coordinates,\cite{McLachlan_accuracysymplecticintegrators_1992,Brewer_Manolopoulos_15dof_1997} and successively calculate the power
spectra in normal-mode coordinates using the semiclassical approximation.

Semiclassical theory approximates the exact quantum mechanical amplitude
by adopting a stationary phase approximation of the Feynman path integral,\cite{feynman_pathintegral_1965}
in the formal limit $\hbar\to0$, implying that the most contributing
paths are those obeying the classical equations of motion. The original
van Vleck formulation of the semiclassical propagator\cite{VanVleck_CorrespondencePrincipleStatistical_1928}
was made more practical via the Semiclassical Initial Value Representation
(SCIVR) theory introduced by Miller,\cite{Miller_Atom-Diatom_1970,miller2001semiclassical}
where a phase space integral over initial conditions $\left(\bm{P}_{0},\bm{X}_{0}\right)$
is performed, instead of a boundary conditions trajectory search.
From now on, a subscript $t$ indicates evolution up to time $t$
from the initial conditions, according to Hamilton equations.

Employing coherent states as proposed by Heller,\cite{Heller_FrozenGaussian_1981,Heller_SCspectroscopy_1981,Heller_Cellulardynamics_1991}
later developed by Herman and Kluk,\cite{Herman_Kluk_SCnonspreading_1984}
and settled on firmer ground by Kay,\cite{Kay_Numerical_1994,Kay_Integralexpression_1994,Kay_Multidim_1994,Kay_SCcorrections_2006}
the quantum propagator in semiclassical approximation is 
\begin{equation}
e^{-i\hat{H}t}=\int\frac{d\bm{P}_{0}d\bm{X}_{0}}{\left(2\pi\right)^{F}}C_{t}\left(\bm{P}_{0},\bm{X}_{0}\right)e^{iS_{t}}\ket{\bm{P}_{t},\bm{X}_{t}}\bra{\bm{P}_{0},\bm{X}_{0}},\label{eq:HHKK_propagator}
\end{equation}
where $F=3N$ and $S_{t}\equiv S_{t}(\bm{P}_{0},\bm{X}_{0})=\int_{0}^{t}dt^{\prime}\left[\frac{1}{2}\sum_{k}P_{t^{\prime}k}^{2}-V(\bm{X}_{t^{\prime}})\right]$
is the classical action of the trajectory, starting from $\left(\bm{P}_{0},\bm{X}_{0}\right)$.
The wavepackets $\ket{\bm{P}_{t},\bm{X}_{t}}$ are coherent states,
displaying a Gaussian shape, in both position and momentum representations,
and saturating the uncertainty bound, thus drawing a link between
the quantum and classical representations of atoms. See Appendix~\ref{app:coherent}
for a recall of basic properties of coherent states. Explicitly: 
\begin{equation}
\braket{\bm{X}|\underline{\bm{P}},\underline{\bm{X}}}=\left|\frac{\bm{\Gamma}}{\pi}\right|^{\frac{1}{4}}e^{-\frac{1}{2}(\bm{X}-\underline{\bm{X}})\bm{\Gamma}(\bm{X}-\underline{\bm{X}})+i\underline{\bm{P}}(\bm{X}-\underline{\bm{X}})}\label{eq:multicoherent_state}
\end{equation}
where $(\underline{\bm{P}},\underline{\bm{X}})$ parametrizes the center of the
Gaussian in the momentum and the position representations and $\bm{\Gamma}$
is a (in principle arbitrary) constant real symmetric positive-definite
matrix. Although a simple approach is to take a diagonal $\bm{\Gamma}$,
implying absence of correlation between Cartesian coordinates, it
is clear that considering a full non-sparse matrix opens up the possibility
of deep optimization of the convergence of Eq.~\eqref{eq:HHKK_propagator}.
We discuss this in detail in the next section.

The prefactor $C_{t}$ in Eq.~\eqref{eq:HHKK_propagator} can be
determined by imposing that the saddle point approximation
of Eq.~\eqref{eq:HHKK_propagator}, in the position
basis $\ket{\bm{X}}$, matches the van Vleck propagator.\cite{Kay_Integralexpression_1994}
$C_{t}$ depends on the full monodromy matrix 
\begin{equation}
\left(\begin{array}{cc}
\boldsymbol{M_{PP}} & \boldsymbol{M_{PX}}\\
\boldsymbol{M_{XP}} & \boldsymbol{M_{XX}}
\end{array}\right)\equiv\left(\begin{array}{cc}
\partial\bm{P}_{t}/\partial\bm{P}_{0} & \partial\bm{P}_{t}/\partial\bm{X}_{0}\\
\partial\bm{X}_{t}/\partial\bm{P}_{0} & \partial\bm{X}_{t}/\partial\bm{X}_{0}
\end{array}\right),\label{eq:monodromy_matrix}
\end{equation}
and the $\bm{\Gamma}$ matrix. The resulting expression is
\begin{equation}
C_{t}=
\left|\frac{1}{2}\left(\boldsymbol{M_{XX}}+\bm{\Gamma}^{-1}\boldsymbol{M_{PP}}\bm{\Gamma}+i\bm{\Gamma}^{-1}\boldsymbol{M_{PX}}-i\bm{M_{XP}}\bm{\Gamma}\right)\right|^{\frac{1}{2}}.\label{eq:pre-exponential_factor}
\end{equation}

The propagator in Eq.~\eqref{eq:HHKK_propagator} can require hundreds of thousands of classical trajectories to converge, when evaluated with Monte Carlo methods, even for relatively small molecules.\cite{Kay_Numerical_1994,Kay_Multidim_1994,Nandini_Church_Mixedqcl_2017,Antipov_Nandini_Mixedqcl_2015,Bonfanti_ComputationS1S0_2018}
To overcome this issue,
Kaledin and Miller\cite{Kaledin_Miller_Timeaveraging_2003,Kaledin_Miller_TAmolecules_2003}
proposed the following time-averaged version (TA SCIVR) of the spectral
density: 
\begin{equation}
I_{\chi}\left(E\right)=\frac{2\pi}{T}\int\frac{d\bm{P}_{0}d\bm{X}_{0}}{\left(2\pi\right)^{F}}\left|\int_{0}^{T}\frac{dt}{2\pi}e^{i\left(S_{t}+\varphi_{t}+Et\right)}\braket{\chi|\bm{P}_{t},\bm{X}_{t}}\right|^{2},\label{eq:TASCIVR_spectral_density}
\end{equation}
where the separable approximation is employed, namely only the complex
phase $\varphi_{t}(\bm{P}_{0},\bm{X}_{0})=\arg{\left[C_{t}(\bm{P}_{0},\bm{X}_{0})\right]}$
of the prefactor is retained, and $T$ is the total duration of the
classical trajectories. The time-averaging procedure (in separable
approximation) acts as a filter on rapidly oscillating phase contributions,
thus strongly dampening noise in the resulting spectra, while still
retaining accuracy on the position of the spectral peaks. Within this
formalism it was possible to reproduce vibrational spectra of small
molecules by evolving roughly only one thousand classical trajectories
per degree of freedom,\cite{Kaledin_Miller_Timeaveraging_2003,Kaledin_Miller_TAmolecules_2003,Tamascelli_Ceotto_GPU_2014,DiLiberto_Ceotto_Prefactors_2016,Zhuang_Ceotto_Hessianapprox_2012,Ceotto_Hase_AcceleratedSC_2013}
 also demonstrating that it does not suffer from ZPE leakage,\cite{buchholz_ivanov_2018_ZPEL} and that the cost of evaluating the Hessian can be reduced by employing a database.\cite{Conte_HessianDatabase_2019}
With a careful choice of initial conditions, it is even possible to employ a single classical trajectory per sought spectral peak, via the Multiple Coherent State (MC SCIVR) approach.\cite{Ceotto_AspuruGuzik_PCCPFirstprinciples_2009,Ceotto_AspuruGuzik_Multiplecoherent_2009,Ceotto_AspuruGuzik_Curseofdimensionality_2011,Ceotto_AspuruGuzik_Firstprinciples_2011,Ceotto_Tantardini_Copper100_2010,Conte_Ceotto_NH3_2013,Gabas_Ceotto_Glycine_2017,Buchholz_Ceotto_MixedSC_2016,Buchholz_Ceotto_applicationMixed_2017,Ceotto_Buchholz_SAM_2018} Impressive results have also been obtained with the Thawed Gaussian approach.\cite{Wehrle_Vanicek_Oligothiophenes_2014,Wehrle_Vanicek_NH3_2015,Begusic_SingleHessianthawedGaussian_2019}
Due to the floppy nature of the Zundel cation, in this work we focus on phase-space integration, that validates our
use of single trajectories in Sec~\ref{subsec:single} for a qualitative study of proton transfer.

\subsection{Roto-translational modes orthonormalization}

\label{subsec:cleannormal}
Even if we perform Cartesian dynamics, we choose to introduce normal modes in the specific choice of the $\bm{\Gamma}$ and $\bm{M}$ matrices. This increases efficiency and allows for a direct term of comparison with classical
normal mode analysis and classification. Normal modes are defined as
$q_{l}=\sum_{j}L_{lj}^{T}\delta X_{j}=\sum_{j}L_{jl}\delta X_{j}$, and conversely
$\delta{X}_{j}=\sum_{l}L_{jl}q_{l}$, namely they are
linear combinations of displacements of mass-scaled Cartesian coordinates
$\delta\bm{X}=\bm{X}-\bm{X}^{eq}$ from the (typically global) equilibrium
molecular geometry $\bm{X}^{eq}$. Analogously, normal momenta
result in $p_{l}=\sum_{j}L_{jl}P_{j}$. The $\bm{L}$ matrix is orthogonal
$\bm{L}^{-1}=\bm{L}^{T}$ and is determined by diagonalizing the (mass-scaled)
Hessian $h_{jk}={\partial^{2}V(\bm{X})}/{\partial X_{j}\partial X_{k}}=\sum_{l}L_{jl}\Omega_{l}L_{lk}^{T}$
of the potential evaluated at $\bm{X}^{eq}$. The diagonal $\bm{\Omega}$ matrix is conventionally ordered
by increasing positive eigenvalues. 
However, due to global rotational and translational invariance, the Hessian at the minimum should display 
 3 null eigenvalues for translations, and 3 (or 2 for linear geometries) null
eigenvalues pertaining to linearized rigid rotations. However, such
eigenvalues are not found to be exactly zero, due to the used
finite differences algorithm, and numerical
precision in the geometry optimization and in the diagonalization
routines. Moreover, the corresponding rows of $\bm{L}^{T}$ typically
result in an arbitrary combination of translations and infinitesimal
rotations, due to their near degeneracy. In this work, we analytically
determine such rows, in the center-of-mass and principal-axes frame. We conventionally assign the roto-translational
normal modes to the last 6 rows of $\bm{L}^{T}$, and obtain (deferring
details to Appendix~\ref{app:cnorm}):\cite{Eckart_StudiesConcerningRotating_1935,wilson1980molecular,Miller_ReactionpathHamiltonian_1980,Jellinek_SeparationEnergyOverall_1989}
\begin{equation}
L_{F-3+\alpha,k\beta}^{T}=\frac{\delta_{\beta,\alpha}\sqrt{m_{k}}}{\sqrt{\sum_{j}m_{j}}}\;,\label{eq:cnormT}
\end{equation}
for the translational modes, where $\alpha=1,2,3$ refer to the $x,y,z$
axes, respectively, and we render the coordinate index $\beta=1,2,3$,
for atom $k$ explicit. For the infinitesimal rotational modes around
the reference geometry, we obtain 
\begin{equation}
L_{F-6+\alpha,k\gamma}^{T}=\frac{\sum_{\beta}\epsilon_{\alpha\beta\gamma}X_{k\beta}^{eq}}{\sqrt{\sum_{j}\left[-(X_{j\alpha}^{eq})^{2}+\sum_{\beta}(X_{j\beta}^{eq})^{2}\right]}}\label{eq:cnormR}
\end{equation}
with $\alpha=1,2,3$, where $\epsilon_{\alpha\beta\gamma}$ is the
Levi-Civita symbol. We enforce the other rows of $\bm{L}^{T}$, pertaining
to internal vibrations, to be orthonormal with each other and with
the roto-translational modes via a Gram-Schmidt procedure.

In the last equation, it is important to notice that we use the coordinates of the reference geometry, since, for efficiency, we want a constant-in-time $\bm{L}$ matrix and we are linearizing
the rotational coordinates at that specific configuration. We enforce
such analytical orthonormalization, because, when performing normal-mode
dynamics,\cite{Kaledin_Miller_Timeaveraging_2003,Ceotto_AspuruGuzik_Curseofdimensionality_2011}
that ignores Watson's coupling between vibrations and global rotations,\cite{Watson_Simplificationmolecularvibrationrotation_1968,wilson1980molecular}
the use of these infinitesimal rotational coordinates, referred to
$\bm{X}^{eq}$, yields small errors for small vibrations, which are
often neglected (see, for example, Ref.~\onlinecite{Avila_Carrington_C2H4_2011}, for a complete treatment).
Symplectic Cartesian dynamics has comparatively the advantage that angular
momentum is exactly conserved and the kinetic term has its simplest
form. Since we remove angular momentum at the beginning of the trajectories,
this is zero along the symplectic dynamics, except for numerical accuracy
errors, which could be removed at each step.\cite{KumarP_Understandinghydrogenscrambling_2006}
We found indeed that the precision of the Zundel cation semiclassical
spectrum was much refined when employing Cartesian dynamics rather
than normal-mode dynamics.

\subsection{Normal mode power spectra formulation}

\label{subsec:Semiclassical-normal-mode}To derive the semiclassical
normal-mode expression for the vibrational density of states calculation,
we choose the $\bm{\Gamma}$ matrix of the widths of the employed
coherent states in Eq.~\eqref{eq:multicoherent_state} to be the
optimal one in the quadratic approximation: 
\begin{equation}
\Gamma_{kj}=\sum_{l}L_{kl}\omega_{l}L_{lj}^{T}\;,\label{eq:gamma}
\end{equation}
where $\omega_{l}=\sqrt{\Omega_{l}}$ for the first $N_{v}=3N-6$
vibrational modes. The eigenvalues of the Hessian for the roto-translational modes are zero and cannot thus provide suitable widths, which we temporarily set at arbitrary positive values. Our divide-and-conquer approach, described below, will allow us to prevent them to affect the results. We also adopt the matrix notation $\bm{\omega}\equiv\text{diag}(\omega_{1},\dots,\omega_{F})$.

By choosing this specific expression for $\bm{\Gamma}$, one is able
to draw a direct relation between approaches employing normal coordinates
only,\cite{Kaledin_Miller_Timeaveraging_2003,Ceotto_AspuruGuzik_PCCPFirstprinciples_2009,Ceotto_AspuruGuzik_Multiplecoherent_2009,Ceotto_Tantardini_Copper100_2010,Ceotto_AspuruGuzik_Curseofdimensionality_2011,Ceotto_AspuruGuzik_Firstprinciples_2011}
and those expressed in Cartesian coordinates.\cite{Harland_Roy_SCIVRconstrained_2003,Issack_Geometricconstraintssemiclassical_2005,Issack_Geometricconstraintssemiclassical_2007,Issack_Semiclassicalinitialvalue_2007,Wong_Roy_Formaldehyde_2011}
While the potential in the classical action is evaluated in Cartesian
notation, since the PES is available in Cartesian coordinates, all
other elements composing Eq.~\eqref{eq:HHKK_propagator} are easily
converted from Cartesian to normal coordinates, using Eq.~\eqref{eq:gamma}. 
Since $\bm{L}$ is orthogonal, the Jacobian
of the transformation from Cartesian to normal coordinates is unity,
so $d\bm{P}_{0}d\bm{X}_{0}=d\bm{p}_{0}d\bm{q}_{0}$. The coherent
states in normal coordinate representation are 
\begin{equation}
\braket{\bm{q}|\underline{\bm{p}},\underline{\bm{q}}}\equiv\left|\frac{\bm{\omega}}{\pi}\right|^{\frac{1}{4}}e^{\sum_{l}^{F}\left[-\frac{\omega_{l}}{2}(q_{l}-\underline{q}_{l})^{2}+i\underline{p}_{l}(q_{l}-\underline{q}_{l})\right]}=\braket{\bm{X}|\underline{\bm{P}},\underline{\bm{X}}}\label{eq:normal_multicoherent_state}
\end{equation}
and they are centered in $(\underline{\bm{p}},\underline{\bm{q}})\equiv(\bm{L}^{T}\underline{\bm{P}},\bm{L}^{T}\delta\underline{\bm{X}})$.
Thanks to the product property of determinants, for a generic matrix
$\bm{{A}}$ we have $\det{\bm{A}}=\det{(\bm{L}^{T}\bm{A}\bm{L})}$,
and the prefactor is simply transformed to: 
\begin{equation}
C_{t}=\left|\frac{1}{2}\left(\boldsymbol{M_{qq}}+\bm{\omega}^{-1}\boldsymbol{M_{pp}}\bm{\omega}+i\bm{\omega}^{-1}\boldsymbol{M_{pq}}-i\boldsymbol{M_{qp}}\bm{\omega}\right)\right|^{\frac{1}{2}},\label{eq:pre-exponential_factor_normal}
\end{equation}
where the notation for $\bm{M}$ is analogous to Eq.~\eqref{eq:monodromy_matrix}. Notice
that the $\bm{\omega}$
matrices multiplying $\boldsymbol{M_{pp}}$ in general do not simplify,
even though they are diagonal, since they are not uniform along the
diagonal.

Recently, some of us have proposed the divide-and-conquer semiclassical
initial value representation method, DC SCIVR, that allows to recover
vibrational power spectra of high-dimensional molecules, as well as
complex systems, like water clusters, protonated glycine molecules
and nucleobases.\cite{ceotto_conte_DCSCIVR_2017,DiLiberto_Ceotto_Jacobiano_2018,DiLiberto_Divideandconquersemiclassicalmolecular_2018,Gabas_Protonatedglycinesupramolecular_2018,Gabas_nucleobases2019}
The very basic idea of this method is to exploit the usual full-dimensional
dynamics, but applying the semiclassical formalism each time to a
subspace $\mathcal{S}$ of reduced dimensionality $\tilde{F}$, to enhance the Fourier
signal pertaining to the states of interest. The sum of the spectra
of each subspace provides the full-dimensional spectrum.\cite{ceotto_conte_DCSCIVR_2017} A related method was devised in Ref.~\onlinecite{Wehrle_Vanicek_Oligothiophenes_2014}. We denote quantities projected to the subspace by $\sim$. In our case, we only consider subspaces made of collections of normal
modes: in practice, this results in the action of the projection
simply being the removal of rows and columns pertaining to excluded
modes. The working DC SCIVR formula is then:
\begin{equation}
\tilde{I}_{\chi}\left(E\right)=\frac{2\pi}{T}\int\frac{d\tilde{\bm{p}}_{0}d\tilde{\bm{q}}_{0}}{(2\pi)^{\tilde{F}}}\left|\int_{0}^{T}\frac{dt}{2\pi}e^{i\left(\tilde{S}_{t}+\tilde{\phi}_{t}+Et\right)}\braket{\tilde{\chi}|\tilde{\bm{p}}_{t},\tilde{\bm{q}}_{t}}\right|^{2}.\label{eq:DC-SCIVR_spectral_density}
\end{equation}

Coherent states can be straightforwardly projected as $\ket{\tilde{\bm{p}},\tilde{\bm{q}}}=\prod_{l\in\mathcal{S}}\ket{p_{l},q_{l}}$.
Analogously, the reference state $\ket{\tilde{\chi}}$ is defined
only in the subspace. Employing the $\tilde{\bm{M}}_{\bm{ij}}$
sub-blocks, the pre-exponential factor is analogous to Eq.~\eqref{eq:pre-exponential_factor_normal}.
Notice that we project the monodromy matrix $\bm{M}$ onto the subspace
only after evolving its full-dimensional version. This would be equivalent
to only evolving the subspace monodromy matrix $\tilde{\bm{M}}$,
only in the case of complete decoupling.

The most delicate part, within the DC method, is the calculation of
the projected action, since, for a non-separable potential, the exact
projected potential is in general unknown. While the kinetic term
is obtained by only considering the momenta projected into $\mathcal{S}$,
a suitable choice for an effective potential,\cite{ceotto_conte_DCSCIVR_2017}
which is exact in the separability limit, is $\tilde{V}\left(\tilde{\bm{q}}\right)\equiv V\left(\bm{q}\right)-V\left(\bm{q}_{\mathcal{S}}^{eq};\bm{q}_{\bar{\mathcal{S}}}\right)$,
where, from the full potential at the current configuration, we remove
the potential due to modes belonging to the complementary subspace $\bar{\mathcal{S}}$, while
modes in $\mathcal{S}$ are set at equilibrium.

The phase-space integration in Eq.~\eqref{eq:DC-SCIVR_spectral_density}
is reduced to the degrees of freedom of the subspace, while the other
modes are set initially at their equilibrium geometry position, and
mass-scaled momenta corresponding to their harmonic ZPE,
$\bar{p}_{l}=\sqrt{\omega_{l}}$. The subspaces are chosen in order
to collect together strongly interacting modes and the partition is
devised by taking advantage either of a time-averaged Hessian matrix
along trial trajectories or by looking at the conservation of Liouville
theorem.\cite{ceotto_conte_DCSCIVR_2017,DiLiberto_Ceotto_Jacobiano_2018}
In this work, we always project away the global translational and rotational modes, thus removing any dependence on their arbitrary width (a symbol $\sim$ is understood in all the following definitions). 
This is crucial for avoiding that spurious rotational peaks appear in the spectra. The vibrational modes 
are instead collected in a single 15-dimensional subspace that is used for the evaluation 
of the overlaps, the action and the prefactor. Further partitioning is used only 
for trajectory sampling, as described in Sec.~\ref{subsec:sampling}.

\subsection{Choice of reference states}

\label{subsec:referencestates}

Although the position $E_{n}$ of the peaks in the spectra does not
depend on the reference state $\ket{\chi}$, their height is directly
related to the overlap $c_{n}$ of $\ket{\chi}$ with the corresponding
vibrational eigenstates $\ket{n}$ of the system, namely $I_{\chi}(E)=\sum_{n}|c_{n}|^{2}\delta(E-E_{n})$.
The choice of the reference state is then crucial in obtaining a high
signal-to-noise ratio, and in the correct assignment of the peaks.

In this work we investigate four types of reference states, and show that they can portray useful complementary information: i) (anti)symmetrized
coherent states of normal modes; ii) harmonic states of normal modes;
iii) Cartesian superpositions of harmonic states; and iv) harmonic
states symmetrized according to different molecular symmetric configurations.
As shown below, their implementation is simple, and with a single simulation one can simultaneously evaluate their corresponding correlation functions. Also, their allow for a direct physical insight of one vibrational peak at a time.

i) For harmonic systems, the coherent reference state $\ket{p_{l}^{eq},q_{l}^{eq}}$
of a single mode $l$ in its equilibrium position $q^{eq}=0$ yields
a signal for all spectral peaks, with a height that is most pronounced
at $E\approx(p_{l}^{eq})^{2}/2$, so it is beneficial to choose $(p_{l}^{eq})^{2}=(2n_{l}+1)\omega_{i}$,
when one is interested in the $n_{l}$-th state of mode $l$.\cite{Ceotto_AspuruGuzik_Multiplecoherent_2009}
For anharmonic systems, this harmonic prescription is still efficient
because of the Gaussian delocalization. When phase-space integration
is very computationally demanding, a single coherent reference state
can be used, where all $p_{l}^{eq}$ are set to their harmonic ZPE
value $p_{l}^{eq}=\sqrt{\omega_{l}}$. A more precise characterization
of peaks can be obtained by taking combinations of coherent states
that reproduce relevant symmetries. For example, one can select different
parities related to even/odd harmonic states by considering a superposition
of the following, unnormalized, form:\cite{Kaledin_Miller_TAmolecules_2003,Kaledin_Miller_Timeaveraging_2003,Ceotto_AspuruGuzik_Curseofdimensionality_2011,Conte_Ceotto_NH3_2013}
\begin{equation}
\ket{\chi}=\prod_{l=1}^{F}\left(\ket{p_{l}^{eq},q_{l}^{eq}}+\varepsilon_{l}\ket{-p_{l}^{eq},q_{l}^{eq}}\right)\;.\label{eq:antysimmetric_state}
\end{equation}
By setting $\varepsilon_{l}=1$ for each mode, the ZPE signal (and
even overtones) is enhanced, while setting $\varepsilon_{l}=-1$ for
the l-th degree of freedom, selects its fundamental excitation (and
odd overtones). In the latter case, an even better signal is obtained
if the reference momentum of the l-th mode is set to its harmonic
value $p_{l}^{eq}=\sqrt{3\omega_{l}}$.

ii) Although the semiclassical representation of the propagator is
expanded on a coherent basis set, it may be useful to use harmonic
reference states.\cite{Micciarelli_Anharmonicvibrationaleigenfunctions_2018}
These states are particularly advantageous, when considering multiple excited
states. These multiple harmonic states are more convenient than antisymmetric
combinations of coherent states, because they provide a better defined signal onto the states of interest. By exploiting the property that
the coherent states of the l-th normal mode are eigenstates of the
destruction operator $\hat{a}_{l}=\frac{1}{\sqrt{2}}\left(\sqrt{\omega_{l}}\hat{q}_{l}+i\frac{\hat{p}_{l}}{\sqrt{\omega_{l}}}\right)$,
it is immediate to get the following standard result for the overlap
between a harmonic reference state $\ket{l_{n_{l}}}$, where $n_{l}$
is the excited state quantum number, and the running coherent state
$\ket{p_{l}^{\alpha},q_{l}^{\alpha}}$: 
\begin{equation}
\braket{l_{n_{l}}|p_{l}^{\alpha},q_{l}^{\alpha}}=\braket{0|p_{l}^{\alpha},q_{l}^{\alpha}}\frac{\alpha_{l}^{n_{l}}}{\sqrt{n_{l}!}}\;,
\end{equation}
where $\braket{0|p_{l}^{\alpha},q_{l}^{\alpha}}=\exp{\left(-\frac{\omega_{l}(q_{l}^{\alpha})^{2}}{4}-\frac{(p_{l}^{\alpha})^{2}}{4\omega_{l}}-\frac{ip_{l}^{\alpha}q_{l}^{\alpha}}{2}\right)}$
and $\alpha_{l}=\left(\sqrt{\omega_{l}}q_{l}^{\alpha}+ip_{l}^{\alpha}/\sqrt{\omega_{l}}\right)/\sqrt{2}$.

iii) The third class of reference states that we consider corresponds
to the states resulting from the application of a nuclear Cartesian
coordinate of interest to the harmonic normal-mode ground state, i.e.
$\ket{\chi}=\hat{x}_{j\gamma}\ket{\bm{0}}$.\cite{vendrell_Meyer_zundelspectra_2007}
This reference state highlights multiple spectral peaks corresponding
to the displacement of that Cartesian coordinate, and it is useful
for considering the contribution to that displacement from all normal
modes, mostly in their fundamental excitations. Of course, it is also related to an element of the nuclear
dipole-dipole correlation function. In this work, we consider, in
particular, the projection of the position of the shared proton on
the axis connecting the oxygen atoms, which is conventionally called
$z$. Close to the reference geometry, this projection may be approximated
by $z=x_{H,z}-(x_{O_{1},z}+x_{O_{2},z})/2$, where $\bm{x}_{H}$ is
the position of the shared proton, and $\bm{x}_{O_{i}}$is the position
of the i-th oxygen nucleus. The projection is obtained by observing
that, after Cartesian coordinates are expanded onto normal modes,
we can use the standard result (see Appendix~\ref{app:coherent}): 
\begin{equation}
\braket{0|\hat{q}_{l}|p_{l}^{\alpha},q_{l}^{\alpha}}=\left(\frac{q_{l}^{\alpha}}{2}+i\frac{p_{l}^{\alpha}}{2\omega_{l}}\right)\braket{0|p_{l}^{\alpha},q_{l}^{\alpha}}\;,
\end{equation}
and we trace back to the previous harmonic-state case. Since in principle
all normal modes are necessary, to reconstruct the full $z$ coordinate,
in the DC approach either one considers the full-dimensional set of
normal modes, or the coordinate is expanded only onto a subset of
normal modes.

iv) The last class of states that we consider is specific to fluxional
molecules, where different versions of the reference geometry, related
by global rotations, reflections and permutations, are relevant. These
states are a suitable combination of harmonic states and we describe
them in detail in Sec.~\ref{subsec:perp}, where we apply them to
the study of the perpendicular motion of the shared proton.

\subsection{Trajectory length and phase space sampling criteria}

\label{subsec:sampling} Since we perform a Fourier transform of the
survival amplitude, to get the vibrational spectra, there is an intrinsic
width $\pi/T$ in the spectral peaks, depending on the total time
of the trajectories $T$. One would then aim at evolving long trajectories,
to reduce the peak width. When complex systems are under investigation,
however, some monodromy matrix eigenvalues increase exponentially
along the dynamics, causing issues in the evaluation of the pre-exponential
factor $C_{t}$, and, consequently, on the spectral density in Eq.~\eqref{eq:TASCIVR_spectral_density}.
A number of approaches has been devised to tackle this issue, including
the use of approximate $C_{t}$,\cite{DiLiberto_Ceotto_Prefactors_2016}
or the use of the original Eq.~\eqref{eq:pre-exponential_factor_normal}
and rejecting the trajectories such that $\left|1-\left|\bm{M}^{T}\bm{M}\right|\right|\geq\epsilon$,
with the arbitrary threshold $\epsilon$ usually in the range 10$^{-5}$--10$^{-3}$.

The drawback of this approach is that, if the rejection rate is higher
than 90\%, there is an order of magnitude ratio between the propagated
trajectories and those effectively contributing to Eq.~\eqref{eq:TASCIVR_spectral_density}.
In the case of the Zundel cation, we typically consider trajectories
as long as $T=2\cdot10^{4}\text{au}$ ($0.5\text{ps}$), a duration
that corresponds to a Fourier width of $35\text{cm}^{-1}$ (a resolution
analog to the one of the MCTDH calculations in Ref.~\onlinecite{vendrell_Meyer_zundelspectra_2007}).
We found that this would typically correspond to a rejection rate
higher than 95\%, making almost unfeasible to converge Eq.~\eqref{eq:TASCIVR_spectral_density}.

In this work, we aim at obtaining the best possible performance of
the TA SCIVR method applied to the Zundel cation, while still retaining
the original Herman-Kluk prefactor within the separable approximation.
We then modify an approach by Kay,\cite{Kay_Multidim_1994} to grasp
all the possible information by each trajectory run, before they become
too much chaotic. In performing the initial representation phase-space
integral, the contribution of each trajectory is accounted for by
a weight $w$, depending on the time $T_{\epsilon}$ at which the
$\epsilon$ threshold is crossed, and defined as: 
\begin{equation}
w=\begin{cases}
0, & \text{if}\ T_{\epsilon}<T_{m}/2\\
\left(\frac{T_{\epsilon}}{T_{m}/2}-1\right)^{2} & \text{otherwise}
\end{cases}\label{eq:weigth}
\end{equation}
where $T_{m}$ is the duration of the longest non-chaotic trajectory.
The total spectrum is then the weighted average of the spectra corresponding
to all trajectories. This strategy allows to significantly increase
the number of contributing trajectories (albeit shorter than the longest
ones). Too short trajectories, which yield a broad contribution to
the spectrum, are not contributing anyway. We observed that the main effect of this approach (also on smaller molecules such as methane, not shown here) is to smoothen the resulting spectra, without the need of a damping factor, while the position of the spectral peaks is not affected, being dominated by the longest trajectories.

These improvements in the TA SCIVR methodology allow
us to employ an hybrid approach between the full-dimensional and the
DC methods. On the one hand, we retain the full vibrational subspace $\mathcal{S}$, 
projecting away only the global rotations and translations, when evaluating the prefactor, 
action and overlaps, like in the standard TA SCIVR method.
On the other hand, we restrict the initial phase-space sampling to subspaces $\mathcal{S}^\prime\subset\mathcal{S}$ of normal modes, depending on the vibrational states of interest, like in the DC SCIVR method. 
In particular, we find it important to assign initial
zero momentum to modes outside the considered subspaces $\mathcal{S}^\prime$, especially the low-frequency ones, that correspond to torsion, wagging and rocking. This prescription
is crucial in order to remove the appearance of secondary
peaks in the spectra,\cite{DiLiberto_Divideandconquersemiclassicalmolecular_2018} 
which would naturally occur due to coupling.
In this way, we can also avoid to introduce a damping factor in the
Fourier transform, which would produce artificial broadening of the
spectral features. To justify this approach, one has to consider
that classical dynamics transfers energy also to such modes, but not
sufficiently so as to introduce noise in the resulting spectra,
and that the harmonic estimate is far above the actual ZPE. Notice, moreover, 
that the typical classical energy of the trajectories that we sample is of the 
order of $1000\text{cm}^{-1}\approx1500K$,
depending on the considered normal modes subspace, and would still
correspond, in a classical molecular dynamics simulation, to very
high temperatures. This explains why the anharmonic part of the potential
is explored, even when we adopt the partial sampling procedure described
in this section. This also indicates that, in the Zundel cation case, the crucial benefit of the DC approach used in Ref.~\onlinecite{DiLiberto_Ceotto_Jacobiano_2018}
was not the projection \emph{per se}, but the careful choice of the
initial conditions for the modes weakly involved in the spectral peaks
of interest. This approach should be considered the new standard for the semiclassical study of water systems. 

When performing integration of the initial phase-space coordinates
in Eq.~\eqref{eq:DC-SCIVR_spectral_density} for modes in a subspace $\mathcal{S}^\prime$,
we employ a Monte Carlo method with importance sampling. The roto-translational 
modes are set at $q_{l,0}=0$ and $p_{l,0}=0$. Modes in the full vibrational subspace $\mathcal{S}$ are indicated by $\tilde{q}_l$, while those belonging to the sampling subset $\mathcal{S}^\prime$ by $\bar{q}_l$. Vibrational modes in $\mathcal{S}$, but not in $\mathcal{S}^\prime$, are initially set at momenta equal to zero or to the ZPE prescription $p_{l,0}=\sqrt{\omega_l}$. We consider
the distribution $|\braket{\tilde{\chi}|\tilde{\bm{p}}_{0},\tilde{\bm{q}}_{0}}|^{2}$
at time $t=0$, which contains the factor 
\begin{equation}
g_{q}(\bar{\bm{q}}_{0})=\prod_{l\in\mathcal{S}^\prime}\exp{\left(-\omega_{l}\bar{q}_{0,l}^{2}/2\right)}\;,\label{eq:sampleq}
\end{equation}
since $\bm{q}^{eq}=0$, that we use as a distribution for $\bar{\bm{q}}_0$. When the reference state is a coherent one, in addition one analogously
gets the following sampling factor: 
\begin{equation}
g_{p}(\bar{\bm{p}}_{0})=\prod_{l\in\mathcal{S}^\prime}\exp{\left(-\left(\bar{p}_{0,l}-\bar{p}_{l}^{eq}\right)^{2}/(2\omega_{l})\right)}\;,\label{eq:samplep}
\end{equation}
that we use as a distribution for $\bar{\bm{p}}_0$. 
When considering harmonic reference states, we observe that $(\hat{a}_{l}^{\dagger})^{n}\ket{0}$
also has a significant overlap with the coherent state $\ket{p_{l}^{eq},q_{l}^{eq}}$,
with $p_{l}^{eq}=\sqrt{(2n+1)\omega_{l}}$. The integral in Eq.~\eqref{eq:DC-SCIVR_spectral_density}
is thus estimated with 
\begin{equation}
\tilde{I}_{\chi}\left(E\right)\approx\frac{\mathcal{N}}{\sum_{j}w_{j}}\sum_{j}^{N_{T}}w_{j}\left|\sum_{t}^{T_{\epsilon}^{j}}e^{i\left(\tilde{S}_{t}^{j}+\tilde{\phi}_{t}^{j}+Et\right)}\frac{\braket{\tilde{\chi}|\tilde{\bm{p}}_{t}^{j},\tilde{\bm{q}}_{t}^{j}}}{\sqrt{g_{q}^{j}g_{p}^{j}}}\right|^{2},\label{eq:spectrumformula}
\end{equation}
where $\mathcal{N}$ is a normalization factor, $N_{T}$ is the number of trajectories, $j$ is their index and correspondingly $g_{q}^{j}g_{p}^{j}\equiv g_{q}(\bar{\bm{q}}^{j}_{0})g_{p}(\bar{\bm{p}}^{j}_{0})$, while $w_{j}$ is the weight according to Eq.~\eqref{eq:weigth}
for a time evolution $T_{\epsilon}^{j}$. Notice, again, that the overlaps, the action and the pre-exponential factor phase are evaluated using all the vibrational modes $\tilde{\bm{p}}, \tilde{\bm{q}}$ at time $t$. The Monte Carlo uncertainty can be evaluated from the variance of the above expression, using the sum of the weights as a proxy for an effective number of independent trajectories. We employ the timestep
$\Delta t=10\text{au}$ ($0.25\text{fs}$) and the threshold $\epsilon=10^{-3}$.

\section{Results}

\label{sec:results}

\begin{table*}[bth]
\setlength{\tabcolsep}{6pt} \centering %
\begin{tabular}{lcc|ccccc}
\hline 
Description & Symbol & Normal mode & HO & Expt.\cite{hammer_carter_expzundelspectrum_2005} & VCI(DMC)\cite{mccoy_bowman_VCIzundel_2005,hammer_carter_expzundelspectrum_2005} & MCTDH\cite{vendrell_Meyer_zundelspectra_2007,vendrell_Meyer_isotopeffects2_2009} & This work\tabularnewline
\hline 
wagging & $w_{3}$ & $2_{1}$ & 338 &  &  & 374(386)  & 358\tabularnewline
O-O stretching  & $1R$  & $6_{1}$  & 630  &  &  & 550 & 580\tabularnewline
 & $2R$  & $6_{2}$  & 1260  &  &  & 1069  & 1124\tabularnewline
transfer low  & $(w_{3},1R)$  & $(2_{1},6_{1})$  & 968  & 928  &  & 918(913)  & 891\tabularnewline
\phantom{transfer} high  & $z$  & $7_{1}$  & 861  & 1047  & 1070(995)  & 1033(1050)  & 1062\tabularnewline
proton perp.  & $y,x$  & $8_{1},9_{1}$  & 1494,1574  &  &  & 1391  & 1453\tabularnewline
bending modes  &  &  &  &  &  &  & \tabularnewline
\phantom{bending} gerade  & bg  & $10_{1}$  & 1720  &  & 1604  & 1606  & 1678\tabularnewline
\phantom{bending} ungerade  & bu  & $11_{1}$  & 1770  & 1763  & 1781  & 1741(1756)  & 1751\tabularnewline
O-H stretch &  &  &  &  &  &  & \tabularnewline
\phantom{1} s(m)as(d)  & s-as  & $12_{1}$  & 3744  & 3603  & 3610(3511)  & 3607  & 3607\tabularnewline
\phantom{1} s(m)s(d)  & s-s  & $13_{1}$  & 3750  &  & 3625(3553)  & 3614(3618)  & 3609\tabularnewline
\phantom{1} as(m)as(d),as(m)s(d)  & as  & $14_{1},15_{1}$  & 3832  & 3683  & 3698(3652) & 3689(3680) & 3679,3690\tabularnewline
\hline 
\end{tabular}\caption{Semiclassical vibrational energy levels of the Zundel cation. 
Direct comparison to available MCTDH levels yields $\text{MAE}=29\text{cm}^{-1}$.
Notice that both the TA SCIVR and the MCTDH\cite{vendrell_Meyer_zundelspectra_2007}
methods have a similar Fourier resolution of about $30\text{cm}^{-1}$.
The results in this work, compared to available experimental data, yield
a MAE of $14\text{cm}^{-1}$, while the MAE of the MCTDH results reported
in Ref.~\onlinecite{vendrell_Meyer_zundelspectra_2007} is $11\text{cm}^{-1}$.
See text for a critical assessment of other systematic errors. Symbols
in parentheses indicate combined excitations, while lists of results
separated by commas indicate alternate estimations of spectral peaks
using different reference states or trajectory sampling.}
\label{tab:peaks} 
\end{table*}

In this Section, we describe the results obtained for the vibrational
spectrum of the Zundel cation, using the hybrid full-dimensional/DC
SCIVR approach described above. Although we evaluate the full pre-exponential
factor, action and overlaps (excluded global translational and rotational
modes), we sample phase space according to the subspace partitioning
introduced in Ref.~\onlinecite{DiLiberto_Ceotto_Jacobiano_2018} for the
Zundel cation, where the magnitude of the off-diagonal Hessian elements
along a representative trajectory was monitored. In each following
Section, we indicate the corresponding sampling subspace $\mathcal{S}^\prime$. Results are
typically obtained with $12000$ sampled trajectories, which are sufficient
to reach convergence for the positions of the peaks with a Monte Carlo uncertainty
of $10\text{cm}^{-1}$, which is lower than the typical Fourier width
of $35\text{cm}^{-1}$ and the typical accuracy of SCIVR methods which
is $\sim20\text{cm}^{-1}$.\cite{Ma_quantummechanicalinsight_2018}
We also draw a gray error band behind the spectral profiles in the
figures, indicating the estimation of the standard deviation of the
Monte Carlo mean evaluated by Eq.~\eqref{eq:spectrumformula}, conditioned to the 
choice that modes belonging to the complementary subspaces $\bar{\mathcal{S}}^\prime$
are initialized at their equilibrium positions and with momenta corresponding to either 
their harmonic ZPE or to zero. Spectra
are shifted with respect to the ZPE value of the subspace.\cite{ceotto_conte_DCSCIVR_2017}
We normalize each spectrum to its maximum amplitude, since we do not
evaluate absorption spectra, but power spectra of relevant reference
states. We also notice that the relative height of secondary peaks,
while informative, may be particularly affected by the sampling phase
space center, contrarily to the main peaks, whose energies are close
to the typical kinetic energies distributed in the sampling of initial
momenta.\cite{DeLeon_Heller_SCeigenfunctions_1983}

According to Ref.~\onlinecite{DiLiberto_Ceotto_Jacobiano_2018}, sampling subspaces are chosen 
to be the O-H stretching sector, the bending sector, the proton transfer mode, the proton perpendicular sector, 
the O-O stretching mode. Variants of these choices are indicated when discussing the results.

Quite generically, we are able to recover good accuracy for the fundamental
transfer, bending, and O-H stretching modes, which are also the most
significant states in experimental absorption spectra. However, the
convergence in the number of trajectories for the states in the frequency
region $1100\div1700\text{cm}^{-1}$ is quite difficult to achieve.
This especially affects the overtones of the O-O stretching and proton
transfer modes, and the fundamental excitation of the shared proton
perpendicular motion. The strong coupling of these modes was already
observed in a classical analysis.\cite{Kaledin_VibrationalAnalysisH5O2_2006}
Moreover, we do not show results for states at frequencies below $500\text{cm}^{-1}$,
except for a wagging state. For these modes a monodimensional sampling
would strongly affect the position of the peaks, while attempting
to extend the dimension of the sampling subspaces renders the trajectories
so chaotic that only broad features are recovered. In particular,
the lowest torsional mode, at a harmonic frequency equal to $170\text{cm}^{-1}$,
but much lower in frequency at anharmonic level,\cite{vendrell_Meyer_zundelspectra_2007}
is so easily excitable along the classical trajectories of other low
frequency modes that it would jeopardize any spectroscopic signal
resolution.

In Table \ref{tab:peaks}, we collect the positions of the peaks of
the various vibrational modes, as extracted from the semiclassical
spectra, and compare them to the harmonic frequencies, the experimental
results for Ne-tagged molecules from Ref.~\onlinecite{hammer_carter_expzundelspectrum_2005},
the VCI and DMC results of Refs.~\onlinecite{mccoy_bowman_VCIzundel_2005,hammer_carter_expzundelspectrum_2005},
and the MCTDH results from Refs.~\onlinecite{vendrell_Meyer_zundelspectra_2007,vendrell_Meyer_isotopeffects2_2009}.
We adopt the nomenclature used in Ref.~\onlinecite{vendrell_Meyer_zundelspectra_2007}.
We take the MCTDH results of Ref.~\onlinecite{vendrell_Meyer_zundelspectra_2007} as a benchmark for our calculations, since it uses the same PES as in our work, even though some of those findings have been updated.\cite{vendrell_Meyer_isotopeffects2_2009}

\subsection{O-H stretching modes}

\label{subsec:stretching}

\begin{figure}[tb]
\includegraphics[width=1\columnwidth]{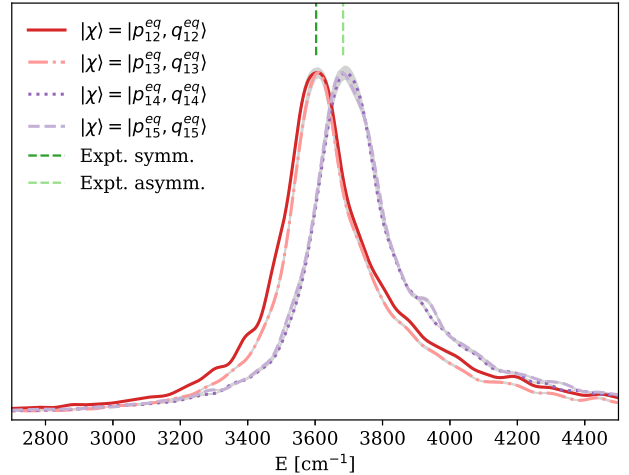} \caption{Semiclassical power spectra in the O-H stretching region, for different reference coherent states and corresponding trajectory
sampling, as described in Sec.~\ref{subsec:stretching}. The straight dashed
lines correspond to the experimental results of Ref.~\onlinecite{hammer_carter_expzundelspectrum_2005}
for the (monomer) symmetric and (monomer) asymmetric bands. Gray bands behind the curves
indicate the standard deviation of the estimated spectra.}
\label{fig:stretchings_Zundel} 
\end{figure}

In Fig.~\ref{fig:stretchings_Zundel}, we show the results for the
fundamental O-H stretching modes. Experimentally, a doublet is observed.\cite{hammer_carter_expzundelspectrum_2005}
MCTDH \cite{vendrell_Meyer_zundelspectra_2007} indicated that each
peak of the doublet comprises two fundamental states, and that the
degeneracy of the lower energy peak is slightly lifted.

We consider the set of vibrational stretchings (modes $12-15$), and
sample their initial momenta with the distribution in Eq.~\eqref{eq:samplep}
centered at their ZPE momenta $p_{l}^{eq}=\sqrt{\omega_{l}}$. All
other modes have initial null velocity. We employ antisymmetrized
coherent states, of type (i), centered at the same momenta as the
relevant reference states. These stretching modes are usually fairly
decoupled from the rest of the dynamics, due to their high energy,
in the $3000\div4000~\text{cm}^{-1}$ range. We find very good agreement
with both the MCTDH results and the experimental observations. We
adopt the nomenclature of Ref.~\onlinecite{Pitsevich_MP4StudyAnharmonic_2017},
$a(m)b(d)$, simplified into $a-b$, where $a=s,as$ and $b=s,as$.
$s(m)$ and $as(m)$ indicate the symmetry/asymmetry of the monomer
stretchings and $s(d)$, $as(d)$ indicate the in- or out-of-phase
combination of the monomer stretchings in the dimer. The $s-as$
and $s-s$ modes are essentially degenerate. The optically active
one ($s-as$), at $3607\text{cm}^{-1}$, is compatible with the experimental
observation at $3603\text{cm}^{-1}$ and with the MCTDH result at
$3607\text{cm}^{-1}$. The fully symmetric optically inactive mode
($s-s$), found at $3609\text{cm}^{-1}$, is compatible with the MCTDH
result at $3614\text{cm}^{-1}$. For the higher frequency peak,
which has double degeneracy and corresponds to the optically active
monomer-asymmetric modes, we obtain two estimates, $3679\text{cm}^{-1}$
for $as-as$ and $3690\text{cm}^{-1}$ for $as-s$, that differ by
only $\sim10\text{cm}^{-1}$. They are equivalent within the uncertainty
given by the finite number of trajectories and compare very well to
the MCTDH result at $3689\text{cm}^{-1}$ and the experimental observation
at $3683\text{cm}^{-1}$.

\subsection{Bending modes}

\label{subsec:bending}

\begin{figure}[tb]
\includegraphics[width=1\columnwidth]{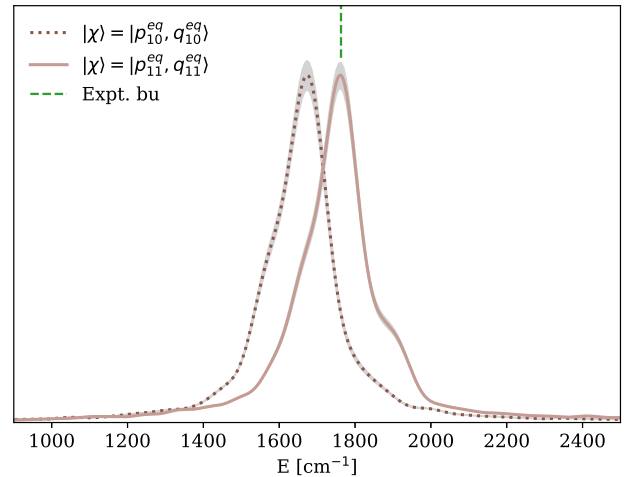} \caption{Semiclassical power spectra in the bendings' region. Curves refer to different reference coherent states and trajectory sampling,
as explained in Sec.~\ref{subsec:bending}. The straight dashed
line corresponds to the experimental result of Ref.~\onlinecite{hammer_carter_expzundelspectrum_2005}
for optically active bending band. Gray bands behind the curves
indicate the standard deviation of the estimated spectra.}
\label{fig:bendings_Zundel} 
\end{figure}

In Fig.~\ref{fig:bendings_Zundel}, we show our results for the subspace
of the two in-plane water bendings (normal modes $10,11$). We sample
their momenta around the ZPE harmonic prescription $p_{l}^{eq}=\sqrt{\omega_{l}}$,
together with the O-H stretching modes. The stretching modes are initialized at their ZPE momenta, while all other modes 
are initialized at zero momentum. We employ antisymmetrized
coherent states of type (i) centered at the same momenta as the relevant
reference states. We find very good agreement with MCTDH ($1741\text{cm}^{-1}$)
and the experimental values ($1763\text{cm}^{-1}$) for the higher
energy mode, found at $1751\text{cm}^{-1}$. This mode is optically active, because the corresponding normal mode contains a significant contribution from proton transfer. The lower frequency
mode, at $1668\text{cm}^{-1}$, is not optically active, since the corresponding normal mode involving a perpendicular shared proton fluctuation, and we cannot thus compare it to experiment. Notice that the height of the power spectrum peak is arbitrarily normalized, and not directly comparable to absorption spectra. Our result is half way between the harmonic
($1720\text{cm}^{-1}$) and the MCTDH ($1606\text{cm}^{-1}$) ones.

\subsection{Proton transfer modes}

\label{subsec:transfer}

\begin{figure}[tb]
\includegraphics[width=1\columnwidth]{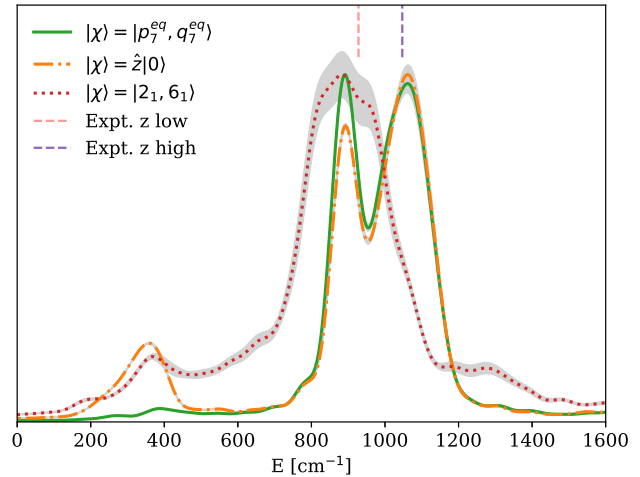} 
\caption{Semiclassical power spectra in the proton transfer
region. Curves refer to different reference states, of types (i), (iii), and (ii), respectively. The corresponding trajectory sampling is 
described in Sec.~\ref{subsec:transfer}. The straight dashed
lines correspond to the experimental result of Ref.~\onlinecite{hammer_carter_expzundelspectrum_2005}
for the proton transfer doublet. Gray bands behind the curves
indicate the standard deviation of the estimated spectra.}
\label{fig:transfer_Zundel} 
\end{figure}

In Fig.~\ref{fig:transfer_Zundel}, we show our results for the proton
transfer excitation. Experimentally, the proton transfer mode shows
a very neat doublet, especially when the protonated water dimer is
tagged with Ne.\cite{hammer_carter_expzundelspectrum_2005} These
two states have been investigated in many articles, due to their controversial
nature.\cite{Kaledin_VibrationalAnalysisH5O2_2006,vendrell_Meyer_zundelspectra_2007,vendrell_meyer_zundelquantumdynamics_2008}
Initially, the doublet was associated to tunneling splitting similar
to hydrogen bonding. However, it is now clear that, in most of the
relevant configuration space close to equilibrium, the shared proton
only visits a shallow single minimum (see Sec.~\ref{subsec:single}),\cite{Schran_ConvergedColoredNoise_2018,marx_abinitio_200years_2006}
and the doublet structure is due to a Fermi resonance involving the
bare proton-transfer mode and a combination of O-O stretching and
wagging modes.\cite{vendrell_Meyer_zundelspectra_2007,vendrell_Meyer_isotopeffects2_2009}
Due to the very shallow shape of the PES in this region, this is a
very tough calculation for a semiclassical approach, which relies
of the evaluation of the Hessian matrix along classical trajectories.
In Ref.~\onlinecite{DiLiberto_Ceotto_Jacobiano_2018}, some of us proposed
an interpretation in terms of a combination of the fundamental transition
of the $7$-th mode (the bare proton transfer mode) and the first
overtone of mode $3$ (namely, the second wagging mode). Here, we
investigate more thoroughly this issue.

In the proton transfer dynamics, we identify three main players.
One is the wagging normal mode $2$, which is \emph{per se} a collective
mode describing the out-of-phase combination of the monomer-localized
wagging modes. In the notation of Ref.~\onlinecite{vendrell_Meyer_zundelspectra_2007},
this state can be identified with $w_{3}$, namely an overtone
in the wagging subspace. The second one is mode 7, which describes
the bare proton transfer. By inspecting matrix $\bm{L}$, normal modes
7 and 2 are those which give the largest contribution to the projection
of the shared proton on the O-O axis. In particular, both modes involve
asymmetric wagging of the water monomers, synchronized with shared
proton transfer, and they essentially differ by the sign and amplitude
of the proton transfer component. The third relevant player is normal
mode 6, describing O-O stretching, which couples very anharmonically
to proton transfer.

To avoid too chaotic trajectory dynamics, we choose to sample the
initial momentum of mode $7$ only, around its fundamental energy,
$p_{7}^{eq}=\sqrt{3\omega_{7}}$. All the other modes are initialized at zero momentum, but
obviously get excited along the trajectories because of the quick energy transfer.
Other choices, in which the sampling subspace $\mathcal{S}^\prime$ is extended, 
would deteriorate the signal, since trajectories would soon become unstable and thus yield a broader spectrum. 
This refined methodology allows us to observe a clean spectral feature
in the $800-1100\text{cm}^{-1}$ region, with two peaks at $891\text{cm}^{-1}$
and $1062\text{cm}^{-1}$, to be compared with the experimental doublet
for the Ne-tagged molecule at $928\text{cm}^{-1}$ and $1047\text{cm}^{-1}$
and the MCTDH results at $918\text{cm}^{-1}$ and $1033\text{cm}^{-1}$.

Now, we want to understand each mode contribution to the doublet peaks.
For this goal, we employ three different reference states. The first
one is the coherent state $\ket{\chi}=\ket{p_{7}^{eq},q_{7}^{eq}}$.
Once properly antisymmetrized according to Eq.~\eqref{eq:antysimmetric_state},
it directly shows a neat doublet structure, where the higher frequency
peak is very close to experiment, while the lower frequency peak is
half way between the experimental peak and the harmonic $7_{1}$ excitation.
Given the strongly chaotic nature of these modes dynamics, we observe
the peak width to depend on the maximum number of time-steps of the
trajectories. When choosing a looser $\epsilon$ threshold, the main
contribution is given by few long-time trajectories and the Monte
Carlo convergence for the higher frequency peak is very tough. Instead,
if one chooses a smaller value of $\epsilon$, trajectories are too
short to yield a well-resolved and accurate enough signal. For these
reasons, the Monte Carlo uncertainty for the two peaks is$~\sim30\text{cm}^{-1}$.

The second reference that we employ is the harmonic ground state,
excited by the $z$ coordinate $\ket{\chi}=\hat{z}\ket{\bm{0}}$,
evaluated as described in Sec.~\ref{subsec:referencestates} at iii).
Since all normal modes contribute to the composition of this reference
state, we can find a third peak at $358\text{cm}^{-1}$. We assign
it to $w_{3}$, corroborating the view that the wagging mode is intimately
linked to proton transfer.

The third reference state is of type (ii), namely the combined harmonic
overtone of the wagging and the O-O stretching modes $\ket{\chi}=\ket{2_{1},6_{1}}$,
indicated as $(w_{3},1R)$ in Ref.~\onlinecite{vendrell_Meyer_zundelspectra_2007}.
In this case, we modify the sampling subspace $\mathcal{S}^\prime$ and sample the initial momenta of the $2$nd and $6$th
normal-modes around their harmonic approximations $p_{l}=\sqrt{\omega_{l}}$.
The resulting semiclassical power spectrum manifests again a small
feature at a frequency $~380\text{cm}^{-1}$, which can be assigned
an uncertainty of $~\sim25{cm}^{-1}$, and a very prominent peak situated
in the region of the lower-energy peak of the proton-transfer doublet.
Here, the Monte Carlo uncertainty is higher because of the slow convergence
in the trajectory number due to the O-O stretching initial excitation.
Then, it is reasonable to consider this peak as compatible with the
lower frequency peak originated from the $\ket{\chi}=\ket{p_{7}^{eq},q_{7}^{eq}}$
spectrum and to infer a major contribution of the $(w_{3},1R)$
to the lower-energy peak.

\subsection{Proton perpendicular modes}

\label{subsec:perp}

\begin{figure}[tb]
\includegraphics[width=1\columnwidth]{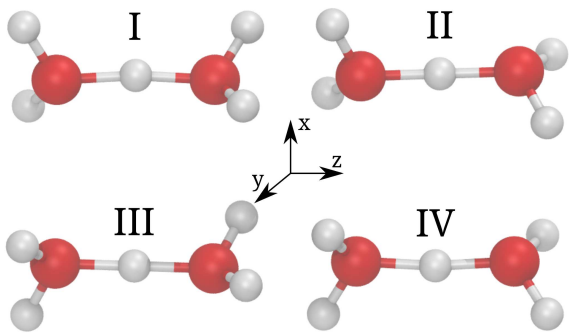} \caption{Four minima of the Zundel cation geometry that are relevant for the
description of the perpendicular motion of the shared proton.}
\label{fig:versions} 
\end{figure}

The vibrational states mainly describing the motion of the shared
proton perpendicular to the O-O segment ($z$ direction) are conventionally
called $x$ and $y$. In Fig.~\ref{fig:versions}, we call version
I the reference geometry minimum of the PES employed in this work.
Such geometry minimum manifests $C_{2}$ symmetry, and the normal
modes which mostly pertain to the shared-proton $x$ and $y$ states
are the $9$th and the $8$th, corresponding to very different harmonic
frequencies $\omega_{9}=1574\text{cm}^{-1}$ and $\omega_{8}=1494\text{cm}^{-1}$.
However, for fluxional molecules, a single global minimum is not sufficient
to describe the relevant symmetries, since very anharmonic low-frequency
modes, like internal rotational modes (torsions and waggings), experience
relatively low-energy barriers towards other equivalent global minima.
These minima are related to the reference one by a combination of
global rotations, reflections and permutations. For these systems,
the relevant symmetry group is a permutation-inversion group, as introduced
by Longuet-Higgins in Ref.~\onlinecite{Longuet-Higgins_symmetrygroupsnonrigid_1963}.
The barrier for wagging is particularly small ($\sim100\text{cm}^{-1}$),\cite{Wales_Rearrangementstunnelingsplittings_1999}
and the Zundel cation belongs to the $\mathcal{G}_{16}$ group, once
torsion and wagging of the water monomers are properly considered.
Within this extended group, the $x$ and $y$ states must be degenerate,
being of $E$ symmetry. This can be seen in Fig.~\ref{fig:versions},
where we indicated by I, II, III, IV the four global minima of the
molecule, that can be reached when $0$, $1$, or 2 monomer inversions
(wagging) are considered (we do not consider the feasibility of overcoming
the torsional barrier). In particular, version II is accessible via
the right monomer wagging, version III via the left monomer wagging,
version IV via both waggings. In terms of permutation-inversions,
version II can be obtained by first a $-\pi/2$ rotation around the
$z$-axis, then a reflection $x\to-x$ and finally the permutation
of the hydrogens of the left monomer. Version III is instead obtained
by a $\pi/2$ rotation around the $z$-axis, then a reflection $x\to-x$
and finally the permutation of the hydrogens of the right monomer.
Version IV is obtained by a $\pi$ rotation around the $z$-axis and
the permutation of both the left hydrogens and, separately, of the
right hydrogens. We call these operations $\mathcal{P}_{A}$, where
$A=$I, II, III, IV and $\mathcal{P}_{I}$ is the identity.

When viewed from versions II or III, the shared-proton $x$ motion
is dominated by normal mode $8$, instead of $9$, at variance with
versions I and IV. The opposite is true for the $y$ motion. By properly
taking into account sign changes, we then introduce two symmetrized
states of harmonic modes of the type (ii), that describe the perpendicular
motion: 
\begin{equation}
\ket{y_{S}}=\ket{8_{1}}_{\text{I}}+\ket{9_{1}}_{\text{II}}-\ket{9_{1}}_{\text{III}}-\ket{8_{1}}_{\text{IV}}
\end{equation}
and 
\begin{equation}
\ket{x_{S}}=\ket{9_{1}}_{\text{I}}-\ket{8_{1}}_{\text{II}}+\ket{8_{1}}_{\text{III}}-\ket{9_{1}}_{\text{IV}}
\end{equation}
where the indexes I, II, III, IV indicate in which geometry the normal
modes are defined. Equivalently, one may make linear combinations
of properly antisymmetrized coherent states of type (i). The two states
are linearly independent, and have the same energy, since one can
apply the operators $\mathcal{P}_{A}$, which commute with the Hamiltonian,
to convert one into the other.

According to MCTDH estimates, the energy of the proton-perpendicular
subspace is $1391\text{cm}^{-1}$. If we naively sampled separately normal modes $8$ and $9$, only
referred to version I and with momenta centered at their respective
first excited harmonic energies, we would obtain non-degenerate peaks,
although strongly red-shifted from the harmonic values.

\begin{figure}[tb]
\includegraphics[width=1\columnwidth]{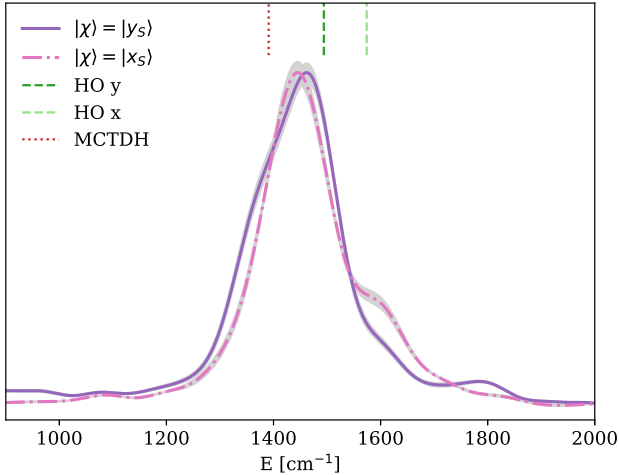} \caption{Semiclassical power spectra in the shared proton perpendicular
mode region, employing symmetrized reference states of type (iv) and multi-reference trajectory
sampling, as explained in Sec.~\ref{subsec:perp}.
The straight dashed lines correspond to the non degenerate harmonic frequencies of the 8th and 9th normal modes. The dotted line indicates the MCTDH result from Ref.~\onlinecite{vendrell_Meyer_zundelspectra_2007} . Gray bands behind the curves
indicate the standard deviation of the estimated spectra.}
\label{fig:perp_Zundel} 
\end{figure}

The correct procedure is, on the contrary, to enforce the described
symmetry and consider the reference states $\ket{y_{S}}$ and $\ket{x_{S}}$.
Moreover, also the sampling of trajectories is to be symmetrized,
in principle by launching ensembles of trajectories starting from
the four versions of the molecule and with the momentum of the shared
proton in the $x$ direction.\cite{Conte_Ceotto_NH3_2013} However,
a more efficient procedure exploits the symmetry operations $\mathcal{P}_{A}$,
by sampling trajectories centered on version I only, but by summing
the contributions of 4 different samples of momenta: two centered
on the fundamental excitation of mode $8$, but with opposite signs,
and two centered on the fundamental excitation of mode $9$, with
both signs, for a total of $96000$ trajectories. To better explore
the regions of phase space close to monomer inversions, we also give
initial momentum to the other normal modes which facilitate such motions,
namely to modes 2 and 7, with an energy corresponding to their harmonic
ZPE. Moreover, the calculation of the overlap between the running
coherent states and the reference states $\ket{y_{S}}$ and $\ket{x_{S}}$
may use the relations $\langle\bm{x}|8_{1}\rangle_{A}=\langle\mathcal{P}_{A}^{-1}(\bm{x})|8_{1}\rangle_{I}$
that relate the Cartesian coordinates referred to geometry A to the
coordinates referred to geometry I by using the inverse operators
$\mathcal{P}_{A}^{-1}$. The resulting spectra are showed in Fig.~\ref{fig:perp_Zundel}.
One can see that degeneracy for the main peaks is essentially recovered, within errorbars,
and that the semiclassical estimate for the perpendicular states is
$1453\text{cm}^{-1}$, given by the average value of the two spectral
peaks. We are unable to assign the minor features that appear far from the main peaks, due to insufficient trajectory sampling in that region.

\subsection{O-O stretching related states}

\label{subsec:other}

\begin{figure}[tb]
\includegraphics[width=1\columnwidth]{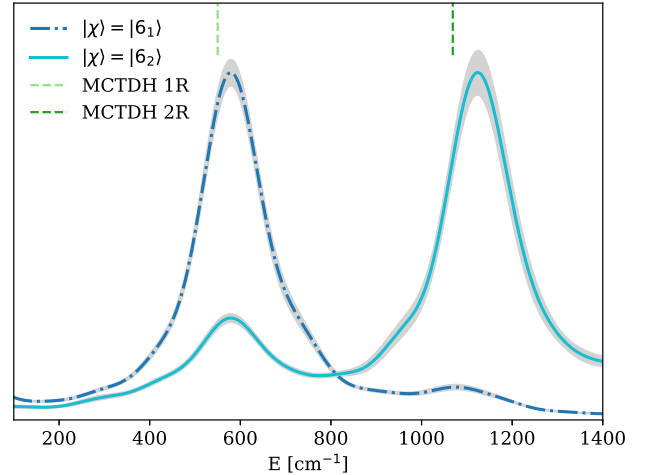} \caption{Semiclassical power spectra in the O-O stretching region, employing harmonic reference states of type (ii), and trajectory
sampling as explained in Sec.~\ref{subsec:other}. The straight dashed lines indicate the MCTDH results from Ref.~\onlinecite{vendrell_Meyer_zundelspectra_2007}. Gray bands behind the curves
show the standard deviation of the estimated spectra.}
\label{fig:OOstretchings_Zundel} 
\end{figure}

We now perform our TA SCIVR calculation for the strong anharmonic
modes involving the O-O stretching. In Fig.~\ref{fig:OOstretchings_Zundel},
we report the power spectra of the harmonic states $\ket{6_{1}}$
and $\ket{6_{2}}$, namely those containing one or two excitations
of $R$, according to Ref.~\onlinecite{vendrell_Meyer_zundelspectra_2007}
nomenclature. We employ a subspace $\mathcal{S}^\prime$ momentum sampling involving normal
modes $2$ (wagging), $6$ (O-O stretch), and $7$ (proton transfer).
We sample modes $6$ and $7$ by centering the distribution of their
momenta around their fundamental energies. Mode $2$ is also sampled,
around its ZPE-associated momentum, because of its strong coupling
to both other states. We noticed, indeed, that especially the estimation
of the correlated state $1R$ is influenced by the sampling of mode
$2$. However, exciting the wagging and the O-O stretching modes together
results in strongly chaotic trajectories. This implies an amplified
Fourier uncertainty of $60\text{cm}^{-1}$. Thus, for these spectra,
we employ $34000$ trajectories. The relatively large statistical
uncertainties present in Fig.~\ref{fig:OOstretchings_Zundel} indicate
the slow convergence due to the strong anharmonicity of these modes.
We did not pursue calculations with even more trajectories, given
the expected uncertainty due to the Fourier transform. However, we
notice a good estimate of the $1R$ frequency at $580\text{cm}^{-1}$,
to be compared to the MCTDH result at $550\text{cm}^{-1}$, and an
acceptable evaluation of the $2R$ frequency at $1124\text{cm}^{-1}$,
which is strongly red-shifted from the harmonic expectation, but partially
blue-shifted with respect to the MCTDH calculation, at $1069\text{cm}^{-1}$.

\subsection{Insight from single trajectories} \label{subsec:single}

\begin{figure*}[t]
\includegraphics[width=1\textwidth]{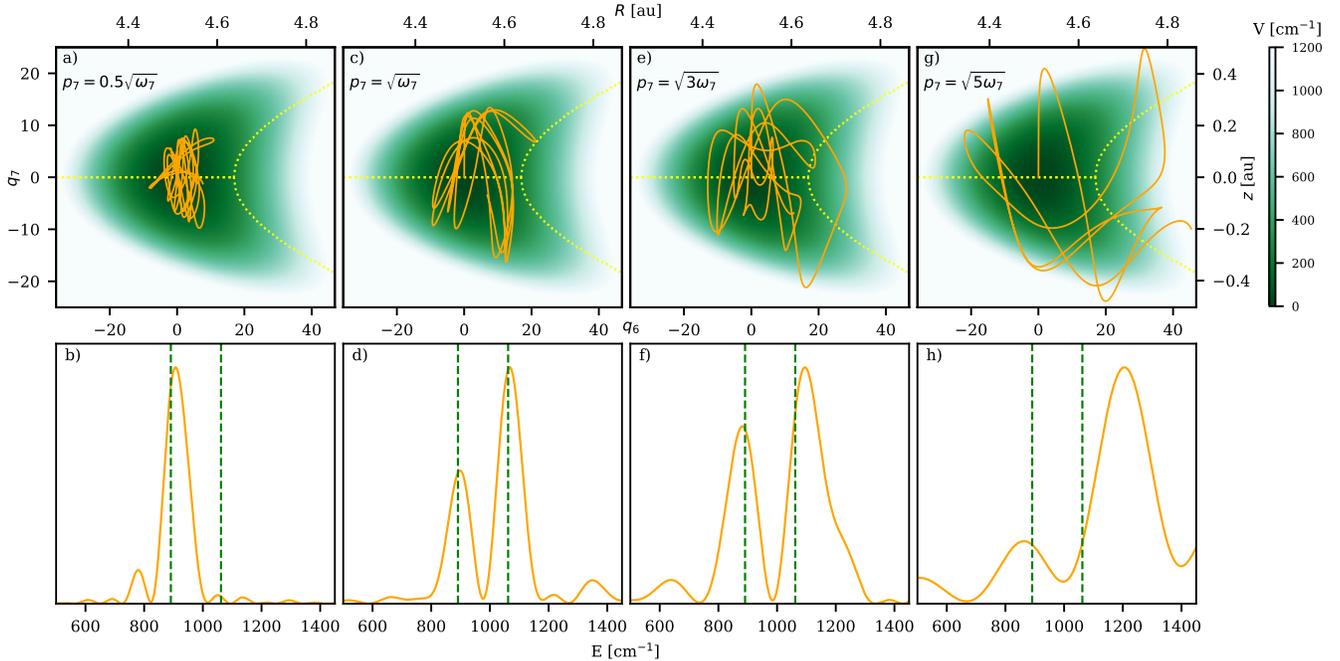} \caption{Upper panels: density plot of a section of the PES with all normal
modes set to equilibrium except for modes $q_{6}$ (O-O stretching)
and $q_{7}$ (proton transfer); top and right axes show the corresponding
$R$ and $z$ coordinates; for each value of $q_{6}$, the minima
of the PES with varying $q_{7}$ are indicated (dotted lines); the
projections of the trajectories, where only $p_{7}$ is initially
excited, are indicated (solid line). Lower panels: corresponding single
trajectory spectra from Eq.~\eqref{eq:spectrumformula} employing
the $\ket{p_{7}^{eq},q_{7}^{eq}}$ reference state; the position of the peaks
from phase-space integration (Table~\ref{tab:peaks}) is indicated
(dashed lines).}
\label{fig:trajectories} 
\end{figure*}

One may ask to what extent single classical trajectories are informative
of the quantum dynamics of the system, in the spirit of the MC SCIVR approach.\cite{Ceotto_AspuruGuzik_Multiplecoherent_2009}
For example, tunneling splitting in ammonia has been resolved with
few trajectories.\cite{Conte_Ceotto_NH3_2013} Here, we focus on the
proton transfer mode, and deepen our investigation on the role of possible
tunneling splitting versus anharmonicity.

We give initial kinetic energy only to the 7-th normal mode, varying
its initial momentum, while setting all the initial coordinates at
the reference geometry. In Fig.~\ref{fig:trajectories}, we observe
that some single trajectories yield the expected doublet spectral
structure, and that this feature strongly and quantitatively depends
on the initial $p_{7}$ momentum. To gain more physical insight, we
draw the projection of the trajectories onto the subspace spanned
by modes 7 (proton transfer) and 6 (O-O stretching), which are close
enough to the global minimum and essentially dominated by the coordinates
$z=x_{H,z}-(x_{O_{1},z}+x_{O_{2},z})/2$ and $R=x_{O_{2},z}-x_{O_{1},z}$.
We also draw the corresponding section of the PES (when all other
normal modes are kept at equilibrium), and indicate the minimum of
the PES along the proton transfer direction, with varying $R$ distance.
One can see that, only for $R\gtrsim4.63a.u.$, a double minimum shape
is acquired by the potential restricted to the $q_{7}$ coordinate.
When $p_{7}\ll\sqrt{\omega_{7}}$ (panels a,b), the trajectory does
not leave the shallow global well and a single peak appears in the
spectrum, close to the harmonic result, because only the bottom of
the well is sampled. For $p_{7}=\sqrt{\omega_{7}}$ (panels c,d),
a double-peak structure is obtained, which is surprisingly close to
the integrated spectrum, even though the trajectory seems to be mostly
confined to the region where a single well is present. This is consistent
with the already discussed interpretation of the doublet as arising
not from tunneling, but from strong anharmonicity and coupling to
O-O stretching (for simplicity, we do not discuss here the wagging
coordinate).\cite{Schran_ConvergedColoredNoise_2018} Notice, also,
that here we have used a single reference coherent state centered
at the geometry minimum, so the coherent states following each classical
trajectory are not overlapping with a superposition of localized states,
differently than the case of ammonia of Ref.~\onlinecite{Conte_Ceotto_NH3_2013}.

For higher momentum $p_{7}=\sqrt{3\omega_{7}}$ (panels e,f), corresponding
to the center of the sampling described in Sec.~\ref{subsec:transfer},
a small part of the trajectory is indeed exploring the double-well
region. The doublet is still present, but the trajectory is apparently
less recurring on already visited space, probably due to excitation
of many other coordinates (not visible in the projection). At even
higher initial momentum (panels g,h), energy transfer is so high that
the resulting spectrum is quite broadened and the doublet is essentially
lost. Notice that the integrated spectra showed in Sec.~\ref{subsec:transfer}
contain all these contributions, properly weighted by the reference
state.

We draw again the conclusion that the double-well structure, and a
related tunneling splitting, is not relevant in quantitatively explaining
the proton-transfer doublet. An extensive analysis of the delocalization
of the shared proton as a function of the relative O-O distance has
been recently performed using path-integral methods.\cite{Schran_ConvergedColoredNoise_2018}

\section{Conclusions}

\label{sec:conclusions}

In this work, we tackled the challenging problem of the estimation
of the vibrational spectrum of a strongly anharmonic molecule, such
as the protonated water dimer, by means of a TA SCIVR method. We showed
that it is currently possible to reach a $20\text{cm}^{-1}$ spectroscopic
accuracy, similarly to what has already been established for smaller
molecules. The crucial improvements, that rendered this calculation
possible, are: (i) the analytical definition of global translational
and rotational coordinates at equilibrium; (ii) the accounting for
the contribution of all classical trajectories, with a weight depending
on the duration of their stability; (iii) the application of various
reference states, highlighting the nature of the states of interest;
(iv) the tailored initial distribution of kinetic energy to the modes
that form the subspaces of the DC approach, avoiding the initial excitation of floppy modes, and (v) the proper symmetrization
of both the reference state and trajectory sampling, when dealing
with the full symmetry group of fluxional molecules.

We believe this work paves the way toward different research directions.
In particular, due to the strongly anharmonic nature of acid solutions,
and of water clusters, integration over phase space of at least some
relevant modes is probably necessary to obtain valuable quantitative, and not
only qualitative, comparison to the experimental results. To be competitive,
phase-space integration necessitates high-dimensional PES. Nevertheless, as
the analysis of the proton transfer mechanism demonstrates, valuable
qualitative information can be extracted also from a single suitably
chosen trajectory, and this should be investigated more. Moreover, the chaotic nature of most of the trajectories
stimulates an effort in extracting useful information from short time
duration, employing estimation techniques other than the Fourier transform, that allow
to reduce the width of the reconstructed spectral peaks by means of, for example, filter diagonalization, compressed sensing or super-resolution. The floppiness
of water complexes points at the usefulness of an Eckart frame in
defining normal modes in regions far from the equilibrium geometry\cite{Wehrle_Vanicek_Oligothiophenes_2014}
and to a proper consideration of multiple geometries in the reference
states. The divide-and-conquer approach adopted for the phase space integration showed a reasonable scaling respect to number of degrees of freedom and it opens the route for tackling higher dimension protonated water clusters.
Finally, since an electric dipole surface is available from
Ref.~\onlinecite{huang_Bowman_ZundelPES_2005}, the evaluation of the full
absorption spectrum should be feasible.\cite{Micciarelli_Anharmonicvibrationaleigenfunctions_2018,Micciarelli_effectivesemiclassicalapproach_2019}

\acknowledgments We acknowledge R. Conte, M. Micciarelli and F. Gabas for
useful discussions on the methodology. We thank D. Marx for critically reading the manuscript. We acknowledge financial support
from the European Research Council (ERC) under the European Union
Horizon 2020 research and innovation programme (Grant Agreement No.
{[}647107{]}-SEMICOMPLEX-ERC-2014-CoG). We acknowledge the CINECA
awards IscraB-QUASP-2018 and IscraC-MCSCMD-2018 for the availability
of high performance computing resources and support.

\appendix

\section{Properties of coherent states}

\label{app:coherent}

From now on, we assume that the quantum numbers associated to global
translational and rotational invariance, namely the momentum of the
center of mass and the angular momentum, are set to zero.

The remaining $\tilde{F}$ normal modes' coordinates are promoted
to operators $\hat{q}_{l}$ and $\hat{p}_{l}$, respecting the canonical
commutation relations $\left[\hat{q}_{l},\hat{p}_{l^{\prime}}\right]=i\delta_{ll^{\prime}}$.
Let us introduce the annihilation and construction operators via $\hat{a}_{l}=\frac{1}{\sqrt{2}}\left(\sqrt{\omega_{l}}\hat{q}_{l}+\frac{i}{\sqrt{w_{l}}}\hat{p}_{l}\right)$
and $\hat{a}_{l}^{\dagger}=\frac{1}{\sqrt{2}}\left(\sqrt{\omega_{l}}\hat{q}_{l}-\frac{i}{\sqrt{w_{l}}}\hat{p}_{l}\right)$,
where $\left[\hat{a}_{l},\hat{a}_{l^{\prime}}^{\dagger}\right]=1$.
The harmonic vacuum state for a single normal mode $\ket{0}$ is defined
by $\hat{a}_{l}\ket{0}=0$, and has coordinate representation $\braket{q_{l}|0}=(\omega/\pi)^{1/4}\exp{(-\omega q_{l}^{2}/2)}$.

Two classes of states are of major importance in the context of semiclassical
vibrational spectroscopy: harmonic states and coherent states. The
normalized harmonic basis is defined by the repeated action of the
creation operator $\ket{l_{n}}=\frac{(\hat{a}_{l}^{\dagger})^{n}}{\sqrt{n!}}\ket{0}$,
with the property that $\ket{l_{n}}$ is the eigenvector of the number
operator $\hat{N}_{l}=\hat{a}_{l}^{\dagger}\hat{a}_{l}$ with eigenvalue
$n$. To account for harmonic excitations of different normal modes,
tensor products are formed $\ket{1_{n_{1}},2_{n_{2}},\dots,\tilde{F}_{n_{\tilde{F}}}}\equiv\ket{1_{n_{1}}}\ket{2_{n_{2}}}\dots\ket{\tilde{F}_{n_{\tilde{F}}}}$.

The coherent states are defined to be eigenvectors of the destruction
operator (we omit now the normal mode index $l$ for clarity) $\hat{a}\ket{\alpha}=\alpha\ket{\alpha}$,
where $\alpha$ is a complex number, which is uniquely defined by
the expectation values of position $q^{\alpha}\equiv\braket{\alpha|\hat{q}|\alpha}=\braket{\alpha|\frac{\hat{a}+\hat{a}^{\dagger}}{\sqrt{2\omega}}|\alpha}=\frac{\alpha+\alpha^{*}}{\sqrt{2\omega}}=\Re{(\alpha)}\sqrt{\frac{2}{\omega}}$
and momentum $p^{\alpha}\equiv\braket{\alpha|\hat{p}|\alpha}=\braket{\alpha|\frac{\hat{a}-\hat{a}^{\dagger}}{i}{\sqrt{\frac{\omega}{2}}}|\alpha}=\frac{\alpha-\alpha^{*}}{i}{\sqrt{\frac{\omega}{2}}}=\Im{(\alpha)}\sqrt{2\omega}$.
Thus $\alpha\equiv\frac{1}{\sqrt{2}}\left(\sqrt{\omega}q^{\alpha}+\frac{i}{\sqrt{w}}p^{\alpha}\right)$.
Typically, when $\omega$ is clear from the context, we also denote
$\ket{\alpha}=\ket{p^{\alpha},q^{\alpha}}$.

To obtain the expansion of coherent states in the harmonic basis $\ket{\alpha}=\sum_{n=0}^{\infty}c_{n}\ket{n}$
(where again, we omit the normal mode index $l$ and only indicate
the number of harmonic excitations $n$), we notice that $\braket{n|\alpha}=\braket{0|\frac{\hat{a}^{n}}{\sqrt{n!}}|\alpha}=\frac{\alpha^{n}}{\sqrt{n!}}\braket{0|\alpha}=c_{0}\frac{\alpha^{n}}{\sqrt{n!}}$.
To fix the modulus of $c_{0}$, one evaluates the norm $\braket{\alpha|\alpha}=\sum_{n=0}^{\infty}|c_{n}|^{2}=|c_{0}|^{2}\sum_{n=0}^{\infty}\frac{|\alpha|^{2n}}{n!}=|c_{0}|^{2}e^{|\alpha|^{2}}\equiv1$.
Therefore, one gets the well-known relation: 
\begin{equation}
\ket{\alpha}=e^{-|\alpha|^{2}/2}e^{-i\frac{p^{\alpha}q^{\alpha}}{2}}\sum_{n=0}^{\infty}\frac{\alpha^{n}}{\sqrt{n!}}\ket{n}\;,
\end{equation}
where the arbitrary global complex phase, which is relevant when taking
overlaps of different states, is fixed by adopting the convention
widely used in the context of the semiclassical literature. By using
the Baker-Campbell-Hausdorff formula $\exp{(\hat{A}+\hat{B})}=\exp{(\hat{A})}\exp{(\hat{B})}\exp{(-c/2)}$,
provided $\left[\hat{A},\hat{B}\right]=c$ is a C-number, it is easy
to verify that the above equation is equivalent to the standard relation
\begin{equation}
\ket{\alpha}=e^{-i\frac{p^{\alpha}q^{\alpha}}{2}}e^{\alpha\hat{a}^{\dagger}-\alpha^{*}\hat{a}}\ket{0}=e^{-iq^{\alpha}\hat{p}}e^{i\hat{q}p^{\alpha}}\ket{0}\;.
\end{equation}

The coordinate representation of a coherent state is thus: 
\begin{align}
\braket{q|\alpha}&=\braket{q|e^{-iq^{\alpha}\hat{p}}e^{i\hat{q}p^{\alpha}}|0}=\braket{q-q^{\alpha}|e^{i\hat{q}p^{\alpha}}|0}\nonumber\\
&=e^{i(q-q^{\alpha})p^{\alpha}}\braket{q-q^{\alpha}|0}\nonumber\\
&=\left(\frac{\omega}{\pi}\right)^{\frac{1}{4}}e^{-\frac{\omega}{2}(q-q^{\alpha})^{2}+ip^{\alpha}(q-q^{\alpha})}\label{eq:coherent_state}
\end{align}
having used $e^{iq^{\alpha}\hat{p}}\ket{q}=\ket{q-q^{\alpha}}$.

The overlap of two coherent states is: 
\begin{align}
\braket{\alpha|\beta}
&=e^{-\frac{|\alpha|^{2}+|\beta|^{2}}{2}+i\frac{p^{\alpha}q^{\alpha}-p^{\beta}q^{\beta}}{2}}\sum_{n,n^{\prime}=0}^{\infty}\frac{(\alpha^{*})^{n}\beta^{n^{\prime}}}{\sqrt{n!n^{\prime}!}}\braket{n|n^{\prime}}\nonumber\\
&=e^{-\frac{|\alpha|^{2}+|\beta|^{2}}{2}+i\frac{p^{\alpha}q^{\alpha}-p^{\beta}q^{\beta}}{2}}\sum_{n=0}^{\infty}\frac{(\alpha^{*}\beta)^{n}}{n!}\nonumber\\
&=e^{\alpha^{*}\beta-\frac{|\alpha|^{2}+|\beta|^{2}}{2}+i\frac{p^{\alpha}q^{\alpha}-p^{\beta}q^{\beta}}{2}}\nonumber\\
&=e^{-\frac{\omega(q^{\alpha}-q^{\beta})^{2}}{4}-\frac{(p^{\alpha}-p^{\beta})^{2}}{4\omega}+i\frac{(p^{\alpha}+p^{\beta})(q^{\alpha}-q^{\beta})}{2}}\;.
\end{align}
It is particularly easy to calculate the expectation value of creation
and destruction operators in coherent states: 
\begin{equation}
\braket{\alpha|(\hat{a}^{\dagger})^{n}|\beta}=(\alpha^{*})^{n}\braket{\alpha|\beta},\quad\braket{\alpha|\hat{a}^{n}|\beta}=\beta^{n}\braket{\alpha|\beta}.
\end{equation}

In the calculation of some of the semiclassical survival amplitudes
evaluated in this work, the expectation value of a normal mode coordinate
operator is relevant: 
\begin{align}
\braket{\alpha|\hat{q}|\beta}&=\frac{\braket{\alpha|\hat{a}+\hat{a}^{\dagger}|\beta}}{\sqrt{2\omega}}=\frac{\beta+\alpha^{*}}{\sqrt{2\omega}}\braket{\alpha|\beta}\nonumber\\
&=\left(\frac{q^{\alpha}+q^{\beta}}{2}-i\frac{p^{\alpha}-p^{\beta}}{2\omega}\right)\braket{\alpha|\beta}\;.\label{eq:expectqincoherent}
\end{align}

Notice that in the MC SCIVR approach,\cite{Ceotto_AspuruGuzik_Multiplecoherent_2009}
emphasis on specific modes is put by considering coherent overlaps
of the form: 
\begin{align}
\bra{\alpha}&\left(\ket{p^{\beta},q^{\beta}}-\ket{-p^{\beta},q^{\beta}}\right)=\bra{\alpha}\left(\ket{\beta}-\ket{\beta^{*}}\right)\nonumber\\
&=e^{-\frac{\omega(q^{\alpha}-q^{\beta})^{2}}{4}-\frac{(p^{\alpha}-p^{\beta})^{2}}{4\omega}+i\frac{(p^{\alpha}+p^{\beta})(q^{\alpha}-q^{\beta})}{2}}\nonumber\\
&\phantom{=}-e^{-\frac{\omega(q^{\alpha}-q^{\beta})^{2}}{4}-\frac{(p^{\alpha}+p^{\beta})^{2}}{4\omega}+i\frac{(p^{\alpha}-p^{\beta})(q^{\alpha}-q^{\beta})}{2}}\nonumber\\
&=\braket{\alpha|\beta}\left(1-e^{-\frac{p^{\alpha}p^{\beta}}{\omega}-ip^{\beta}(q^{\alpha}-q^{\beta})}\right)\nonumber\\
&\underset{p^{\beta}\to0}{\sim}2ip^{\beta}\braket{\alpha|\beta}\left(\frac{q^{\alpha}-q^{\beta}}{2}-i\frac{p^{\alpha}}{2\omega}\right)\;,
\end{align}
which is proportional to Eq.~\eqref{eq:expectqincoherent}, provided
momentum $p^{\beta}$ is small and the center of the reference state
is $q^{\beta}=0$.

Finally, the expectation value of a Cartesian coordinate operator,
considering the relevant case $\ket{\bm{\beta}}=\ket{\bm{0}}$, is
simply given by a linear combination of normal modes' expectation
values: 
\begin{equation}
\braket{\bm{\alpha}|\hat{x}_{j\gamma}|\bm{0}}-{x}_{j\gamma}^{eq}
=\braket{\bm{\alpha}|\bm{0}}\sum_{l}\frac{L_{j\gamma,l}}{\sqrt{m_{j}}}\left[\frac{(q^{\alpha})_{l}}{2}-i\frac{(p^{\alpha})_{l}}{2\omega}\right].\label{eq:expectxincoherent}
\end{equation}

\section{Global translational and rotational symmetries}

\label{app:cnorm}

We follow Refs.~\onlinecite{Eckart_StudiesConcerningRotating_1935,wilson1980molecular,Miller_ReactionpathHamiltonian_1980,Jellinek_SeparationEnergyOverall_1989}
in analytically defining the normal modes corresponding to global
translations and infinitesimal rotations at the reference geometry.

We proceed with a classical derivation using Poisson brackets, which
can easily translated to quantum formalism. We generically consider
a function which is a linear superposition of single atom (mass-scaled)
Cartesian momenta: 
\begin{equation}
B=\sum_{k\beta}b_{k\beta}(\bm{X})P_{k\beta}\;,\label{eq:B}
\end{equation}
where, possibly, the coefficients $b_{k\beta}$ depend on the position,
and we render the axis coordinate $\beta$ explicit. We assume that
the Poisson bracket of such function with the interaction potential
is zero: $\{B,V(\bm{X})\}=0$. 
Since the interaction potential does not depend on momentum, we get
\begin{equation}
\sum_{k\beta}b_{k\beta}(\bm{X})\frac{\partial V}{\partial X_{k\beta}}=0\;,
\end{equation}
namely, functions, which are linear combinations of momenta and have
null Poisson bracket (commute) with the interaction potential, simply
correspond to null gradients of the potential along the same linear
combinations of Cartesian coordinates.

From the above commutation relation, it obviously follows that $\{B,\{B,V(\bm{X})\}\}=0$.
We thus get: 
\begin{multline}
\sum_{k\beta,j\alpha}\left(b_{k\beta}(\bm{X})\frac{\partial b_{j\alpha}(\bm{X})}{\partial X_{k\beta}}\frac{\partial V}{\partial X_{j\alpha}}+\right.\\
\left.b_{k\beta}(\bm{X})b_{j\alpha}(\bm{X})\frac{\partial^{2}V}{\partial X_{k\beta}\partial X_{j\alpha}}\right)=0\;.\label{eq:second}
\end{multline}

In the case of global momentum, associated to the translational symmetry
of the molecule as a whole, $\partial b_{j\alpha}/\partial X_{k\beta}=0$.
So, it is guaranteed that also the second derivative of the potential
along the direction corresponding to $\bm{b}$, at any position, is
null, since the first term in Eq.~\eqref{eq:second} drops out. In
order to analytically assign such zero eigenvalues to global translations,
we explicitly construct the corresponding eigenvectors and conventionally
associate them to the last three rows of $\bm{L}^{T}$. Since the
(not mass-scaled) center-of-mass momenta of the molecule are $P_{\alpha}^{CM}={\sum_{k}\sqrt{m_{k}}P_{k\alpha}}$,
we impose $p_{F-3+\alpha}=\sum_{k\beta}L_{F-3+\alpha,k\beta}^{T}P_{k\beta}\propto P_{\alpha}^{CM}$,
yielding Eq.~\eqref{eq:cnormT}. It is clear that these vectors are
orthonormal to each other. 
These rows of $\bm{L}^{T}$ do not depend on $\bm{X}^{eq}$, so they
are independent of the initial position and orientation of the molecule.
Their treatment is therefore analytical.

Rotation around any axis does not change the interaction potential,
however it is beneficial to relate the normal coordinates corresponding
to global rotations to the infinitesimal generators of rotation around
the principal axes of inertia, namely the global angular momentum
components. To render $\bm{L}$ more sparse, it is useful to use the
principal axes' coordinate frame. The inertia tensor of the reference
geometry 
\begin{equation}
\tilde{I}_{\alpha\beta}=\sum_{k}m_{k}\left[-x_{k\alpha}^{eq}x_{k\beta}^{eq}+\delta_{\alpha\beta}\sum_{\gamma}(x_{k\gamma}^{eq})^{2}\right]
\end{equation}
is diagonalized with $\tilde{\bm{I}}=\bm{R}\bm{I}\bm{R}^{T}$, and
the orthogonal matrix $\bm{R}$ is used to rotate the coordinates
of the reference geometry to the principal axes' frame of the reference
geometry $\bm{X}_{i}^{eq}\to\bm{R}^{T}\bm{X}_{i}^{eq}$, which we
then use throughout this article.

Global angular momentum is defined as 
\begin{equation}
J_{\alpha}=\hbar\sum_{k\beta\gamma}\epsilon_{\alpha\gamma\beta}X_{k\gamma}P_{k\beta}\;,\label{eq:angmom}
\end{equation}
where $\epsilon_{\alpha\gamma\beta}$ is the Levi-Civita symbol and
$\hbar$ appears due to our definition of mass-scaled coordinates.
In this case the linear coefficients 
\begin{equation}
b_{k\beta}(\bm{X})=\hbar\sum_{\gamma}\epsilon_{\alpha\gamma\beta}X_{k\gamma}\label{eq:b(x)}
\end{equation}
do depend on position. So the first term in Eq.~\eqref{eq:second}
is in principle not negligible, unless Eq.~\eqref{eq:second} is
evaluated exactly at a stationary point of the potential.\cite{Miller_ReactionpathHamiltonian_1980}
This is what is usually done when considering the equilibrium geometry.
However, it is clear that numerical inaccuracy in calculating $\bm{X}^{eq}$
directly impacts the numerical determination of the rotational modes,
namely, null or almost null eigenvalues of the Hessian, besides those
corresponding to translations, do not necessarily accurately correspond
to global rotations. On the contrary, rotational modes can and should
be analytically determined independently of the evaluated Hessian.
Ideally, one would also evaluate instantaneous rotational and vibrational
modes all along the classical trajectories, especially when considering
finite angular momentum or reaction dynamics,\cite{Eckart_StudiesConcerningRotating_1935,Miller_ReactionpathHamiltonian_1980,Peslherbe_Analysisextensionmodel_1994}
but we prefer to define a constant $\bm{L}$ matrix, and make the
approximation of using the reference geometry coordinates instead
of the instantaneous position coordinates $\bm{X}\to\bm{X}^{eq}$
in Eq.~\eqref{eq:angmom}. Therefore, the introduced rotational modes
are the exact infinitesimal rotations at the reference geometry.\cite{wilson1980molecular} The use of a fixed $\bm{L}$ is probably part of the reason why we are unable to faithfully discriminate the very low energy floppy modes. On the contrary, the relatively higher energy modes can be correctly assigned, since their harmonic widths are smaller and thus their typical classical motion is close to the geometry minimum. Nevertheless, the overlap factors quench the part of the classical motion which explores regions further from the geometry minimum.

We conventionally assign the penultimate three (two, for linear reference
geometries) rows $\bm{L}^{T}$ to the rotational modes $p_{F-6+\alpha}=\sum_{k\beta}L_{F-6+\alpha,k\beta}^{T}P_{k\beta}$,
and use Eqs.~\eqref{eq:B} and \eqref{eq:b(x)}, to obtain $p_{F-6+\alpha}\propto\sum_{k\beta\gamma}\epsilon_{\alpha\gamma\beta}X_{k\gamma}^{eq}P_{i\beta}$ which implies Eq.~\eqref{eq:cnormR}, 
whose denominator simply guarantees normalization. The 9 scalar products
of the infinitesimal rotation modes yield the inertia tensor, so orthogonality
of those vectors is guaranteed when using the principal axes frame.
Finally, the translational and rotational vectors are trivially orthogonal.

The above defined analytical translational and rotational modes are
then projected away from all the internal vibrational modes, as obtained
from the diagonalization of the scaled Hessian, via a Gram-Schmidt
procedure. In the divide-and-conquer approach the rotational modes
are disregarded (set to null values), and this is equivalent to make
the approximations $\sum_{k\gamma\beta}\epsilon_{\alpha\gamma\beta}X_{k\gamma}^{eq}\delta X_{k\beta}\approx0$
($\alpha=1,2,3$). These equations resemble the Eckart conditions,
with the difference that we do not use them to optimally rotate the
reference frame at each step of the trajectories.\cite{Eckart_StudiesConcerningRotating_1935,Wehrle_Vanicek_Oligothiophenes_2014,Wehrle_Vanicek_NH3_2015}
When considering rovibrational spectra, on the contrary, using the
Eckart frame would be crucial.


\begin{thebibliography}{113}%
\makeatletter
\providecommand \@ifxundefined [1]{%
 \@ifx{#1\undefined}
}%
\providecommand \@ifnum [1]{%
 \ifnum #1\expandafter \@firstoftwo
 \else \expandafter \@secondoftwo
 \fi
}%
\providecommand \@ifx [1]{%
 \ifx #1\expandafter \@firstoftwo
 \else \expandafter \@secondoftwo
 \fi
}%
\providecommand \natexlab [1]{#1}%
\providecommand \enquote  [1]{``#1''}%
\providecommand \bibnamefont  [1]{#1}%
\providecommand \bibfnamefont [1]{#1}%
\providecommand \citenamefont [1]{#1}%
\providecommand \href@noop [0]{\@secondoftwo}%
\providecommand \href [0]{\begingroup \@sanitize@url \@href}%
\providecommand \@href[1]{\@@startlink{#1}\@@href}%
\providecommand \@@href[1]{\endgroup#1\@@endlink}%
\providecommand \@sanitize@url [0]{\catcode `\\12\catcode `\$12\catcode
  `\&12\catcode `\#12\catcode `\^12\catcode `\_12\catcode `\%12\relax}%
\providecommand \@@startlink[1]{}%
\providecommand \@@endlink[0]{}%
\providecommand \url  [0]{\begingroup\@sanitize@url \@url }%
\providecommand \@url [1]{\endgroup\@href {#1}{\urlprefix }}%
\providecommand \urlprefix  [0]{URL }%
\providecommand \Eprint [0]{\href }%
\providecommand \doibase [0]{http://dx.doi.org/}%
\providecommand \selectlanguage [0]{\@gobble}%
\providecommand \bibinfo  [0]{\@secondoftwo}%
\providecommand \bibfield  [0]{\@secondoftwo}%
\providecommand \translation [1]{[#1]}%
\providecommand \BibitemOpen [0]{}%
\providecommand \bibitemStop [0]{}%
\providecommand \bibitemNoStop [0]{.\EOS\space}%
\providecommand \EOS [0]{\spacefactor3000\relax}%
\providecommand \BibitemShut  [1]{\csname bibitem#1\endcsname}%
\let\auto@bib@innerbib\@empty
\bibitem [{\citenamefont {Bowman}, \citenamefont {Carrington},\ and\
  \citenamefont {Meyer}(2008)}]{Bowman_Meyer_Polyatomic_2008}%
  \BibitemOpen
  \bibfield  {author} {\bibinfo {author} {\bibfnamefont {J.~M.}\ \bibnamefont
  {Bowman}}, \bibinfo {author} {\bibfnamefont {T.}~\bibnamefont {Carrington}},
  \ and\ \bibinfo {author} {\bibfnamefont {H.-D.}\ \bibnamefont {Meyer}},\
  }\href@noop {} {\bibfield  {journal} {\bibinfo  {journal} {Molecular
  Physics}\ }\textbf {\bibinfo {volume} {106}},\ \bibinfo {pages} {2145}
  (\bibinfo {year} {2008})}\BibitemShut {NoStop}%
\bibitem [{\citenamefont {Thomas}\ and\ \citenamefont {{Carrington
  Jr}}(2015)}]{Thomas_Carrington_SevenAtoms_2015}%
  \BibitemOpen
  \bibfield  {author} {\bibinfo {author} {\bibfnamefont {P.~S.}\ \bibnamefont
  {Thomas}}\ and\ \bibinfo {author} {\bibfnamefont {T.}~\bibnamefont
  {{Carrington Jr}}},\ }\href@noop {} {\bibfield  {journal} {\bibinfo
  {journal} {J. Phys. Chem. A}\ }\textbf {\bibinfo {volume} {119}},\ \bibinfo
  {pages} {13074} (\bibinfo {year} {2015})}\BibitemShut {NoStop}%
\bibitem [{\citenamefont {Bertaina}, \citenamefont {Galli},\ and\ \citenamefont
  {Vitali}(2017)}]{bertaina_statistical_2017}%
  \BibitemOpen
  \bibfield  {author} {\bibinfo {author} {\bibfnamefont {G.}~\bibnamefont
  {Bertaina}}, \bibinfo {author} {\bibfnamefont {D.~E.}\ \bibnamefont {Galli}},
  \ and\ \bibinfo {author} {\bibfnamefont {E.}~\bibnamefont {Vitali}},\ }\href
  {\doibase 10.1080/23746149.2017.1288585} {\bibfield  {journal} {\bibinfo
  {journal} {Adv. Phys. X}\ }\textbf {\bibinfo {volume} {2}},\ \bibinfo {pages}
  {302} (\bibinfo {year} {2017})}\BibitemShut {NoStop}%
\bibitem [{\citenamefont {Miller}(2001)}]{miller2001semiclassical}%
  \BibitemOpen
  \bibfield  {author} {\bibinfo {author} {\bibfnamefont {W.~H.}\ \bibnamefont
  {Miller}},\ }\href@noop {} {\bibfield  {journal} {\bibinfo  {journal} {J.
  Phys. Chem. A}\ }\textbf {\bibinfo {volume} {105}},\ \bibinfo {pages} {2942}
  (\bibinfo {year} {2001})}\BibitemShut {NoStop}%
\bibitem [{\citenamefont {Miller}(2005)}]{Miller_PNAScomplexsystems_2005}%
  \BibitemOpen
  \bibfield  {author} {\bibinfo {author} {\bibfnamefont {W.~H.}\ \bibnamefont
  {Miller}},\ }\href {\doibase 10.1073/pnas.0408043102} {\bibfield  {journal}
  {\bibinfo  {journal} {Proc. Natl. Acad. Sci. USA}\ }\textbf {\bibinfo
  {volume} {102}},\ \bibinfo {pages} {6660} (\bibinfo {year}
  {2005})}\BibitemShut {NoStop}%
\bibitem [{\citenamefont {Yang}\ \emph {et~al.}(2019)\citenamefont {Yang},
  \citenamefont {Duong}, \citenamefont {Kelleher}, \citenamefont {McCoy},\ and\
  \citenamefont {Johnson}}]{Yang_Deconstructingwaterdiffuse_2019}%
  \BibitemOpen
  \bibfield  {author} {\bibinfo {author} {\bibfnamefont {N.}~\bibnamefont
  {Yang}}, \bibinfo {author} {\bibfnamefont {C.~H.}\ \bibnamefont {Duong}},
  \bibinfo {author} {\bibfnamefont {P.~J.}\ \bibnamefont {Kelleher}}, \bibinfo
  {author} {\bibfnamefont {A.~B.}\ \bibnamefont {McCoy}}, \ and\ \bibinfo
  {author} {\bibfnamefont {M.~A.}\ \bibnamefont {Johnson}},\ }\href {\doibase
  10.1126/science.aaw4086} {\bibfield  {journal} {\bibinfo  {journal}
  {Science}\ }\textbf {\bibinfo {volume} {364}},\ \bibinfo {pages} {275}
  (\bibinfo {year} {2019})}\BibitemShut {NoStop}%
\bibitem [{\citenamefont {Gabas}\ \emph {et~al.}(2018)\citenamefont {Gabas},
  \citenamefont {Di~Liberto}, \citenamefont {Conte},\ and\ \citenamefont
  {Ceotto}}]{Gabas_Protonatedglycinesupramolecular_2018}%
  \BibitemOpen
  \bibfield  {author} {\bibinfo {author} {\bibfnamefont {F.}~\bibnamefont
  {Gabas}}, \bibinfo {author} {\bibfnamefont {G.}~\bibnamefont {Di~Liberto}},
  \bibinfo {author} {\bibfnamefont {R.}~\bibnamefont {Conte}}, \ and\ \bibinfo
  {author} {\bibfnamefont {M.}~\bibnamefont {Ceotto}},\ }\href {\doibase
  10.1039/C8SC03041C} {\bibfield  {journal} {\bibinfo  {journal} {Chem. Sci.}\
  }\textbf {\bibinfo {volume} {9}},\ \bibinfo {pages} {7894} (\bibinfo {year}
  {2018})}\BibitemShut {NoStop}%
\bibitem [{\citenamefont {Di~Liberto}, \citenamefont {Conte},\ and\
  \citenamefont
  {Ceotto}(2018{\natexlab{a}})}]{DiLiberto_Divideandconquersemiclassicalmolecular_2018}%
  \BibitemOpen
  \bibfield  {author} {\bibinfo {author} {\bibfnamefont {G.}~\bibnamefont
  {Di~Liberto}}, \bibinfo {author} {\bibfnamefont {R.}~\bibnamefont {Conte}}, \
  and\ \bibinfo {author} {\bibfnamefont {M.}~\bibnamefont {Ceotto}},\ }\href
  {\doibase 10.1063/1.5023155} {\bibfield  {journal} {\bibinfo  {journal} {J.
  Chem. Phys.}\ }\textbf {\bibinfo {volume} {148}},\ \bibinfo {pages} {104302}
  (\bibinfo {year} {2018}{\natexlab{a}})}\BibitemShut {NoStop}%
\bibitem [{\citenamefont {Zundel}\ and\ \citenamefont
  {Metzger}(1968)}]{Zundel_EnergiebandertunnelndenUberschussProtonen_2011}%
  \BibitemOpen
  \bibfield  {author} {\bibinfo {author} {\bibfnamefont {G.}~\bibnamefont
  {Zundel}}\ and\ \bibinfo {author} {\bibfnamefont {H.}~\bibnamefont
  {Metzger}},\ }\href {\doibase 10.1524/zpch.1968.58.5_6.225} {\bibfield
  {journal} {\bibinfo  {journal} {Z. Phys. Chem.}\ }\textbf {\bibinfo {volume}
  {58}},\ \bibinfo {pages} {225} (\bibinfo {year} {1968})}\BibitemShut
  {NoStop}%
\bibitem [{\citenamefont
  {Wales}(1999)}]{Wales_Rearrangementstunnelingsplittings_1999}%
  \BibitemOpen
  \bibfield  {author} {\bibinfo {author} {\bibfnamefont {D.~J.}\ \bibnamefont
  {Wales}},\ }\href {\doibase 10.1063/1.478972} {\bibfield  {journal} {\bibinfo
   {journal} {J. Chem. Phys.}\ }\textbf {\bibinfo {volume} {110}},\ \bibinfo
  {pages} {10403} (\bibinfo {year} {1999})}\BibitemShut {NoStop}%
\bibitem [{\citenamefont {Miyazaki}\ \emph {et~al.}(2004)\citenamefont
  {Miyazaki}, \citenamefont {Fujii}, \citenamefont {Ebata},\ and\ \citenamefont
  {Mikami}}]{miyazaki_mikami_largeprotonatewaterclusters_2004}%
  \BibitemOpen
  \bibfield  {author} {\bibinfo {author} {\bibfnamefont {M.}~\bibnamefont
  {Miyazaki}}, \bibinfo {author} {\bibfnamefont {A.}~\bibnamefont {Fujii}},
  \bibinfo {author} {\bibfnamefont {T.}~\bibnamefont {Ebata}}, \ and\ \bibinfo
  {author} {\bibfnamefont {N.}~\bibnamefont {Mikami}},\ }\href@noop {}
  {\bibfield  {journal} {\bibinfo  {journal} {Science}\ }\textbf {\bibinfo
  {volume} {304}},\ \bibinfo {pages} {1134} (\bibinfo {year}
  {2004})}\BibitemShut {NoStop}%
\bibitem [{\citenamefont {Shin}\ \emph {et~al.}(2004)\citenamefont {Shin},
  \citenamefont {Hammer}, \citenamefont {Diken}, \citenamefont {Johnson},
  \citenamefont {Walters}, \citenamefont {Jaeger}, \citenamefont {Duncan},
  \citenamefont {Christie},\ and\ \citenamefont
  {Jordan}}]{Shin_InfraredSignatureStructures_2004}%
  \BibitemOpen
  \bibfield  {author} {\bibinfo {author} {\bibfnamefont {J.-W.}\ \bibnamefont
  {Shin}}, \bibinfo {author} {\bibfnamefont {N.~I.}\ \bibnamefont {Hammer}},
  \bibinfo {author} {\bibfnamefont {E.~G.}\ \bibnamefont {Diken}}, \bibinfo
  {author} {\bibfnamefont {M.~A.}\ \bibnamefont {Johnson}}, \bibinfo {author}
  {\bibfnamefont {R.~S.}\ \bibnamefont {Walters}}, \bibinfo {author}
  {\bibfnamefont {T.~D.}\ \bibnamefont {Jaeger}}, \bibinfo {author}
  {\bibfnamefont {M.~A.}\ \bibnamefont {Duncan}}, \bibinfo {author}
  {\bibfnamefont {R.~A.}\ \bibnamefont {Christie}}, \ and\ \bibinfo {author}
  {\bibfnamefont {K.~D.}\ \bibnamefont {Jordan}},\ }\href {\doibase
  10.1126/science.1096466} {\bibfield  {journal} {\bibinfo  {journal}
  {Science}\ }\textbf {\bibinfo {volume} {304}},\ \bibinfo {pages} {1137}
  (\bibinfo {year} {2004})}\BibitemShut {NoStop}%
\bibitem [{\citenamefont {Douberly}\ \emph {et~al.}(2010)\citenamefont
  {Douberly}, \citenamefont {Walters}, \citenamefont {Cui}, \citenamefont
  {Jordan},\ and\ \citenamefont
  {Duncan}}]{douberly_duncan_expsmallprotwaterclusters_2010}%
  \BibitemOpen
  \bibfield  {author} {\bibinfo {author} {\bibfnamefont {G.}~\bibnamefont
  {Douberly}}, \bibinfo {author} {\bibfnamefont {R.}~\bibnamefont {Walters}},
  \bibinfo {author} {\bibfnamefont {J.}~\bibnamefont {Cui}}, \bibinfo {author}
  {\bibfnamefont {K.~D.}\ \bibnamefont {Jordan}}, \ and\ \bibinfo {author}
  {\bibfnamefont {M.}~\bibnamefont {Duncan}},\ }\href@noop {} {\bibfield
  {journal} {\bibinfo  {journal} {J. Phys. Chem. A}\ }\textbf {\bibinfo
  {volume} {114}},\ \bibinfo {pages} {4570} (\bibinfo {year}
  {2010})}\BibitemShut {NoStop}%
\bibitem [{\citenamefont {Th{\"a}mer}\ \emph {et~al.}(2015)\citenamefont
  {Th{\"a}mer}, \citenamefont {{De Marco}}, \citenamefont {Ramasesha},
  \citenamefont {Mandal},\ and\ \citenamefont
  {Tokmakoff}}]{Thamer_Tokmakoff_2DIRZundel_2015}%
  \BibitemOpen
  \bibfield  {author} {\bibinfo {author} {\bibfnamefont {M.}~\bibnamefont
  {Th{\"a}mer}}, \bibinfo {author} {\bibfnamefont {L.}~\bibnamefont {{De
  Marco}}}, \bibinfo {author} {\bibfnamefont {K.}~\bibnamefont {Ramasesha}},
  \bibinfo {author} {\bibfnamefont {A.}~\bibnamefont {Mandal}}, \ and\ \bibinfo
  {author} {\bibfnamefont {A.}~\bibnamefont {Tokmakoff}},\ }\href@noop {}
  {\bibfield  {journal} {\bibinfo  {journal} {Science}\ }\textbf {\bibinfo
  {volume} {350}},\ \bibinfo {pages} {78} (\bibinfo {year} {2015})}\BibitemShut
  {NoStop}%
\bibitem [{\citenamefont {Wolke}\ \emph {et~al.}(2016)\citenamefont {Wolke},
  \citenamefont {Fournier}, \citenamefont {Dzugan}, \citenamefont {Fagiani},
  \citenamefont {Odbadrakh}, \citenamefont {Knorke}, \citenamefont {Jordan},
  \citenamefont {McCoy}, \citenamefont {Asmis},\ and\ \citenamefont
  {Johnson}}]{Wolke_Spectroscopicsnapshotsprotontransfer_2016}%
  \BibitemOpen
  \bibfield  {author} {\bibinfo {author} {\bibfnamefont {C.~T.}\ \bibnamefont
  {Wolke}}, \bibinfo {author} {\bibfnamefont {J.~A.}\ \bibnamefont {Fournier}},
  \bibinfo {author} {\bibfnamefont {L.~C.}\ \bibnamefont {Dzugan}}, \bibinfo
  {author} {\bibfnamefont {M.~R.}\ \bibnamefont {Fagiani}}, \bibinfo {author}
  {\bibfnamefont {T.~T.}\ \bibnamefont {Odbadrakh}}, \bibinfo {author}
  {\bibfnamefont {H.}~\bibnamefont {Knorke}}, \bibinfo {author} {\bibfnamefont
  {K.~D.}\ \bibnamefont {Jordan}}, \bibinfo {author} {\bibfnamefont {A.~B.}\
  \bibnamefont {McCoy}}, \bibinfo {author} {\bibfnamefont {K.~R.}\ \bibnamefont
  {Asmis}}, \ and\ \bibinfo {author} {\bibfnamefont {M.~A.}\ \bibnamefont
  {Johnson}},\ }\href {\doibase 10.1126/science.aaf8425} {\bibfield  {journal}
  {\bibinfo  {journal} {Science}\ }\textbf {\bibinfo {volume} {354}},\ \bibinfo
  {pages} {1131} (\bibinfo {year} {2016})}\BibitemShut {NoStop}%
\bibitem [{\citenamefont {Fagiani}\ \emph {et~al.}(2016)\citenamefont
  {Fagiani}, \citenamefont {Knorke}, \citenamefont {Esser}, \citenamefont
  {Heine}, \citenamefont {Wolke}, \citenamefont {Gewinner}, \citenamefont
  {Sch\"ollkopf}, \citenamefont {Gaigeot}, \citenamefont {Spezia},
  \citenamefont {Johnson},\ and\ \citenamefont
  {Asmis}}]{Fagiani_Gasphasevibrational_2016}%
  \BibitemOpen
  \bibfield  {author} {\bibinfo {author} {\bibfnamefont {M.~R.}\ \bibnamefont
  {Fagiani}}, \bibinfo {author} {\bibfnamefont {H.}~\bibnamefont {Knorke}},
  \bibinfo {author} {\bibfnamefont {T.~K.}\ \bibnamefont {Esser}}, \bibinfo
  {author} {\bibfnamefont {N.}~\bibnamefont {Heine}}, \bibinfo {author}
  {\bibfnamefont {C.~T.}\ \bibnamefont {Wolke}}, \bibinfo {author}
  {\bibfnamefont {S.}~\bibnamefont {Gewinner}}, \bibinfo {author}
  {\bibfnamefont {W.}~\bibnamefont {Sch\"ollkopf}}, \bibinfo {author}
  {\bibfnamefont {M.-P.}\ \bibnamefont {Gaigeot}}, \bibinfo {author}
  {\bibfnamefont {R.}~\bibnamefont {Spezia}}, \bibinfo {author} {\bibfnamefont
  {M.~A.}\ \bibnamefont {Johnson}}, \ and\ \bibinfo {author} {\bibfnamefont
  {K.~R.}\ \bibnamefont {Asmis}},\ }\href {\doibase 10.1039/C6CP05217G}
  {\bibfield  {journal} {\bibinfo  {journal} {Phys. Chem. Chem. Phys.}\
  }\textbf {\bibinfo {volume} {18}},\ \bibinfo {pages} {26743} (\bibinfo {year}
  {2016})}\BibitemShut {NoStop}%
\bibitem [{\citenamefont {Singh}\ \emph {et~al.}(2006)\citenamefont {Singh},
  \citenamefont {Park}, \citenamefont {Min}, \citenamefont {Suh},\ and\
  \citenamefont {Kim}}]{singh_kim_magicacid_2006}%
  \BibitemOpen
  \bibfield  {author} {\bibinfo {author} {\bibfnamefont {N.~J.}\ \bibnamefont
  {Singh}}, \bibinfo {author} {\bibfnamefont {M.}~\bibnamefont {Park}},
  \bibinfo {author} {\bibfnamefont {S.~K.}\ \bibnamefont {Min}}, \bibinfo
  {author} {\bibfnamefont {S.~B.}\ \bibnamefont {Suh}}, \ and\ \bibinfo
  {author} {\bibfnamefont {K.~S.}\ \bibnamefont {Kim}},\ }\href {\doibase
  10.1002/anie.200504159} {\bibfield  {journal} {\bibinfo  {journal} {Angew.
  Chem. Int. Ed.}\ }\textbf {\bibinfo {volume} {45}},\ \bibinfo {pages} {3795}
  (\bibinfo {year} {2006})}\BibitemShut {NoStop}%
\bibitem [{\citenamefont {Agostini}, \citenamefont {Vuilleumier},\ and\
  \citenamefont
  {Ciccotti}(2011)}]{agostini_ciccotti_smallprotonatedwaterclus_2011}%
  \BibitemOpen
  \bibfield  {author} {\bibinfo {author} {\bibfnamefont {F.}~\bibnamefont
  {Agostini}}, \bibinfo {author} {\bibfnamefont {R.}~\bibnamefont
  {Vuilleumier}}, \ and\ \bibinfo {author} {\bibfnamefont {G.}~\bibnamefont
  {Ciccotti}},\ }\href@noop {} {\bibfield  {journal} {\bibinfo  {journal} {J.
  Chem. Phys.}\ }\textbf {\bibinfo {volume} {134}},\ \bibinfo {pages} {084302}
  (\bibinfo {year} {2011})}\BibitemShut {NoStop}%
\bibitem [{\citenamefont {Yu}\ and\ \citenamefont
  {Bowman}(2017)}]{Yu_Bowman_protonatedwater_2017}%
  \BibitemOpen
  \bibfield  {author} {\bibinfo {author} {\bibfnamefont {Q.}~\bibnamefont
  {Yu}}\ and\ \bibinfo {author} {\bibfnamefont {J.~M.}\ \bibnamefont
  {Bowman}},\ }\href@noop {} {\bibfield  {journal} {\bibinfo  {journal} {J. Am.
  Chem. Soc.}\ }\textbf {\bibinfo {volume} {139}},\ \bibinfo {pages} {10984}
  (\bibinfo {year} {2017})}\BibitemShut {NoStop}%
\bibitem [{\citenamefont {Yu}\ and\ \citenamefont
  {Bowman}(2019)}]{Yu_ClassicalThermostatedRing_2019}%
  \BibitemOpen
  \bibfield  {author} {\bibinfo {author} {\bibfnamefont {Q.}~\bibnamefont
  {Yu}}\ and\ \bibinfo {author} {\bibfnamefont {J.~M.}\ \bibnamefont
  {Bowman}},\ }\href {\doibase 10.1021/acs.jpca.8b11603} {\bibfield  {journal}
  {\bibinfo  {journal} {J. Phys. Chem. A}\ }\textbf {\bibinfo {volume} {123}},\
  \bibinfo {pages} {1399} (\bibinfo {year} {2019})}\BibitemShut {NoStop}%
\bibitem [{\citenamefont {Egan}\ and\ \citenamefont
  {Paesani}(2019)}]{Egan_AssessingManyBodyEffects_2019}%
  \BibitemOpen
  \bibfield  {author} {\bibinfo {author} {\bibfnamefont {C.~K.}\ \bibnamefont
  {Egan}}\ and\ \bibinfo {author} {\bibfnamefont {F.}~\bibnamefont {Paesani}},\
  }\href {\doibase 10.1021/acs.jctc.9b00418} {\bibfield  {journal} {\bibinfo
  {journal} {J. Chem. Theory Comput.}\ } (\bibinfo {year} {2019}),\
  10.1021/acs.jctc.9b00418}\BibitemShut {NoStop}%
\bibitem [{\citenamefont {{de
  Grotthuss}}(1806)}]{deGrotthuss_decompositioneaucorps_1806}%
  \BibitemOpen
  \bibfield  {author} {\bibinfo {author} {\bibfnamefont {C.~J.~T.}\
  \bibnamefont {{de Grotthuss}}},\ }\href@noop {} {\bibfield  {journal}
  {\bibinfo  {journal} {Annales de chimie}\ }\textbf {\bibinfo {volume} {58}},\
  \bibinfo {pages} {54} (\bibinfo {year} {1806})}\BibitemShut {NoStop}%
\bibitem [{\citenamefont {Agmon}(1995)}]{Agmon_Grotthussmechanism_1995}%
  \BibitemOpen
  \bibfield  {author} {\bibinfo {author} {\bibfnamefont {N.}~\bibnamefont
  {Agmon}},\ }\href@noop {} {\bibfield  {journal} {\bibinfo  {journal} {Chem.
  Phys. Lett.}\ }\textbf {\bibinfo {volume} {244}},\ \bibinfo {pages} {456}
  (\bibinfo {year} {1995})}\BibitemShut {NoStop}%
\bibitem [{\citenamefont {Tuckerman}\ \emph {et~al.}(1995)\citenamefont
  {Tuckerman}, \citenamefont {Laasonen}, \citenamefont {Sprik},\ and\
  \citenamefont {Parrinello}}]{Tuckerman_Parrinello_Abinitiotransport_1995}%
  \BibitemOpen
  \bibfield  {author} {\bibinfo {author} {\bibfnamefont {M.}~\bibnamefont
  {Tuckerman}}, \bibinfo {author} {\bibfnamefont {K.}~\bibnamefont {Laasonen}},
  \bibinfo {author} {\bibfnamefont {M.}~\bibnamefont {Sprik}}, \ and\ \bibinfo
  {author} {\bibfnamefont {M.}~\bibnamefont {Parrinello}},\ }\href@noop {}
  {\bibfield  {journal} {\bibinfo  {journal} {J. Chem. Phys.}\ }\textbf
  {\bibinfo {volume} {103}},\ \bibinfo {pages} {150} (\bibinfo {year}
  {1995})}\BibitemShut {NoStop}%
\bibitem [{\citenamefont {Mohammed}\ \emph {et~al.}(2005)\citenamefont
  {Mohammed}, \citenamefont {Pines}, \citenamefont {Dreyer}, \citenamefont
  {Pines},\ and\ \citenamefont
  {Nibbering}}]{mohammed_Nibbering_waterbrideges_2005}%
  \BibitemOpen
  \bibfield  {author} {\bibinfo {author} {\bibfnamefont {O.~F.}\ \bibnamefont
  {Mohammed}}, \bibinfo {author} {\bibfnamefont {D.}~\bibnamefont {Pines}},
  \bibinfo {author} {\bibfnamefont {J.}~\bibnamefont {Dreyer}}, \bibinfo
  {author} {\bibfnamefont {E.}~\bibnamefont {Pines}}, \ and\ \bibinfo {author}
  {\bibfnamefont {E.~T.~J.}\ \bibnamefont {Nibbering}},\ }\href {\doibase
  10.1126/science.1117756} {\bibfield  {journal} {\bibinfo  {journal}
  {Science}\ }\textbf {\bibinfo {volume} {310}},\ \bibinfo {pages} {83}
  (\bibinfo {year} {2005})}\BibitemShut {NoStop}%
\bibitem [{\citenamefont {Marx}(2006)}]{marx_abinitio_200years_2006}%
  \BibitemOpen
  \bibfield  {author} {\bibinfo {author} {\bibfnamefont {D.}~\bibnamefont
  {Marx}},\ }\href@noop {} {\bibfield  {journal} {\bibinfo  {journal} {Chem.
  Phys. Chem.}\ }\textbf {\bibinfo {volume} {7}},\ \bibinfo {pages} {1848}
  (\bibinfo {year} {2006})}\BibitemShut {NoStop}%
\bibitem [{\citenamefont {Berkelbach}, \citenamefont {Lee},\ and\ \citenamefont
  {Tuckerman}(2009)}]{Berkelbach_Tuckerman_concertedhydrogenbond_2009}%
  \BibitemOpen
  \bibfield  {author} {\bibinfo {author} {\bibfnamefont {T.~C.}\ \bibnamefont
  {Berkelbach}}, \bibinfo {author} {\bibfnamefont {H.-S.}\ \bibnamefont {Lee}},
  \ and\ \bibinfo {author} {\bibfnamefont {M.~E.}\ \bibnamefont {Tuckerman}},\
  }\href@noop {} {\bibfield  {journal} {\bibinfo  {journal} {Phys. Rev. Lett.}\
  }\textbf {\bibinfo {volume} {103}},\ \bibinfo {pages} {238302} (\bibinfo
  {year} {2009})}\BibitemShut {NoStop}%
\bibitem [{\citenamefont {Asmis}\ \emph {et~al.}(2003)\citenamefont {Asmis},
  \citenamefont {Pivonka}, \citenamefont {Santambrogio}, \citenamefont
  {Br\"ummer}, \citenamefont {Kaposta}, \citenamefont {Neumark},\ and\
  \citenamefont {W\"oste}}]{Asmis_GasPhaseInfraredSpectrum_2003}%
  \BibitemOpen
  \bibfield  {author} {\bibinfo {author} {\bibfnamefont {K.~R.}\ \bibnamefont
  {Asmis}}, \bibinfo {author} {\bibfnamefont {N.~L.}\ \bibnamefont {Pivonka}},
  \bibinfo {author} {\bibfnamefont {G.}~\bibnamefont {Santambrogio}}, \bibinfo
  {author} {\bibfnamefont {M.}~\bibnamefont {Br\"ummer}}, \bibinfo {author}
  {\bibfnamefont {C.}~\bibnamefont {Kaposta}}, \bibinfo {author} {\bibfnamefont
  {D.~M.}\ \bibnamefont {Neumark}}, \ and\ \bibinfo {author} {\bibfnamefont
  {L.}~\bibnamefont {W\"oste}},\ }\href {\doibase 10.1126/science.1081634}
  {\bibfield  {journal} {\bibinfo  {journal} {Science}\ }\textbf {\bibinfo
  {volume} {299}},\ \bibinfo {pages} {1375} (\bibinfo {year}
  {2003})}\BibitemShut {NoStop}%
\bibitem [{\citenamefont {Fridgen}\ \emph {et~al.}(2004)\citenamefont
  {Fridgen}, \citenamefont {McMahon}, \citenamefont {MacAleese}, \citenamefont
  {Lemaire},\ and\ \citenamefont
  {Maitre}}]{fridgen2_maitre_expIRMPDzundel_2004}%
  \BibitemOpen
  \bibfield  {author} {\bibinfo {author} {\bibfnamefont {T.~D.}\ \bibnamefont
  {Fridgen}}, \bibinfo {author} {\bibfnamefont {T.~B.}\ \bibnamefont
  {McMahon}}, \bibinfo {author} {\bibfnamefont {L.}~\bibnamefont {MacAleese}},
  \bibinfo {author} {\bibfnamefont {J.}~\bibnamefont {Lemaire}}, \ and\
  \bibinfo {author} {\bibfnamefont {P.}~\bibnamefont {Maitre}},\ }\href@noop {}
  {\bibfield  {journal} {\bibinfo  {journal} {J. Phys. Chem. A}\ }\textbf
  {\bibinfo {volume} {108}},\ \bibinfo {pages} {9008} (\bibinfo {year}
  {2004})}\BibitemShut {NoStop}%
\bibitem [{\citenamefont {Yeh}\ \emph {et~al.}(1989)\citenamefont {Yeh},
  \citenamefont {Okumura}, \citenamefont {Myers}, \citenamefont {Price},\ and\
  \citenamefont {Lee}}]{Yeh_Vibrationalspectroscopyhydrated_1989}%
  \BibitemOpen
  \bibfield  {author} {\bibinfo {author} {\bibfnamefont {L.~I.}\ \bibnamefont
  {Yeh}}, \bibinfo {author} {\bibfnamefont {M.}~\bibnamefont {Okumura}},
  \bibinfo {author} {\bibfnamefont {J.~D.}\ \bibnamefont {Myers}}, \bibinfo
  {author} {\bibfnamefont {J.~M.}\ \bibnamefont {Price}}, \ and\ \bibinfo
  {author} {\bibfnamefont {Y.~T.}\ \bibnamefont {Lee}},\ }\href {\doibase
  10.1063/1.457305} {\bibfield  {journal} {\bibinfo  {journal} {J. Chem.
  Phys.}\ }\textbf {\bibinfo {volume} {91}},\ \bibinfo {pages} {7319} (\bibinfo
  {year} {1989})}\BibitemShut {NoStop}%
\bibitem [{\citenamefont {Headrick}, \citenamefont {Bopp},\ and\ \citenamefont
  {Johnson}(2004)}]{headrick_johnson_expartaggedzundel_2004}%
  \BibitemOpen
  \bibfield  {author} {\bibinfo {author} {\bibfnamefont {J.~M.}\ \bibnamefont
  {Headrick}}, \bibinfo {author} {\bibfnamefont {J.~C.}\ \bibnamefont {Bopp}},
  \ and\ \bibinfo {author} {\bibfnamefont {M.~A.}\ \bibnamefont {Johnson}},\
  }\href@noop {} {\bibfield  {journal} {\bibinfo  {journal} {J. Chem. Phys.}\
  }\textbf {\bibinfo {volume} {121}},\ \bibinfo {pages} {11523} (\bibinfo
  {year} {2004})}\BibitemShut {NoStop}%
\bibitem [{\citenamefont {Hammer}\ \emph {et~al.}(2005)\citenamefont {Hammer},
  \citenamefont {Diken}, \citenamefont {Roscioli}, \citenamefont {Johnson},
  \citenamefont {Myshakin}, \citenamefont {Jordan}, \citenamefont {McCoy},
  \citenamefont {Huang}, \citenamefont {Bowman},\ and\ \citenamefont
  {Carter}}]{hammer_carter_expzundelspectrum_2005}%
  \BibitemOpen
  \bibfield  {author} {\bibinfo {author} {\bibfnamefont {N.~I.}\ \bibnamefont
  {Hammer}}, \bibinfo {author} {\bibfnamefont {E.~G.}\ \bibnamefont {Diken}},
  \bibinfo {author} {\bibfnamefont {J.~R.}\ \bibnamefont {Roscioli}}, \bibinfo
  {author} {\bibfnamefont {M.~A.}\ \bibnamefont {Johnson}}, \bibinfo {author}
  {\bibfnamefont {E.~M.}\ \bibnamefont {Myshakin}}, \bibinfo {author}
  {\bibfnamefont {K.~D.}\ \bibnamefont {Jordan}}, \bibinfo {author}
  {\bibfnamefont {A.~B.}\ \bibnamefont {McCoy}}, \bibinfo {author}
  {\bibfnamefont {X.}~\bibnamefont {Huang}}, \bibinfo {author} {\bibfnamefont
  {J.~M.}\ \bibnamefont {Bowman}}, \ and\ \bibinfo {author} {\bibfnamefont
  {S.}~\bibnamefont {Carter}},\ }\href@noop {} {\bibfield  {journal} {\bibinfo
  {journal} {J. Chem. Phys.}\ }\textbf {\bibinfo {volume} {122}},\ \bibinfo
  {pages} {244301} (\bibinfo {year} {2005})}\BibitemShut {NoStop}%
\bibitem [{\citenamefont {Huang}, \citenamefont {Braams},\ and\ \citenamefont
  {Bowman}(2005)}]{huang_Bowman_ZundelPES_2005}%
  \BibitemOpen
  \bibfield  {author} {\bibinfo {author} {\bibfnamefont {X.}~\bibnamefont
  {Huang}}, \bibinfo {author} {\bibfnamefont {B.~J.}\ \bibnamefont {Braams}}, \
  and\ \bibinfo {author} {\bibfnamefont {J.~M.}\ \bibnamefont {Bowman}},\
  }\href@noop {} {\bibfield  {journal} {\bibinfo  {journal} {J. Chem. Phys.}\
  }\textbf {\bibinfo {volume} {122}},\ \bibinfo {pages} {044308} (\bibinfo
  {year} {2005})}\BibitemShut {NoStop}%
\bibitem [{\citenamefont {Dai}\ \emph {et~al.}(2003)\citenamefont {Dai},
  \citenamefont {Ba{\v c}i{\'c}}, \citenamefont {Huang}, \citenamefont
  {Carter},\ and\ \citenamefont {Bowman}}]{Dai_Bowman_Multimode_Zundel_2003}%
  \BibitemOpen
  \bibfield  {author} {\bibinfo {author} {\bibfnamefont {J.}~\bibnamefont
  {Dai}}, \bibinfo {author} {\bibfnamefont {Z.}~\bibnamefont {Ba{\v c}i{\'c}}},
  \bibinfo {author} {\bibfnamefont {X.}~\bibnamefont {Huang}}, \bibinfo
  {author} {\bibfnamefont {S.}~\bibnamefont {Carter}}, \ and\ \bibinfo {author}
  {\bibfnamefont {J.~M.}\ \bibnamefont {Bowman}},\ }\href {\doibase
  10.1063/1.1603220} {\bibfield  {journal} {\bibinfo  {journal} {J. Chem.
  Phys.}\ }\textbf {\bibinfo {volume} {119}},\ \bibinfo {pages} {6571}
  (\bibinfo {year} {2003})}\BibitemShut {NoStop}%
\bibitem [{\citenamefont {McCoy}\ \emph {et~al.}(2005)\citenamefont {McCoy},
  \citenamefont {Huang}, \citenamefont {Carter}, \citenamefont {Landeweer},\
  and\ \citenamefont {Bowman}}]{mccoy_bowman_VCIzundel_2005}%
  \BibitemOpen
  \bibfield  {author} {\bibinfo {author} {\bibfnamefont {A.~B.}\ \bibnamefont
  {McCoy}}, \bibinfo {author} {\bibfnamefont {X.}~\bibnamefont {Huang}},
  \bibinfo {author} {\bibfnamefont {S.}~\bibnamefont {Carter}}, \bibinfo
  {author} {\bibfnamefont {M.~Y.}\ \bibnamefont {Landeweer}}, \ and\ \bibinfo
  {author} {\bibfnamefont {J.~M.}\ \bibnamefont {Bowman}},\ }\href {\doibase
  10.1063/1.1857472} {\bibfield  {journal} {\bibinfo  {journal} {J. Chem.
  Phys.}\ }\textbf {\bibinfo {volume} {122}},\ \bibinfo {pages} {061101}
  (\bibinfo {year} {2005})}\BibitemShut {NoStop}%
\bibitem [{\citenamefont {Kaledin}, \citenamefont {Kaledin},\ and\
  \citenamefont {Bowman}(2006)}]{Kaledin_VibrationalAnalysisH5O2_2006}%
  \BibitemOpen
  \bibfield  {author} {\bibinfo {author} {\bibfnamefont {M.}~\bibnamefont
  {Kaledin}}, \bibinfo {author} {\bibfnamefont {A.~L.}\ \bibnamefont
  {Kaledin}}, \ and\ \bibinfo {author} {\bibfnamefont {J.~M.}\ \bibnamefont
  {Bowman}},\ }\href {\doibase 10.1021/jp054374w} {\bibfield  {journal}
  {\bibinfo  {journal} {J. Phys. Chem. A}\ }\textbf {\bibinfo {volume} {110}},\
  \bibinfo {pages} {2933} (\bibinfo {year} {2006})}\BibitemShut {NoStop}%
\bibitem [{\citenamefont {Kaledin}\ \emph {et~al.}(2009)\citenamefont
  {Kaledin}, \citenamefont {Kaledin}, \citenamefont {Bowman}, \citenamefont
  {Ding},\ and\ \citenamefont {Jordan}}]{kaledin_jordan_CPMDzundel_2009}%
  \BibitemOpen
  \bibfield  {author} {\bibinfo {author} {\bibfnamefont {M.}~\bibnamefont
  {Kaledin}}, \bibinfo {author} {\bibfnamefont {A.~L.}\ \bibnamefont
  {Kaledin}}, \bibinfo {author} {\bibfnamefont {J.~M.}\ \bibnamefont {Bowman}},
  \bibinfo {author} {\bibfnamefont {J.}~\bibnamefont {Ding}}, \ and\ \bibinfo
  {author} {\bibfnamefont {K.~D.}\ \bibnamefont {Jordan}},\ }\href@noop {}
  {\bibfield  {journal} {\bibinfo  {journal} {J. Phys. Chem. A}\ }\textbf
  {\bibinfo {volume} {113}},\ \bibinfo {pages} {7671} (\bibinfo {year}
  {2009})}\BibitemShut {NoStop}%
\bibitem [{\citenamefont {Huang}, \citenamefont {Habershon},\ and\
  \citenamefont {Bowman}(2008)}]{huang_bowman_RPMDzundel_2008}%
  \BibitemOpen
  \bibfield  {author} {\bibinfo {author} {\bibfnamefont {X.}~\bibnamefont
  {Huang}}, \bibinfo {author} {\bibfnamefont {S.}~\bibnamefont {Habershon}}, \
  and\ \bibinfo {author} {\bibfnamefont {J.~M.}\ \bibnamefont {Bowman}},\
  }\href {\doibase 10.1016/j.cplett.2007.11.048} {\bibfield  {journal}
  {\bibinfo  {journal} {Chem. Phys. Lett.}\ }\textbf {\bibinfo {volume}
  {450}},\ \bibinfo {pages} {253} (\bibinfo {year} {2008})}\BibitemShut
  {NoStop}%
\bibitem [{\citenamefont {Rossi}, \citenamefont {Ceriotti},\ and\ \citenamefont
  {Manolopoulos}(2014)}]{rossi_manolopoulos_TRPDM_2014}%
  \BibitemOpen
  \bibfield  {author} {\bibinfo {author} {\bibfnamefont {M.}~\bibnamefont
  {Rossi}}, \bibinfo {author} {\bibfnamefont {M.}~\bibnamefont {Ceriotti}}, \
  and\ \bibinfo {author} {\bibfnamefont {D.~E.}\ \bibnamefont {Manolopoulos}},\
  }\href@noop {} {\bibfield  {journal} {\bibinfo  {journal} {J. Chem. Phys.}\
  }\textbf {\bibinfo {volume} {140}},\ \bibinfo {pages} {234116} (\bibinfo
  {year} {2014})}\BibitemShut {NoStop}%
\bibitem [{\citenamefont {Rossi}, \citenamefont {Kapil},\ and\ \citenamefont
  {Ceriotti}(2017)}]{rossi_ceriotti_TRPMDLangevin_2017}%
  \BibitemOpen
  \bibfield  {author} {\bibinfo {author} {\bibfnamefont {M.}~\bibnamefont
  {Rossi}}, \bibinfo {author} {\bibfnamefont {V.}~\bibnamefont {Kapil}}, \ and\
  \bibinfo {author} {\bibfnamefont {M.}~\bibnamefont {Ceriotti}},\ }\href
  {\doibase 10.1063/1.4990536} {\bibfield  {journal} {\bibinfo  {journal} {J.
  Chem. Phys.}\ }\textbf {\bibinfo {volume} {148}},\ \bibinfo {pages} {102301}
  (\bibinfo {year} {2017})}\BibitemShut {NoStop}%
\bibitem [{\citenamefont {Di~Liberto}, \citenamefont {Conte},\ and\
  \citenamefont
  {Ceotto}(2018{\natexlab{b}})}]{DiLiberto_Ceotto_Jacobiano_2018}%
  \BibitemOpen
  \bibfield  {author} {\bibinfo {author} {\bibfnamefont {G.}~\bibnamefont
  {Di~Liberto}}, \bibinfo {author} {\bibfnamefont {R.}~\bibnamefont {Conte}}, \
  and\ \bibinfo {author} {\bibfnamefont {M.}~\bibnamefont {Ceotto}},\
  }\href@noop {} {\bibfield  {journal} {\bibinfo  {journal} {J. Chem. Phys.}\
  }\textbf {\bibinfo {volume} {148}},\ \bibinfo {pages} {014307} (\bibinfo
  {year} {2018}{\natexlab{b}})}\BibitemShut {NoStop}%
\bibitem [{\citenamefont {Vendrell}\ \emph {et~al.}(2007)\citenamefont
  {Vendrell}, \citenamefont {Gatti}, \citenamefont {Lauvergnat},\ and\
  \citenamefont {Meyer}}]{vendrell_Meyer_ZundelHamiltonian_2007}%
  \BibitemOpen
  \bibfield  {author} {\bibinfo {author} {\bibfnamefont {O.}~\bibnamefont
  {Vendrell}}, \bibinfo {author} {\bibfnamefont {F.}~\bibnamefont {Gatti}},
  \bibinfo {author} {\bibfnamefont {D.}~\bibnamefont {Lauvergnat}}, \ and\
  \bibinfo {author} {\bibfnamefont {H.-D.}\ \bibnamefont {Meyer}},\ }\href@noop
  {} {\bibfield  {journal} {\bibinfo  {journal} {J. Chem. Phys.}\ }\textbf
  {\bibinfo {volume} {127}},\ \bibinfo {pages} {184302} (\bibinfo {year}
  {2007})}\BibitemShut {NoStop}%
\bibitem [{\citenamefont {Vendrell}, \citenamefont {Gatti},\ and\ \citenamefont
  {Meyer}(2007{\natexlab{a}})}]{vendrell_Meyer_Zundeldynamics_2007}%
  \BibitemOpen
  \bibfield  {author} {\bibinfo {author} {\bibfnamefont {O.}~\bibnamefont
  {Vendrell}}, \bibinfo {author} {\bibfnamefont {F.}~\bibnamefont {Gatti}}, \
  and\ \bibinfo {author} {\bibfnamefont {H.-D.}\ \bibnamefont {Meyer}},\ }\href
  {\doibase 10.1002/anie.200702201} {\bibfield  {journal} {\bibinfo  {journal}
  {Angew. Chem. Int. Ed.}\ }\textbf {\bibinfo {volume} {46}},\ \bibinfo {pages}
  {6918} (\bibinfo {year} {2007}{\natexlab{a}})}\BibitemShut {NoStop}%
\bibitem [{\citenamefont {Vendrell}, \citenamefont {Gatti},\ and\ \citenamefont
  {Meyer}(2007{\natexlab{b}})}]{vendrell_Meyer_zundelspectra_2007}%
  \BibitemOpen
  \bibfield  {author} {\bibinfo {author} {\bibfnamefont {O.}~\bibnamefont
  {Vendrell}}, \bibinfo {author} {\bibfnamefont {F.}~\bibnamefont {Gatti}}, \
  and\ \bibinfo {author} {\bibfnamefont {H.-D.}\ \bibnamefont {Meyer}},\
  }\href@noop {} {\bibfield  {journal} {\bibinfo  {journal} {J. Chem. Phys.}\
  }\textbf {\bibinfo {volume} {127}},\ \bibinfo {pages} {184303} (\bibinfo
  {year} {2007}{\natexlab{b}})}\BibitemShut {NoStop}%
\bibitem [{\citenamefont {Vendrell}\ and\ \citenamefont
  {Meyer}(2008)}]{vendrell_meyer_zundelquantumdynamics_2008}%
  \BibitemOpen
  \bibfield  {author} {\bibinfo {author} {\bibfnamefont {O.}~\bibnamefont
  {Vendrell}}\ and\ \bibinfo {author} {\bibfnamefont {H.-D.}\ \bibnamefont
  {Meyer}},\ }\href@noop {} {\bibfield  {journal} {\bibinfo  {journal} {Phys.
  Chem. Chem. Phys.}\ }\textbf {\bibinfo {volume} {10}},\ \bibinfo {pages}
  {4692} (\bibinfo {year} {2008})}\BibitemShut {NoStop}%
\bibitem [{\citenamefont {Vendrell}\ \emph {et~al.}(2009)\citenamefont
  {Vendrell}, \citenamefont {Brill}, \citenamefont {Gatti}, \citenamefont
  {Lauvergnat},\ and\ \citenamefont
  {Meyer}}]{vendrell_Meyer_Jacobianparametriz_2009}%
  \BibitemOpen
  \bibfield  {author} {\bibinfo {author} {\bibfnamefont {O.}~\bibnamefont
  {Vendrell}}, \bibinfo {author} {\bibfnamefont {M.}~\bibnamefont {Brill}},
  \bibinfo {author} {\bibfnamefont {F.}~\bibnamefont {Gatti}}, \bibinfo
  {author} {\bibfnamefont {D.}~\bibnamefont {Lauvergnat}}, \ and\ \bibinfo
  {author} {\bibfnamefont {H.-D.}\ \bibnamefont {Meyer}},\ }\href@noop {}
  {\bibfield  {journal} {\bibinfo  {journal} {J. Chem. Phys.}\ }\textbf
  {\bibinfo {volume} {130}},\ \bibinfo {pages} {234305} (\bibinfo {year}
  {2009})}\BibitemShut {NoStop}%
\bibitem [{\citenamefont {Vendrell}, \citenamefont {Gatti},\ and\ \citenamefont
  {Meyer}(2009{\natexlab{a}})}]{vendrell_Meyer_isotopeeffects_2009}%
  \BibitemOpen
  \bibfield  {author} {\bibinfo {author} {\bibfnamefont {O.}~\bibnamefont
  {Vendrell}}, \bibinfo {author} {\bibfnamefont {F.}~\bibnamefont {Gatti}}, \
  and\ \bibinfo {author} {\bibfnamefont {H.-D.}\ \bibnamefont {Meyer}},\ }\href
  {\doibase 10.1002/anie.200804646} {\bibfield  {journal} {\bibinfo  {journal}
  {Angew. Chem. Int. Ed.}\ }\textbf {\bibinfo {volume} {48}},\ \bibinfo {pages}
  {352} (\bibinfo {year} {2009}{\natexlab{a}})}\BibitemShut {NoStop}%
\bibitem [{\citenamefont {Vendrell}, \citenamefont {Gatti},\ and\ \citenamefont
  {Meyer}(2009{\natexlab{b}})}]{vendrell_Meyer_isotopeffects2_2009}%
  \BibitemOpen
  \bibfield  {author} {\bibinfo {author} {\bibfnamefont {O.}~\bibnamefont
  {Vendrell}}, \bibinfo {author} {\bibfnamefont {F.}~\bibnamefont {Gatti}}, \
  and\ \bibinfo {author} {\bibfnamefont {H.-D.}\ \bibnamefont {Meyer}},\
  }\href@noop {} {\bibfield  {journal} {\bibinfo  {journal} {J. Chem. Phys.}\
  }\textbf {\bibinfo {volume} {131}},\ \bibinfo {pages} {034308} (\bibinfo
  {year} {2009}{\natexlab{b}})}\BibitemShut {NoStop}%
\bibitem [{\citenamefont {Baer}, \citenamefont {Marx},\ and\ \citenamefont
  {Mathias}(2010)}]{baer_mathias_AIMDzundel_2010}%
  \BibitemOpen
  \bibfield  {author} {\bibinfo {author} {\bibfnamefont {M.}~\bibnamefont
  {Baer}}, \bibinfo {author} {\bibfnamefont {D.}~\bibnamefont {Marx}}, \ and\
  \bibinfo {author} {\bibfnamefont {G.}~\bibnamefont {Mathias}},\ }\href@noop
  {} {\bibfield  {journal} {\bibinfo  {journal} {Angew. Chem. Int. Ed.}\
  }\textbf {\bibinfo {volume} {49}},\ \bibinfo {pages} {7346} (\bibinfo {year}
  {2010})}\BibitemShut {NoStop}%
\bibitem [{\citenamefont {Pitsevich}\ \emph {et~al.}(2017)\citenamefont
  {Pitsevich}, \citenamefont {Malevich}, \citenamefont {Kozlovskaya},
  \citenamefont {Mahnach}, \citenamefont {Doroshenko}, \citenamefont
  {Pogorelov}, \citenamefont {Pettersson}, \citenamefont {Sablinskas},\ and\
  \citenamefont {Balevicius}}]{Pitsevich_MP4StudyAnharmonic_2017}%
  \BibitemOpen
  \bibfield  {author} {\bibinfo {author} {\bibfnamefont {G.}~\bibnamefont
  {Pitsevich}}, \bibinfo {author} {\bibfnamefont {A.}~\bibnamefont {Malevich}},
  \bibinfo {author} {\bibfnamefont {E.}~\bibnamefont {Kozlovskaya}}, \bibinfo
  {author} {\bibfnamefont {E.}~\bibnamefont {Mahnach}}, \bibinfo {author}
  {\bibfnamefont {I.}~\bibnamefont {Doroshenko}}, \bibinfo {author}
  {\bibfnamefont {V.}~\bibnamefont {Pogorelov}}, \bibinfo {author}
  {\bibfnamefont {L.~G.~M.}\ \bibnamefont {Pettersson}}, \bibinfo {author}
  {\bibfnamefont {V.}~\bibnamefont {Sablinskas}}, \ and\ \bibinfo {author}
  {\bibfnamefont {V.}~\bibnamefont {Balevicius}},\ }\href {\doibase
  10.1021/acs.jpca.7b00536} {\bibfield  {journal} {\bibinfo  {journal} {J.
  Phys. Chem. A}\ }\textbf {\bibinfo {volume} {121}},\ \bibinfo {pages} {2151}
  (\bibinfo {year} {2017})}\BibitemShut {NoStop}%
\bibitem [{\citenamefont {Spura}, \citenamefont {Elgabarty},\ and\
  \citenamefont {D.~K\"uhne}(2015)}]{Spura_Ontheflycoupledcluster_2015}%
  \BibitemOpen
  \bibfield  {author} {\bibinfo {author} {\bibfnamefont {T.}~\bibnamefont
  {Spura}}, \bibinfo {author} {\bibfnamefont {H.}~\bibnamefont {Elgabarty}}, \
  and\ \bibinfo {author} {\bibfnamefont {T.}~\bibnamefont {D.~K\"uhne}},\
  }\href {\doibase 10.1039/C4CP05192K} {\bibfield  {journal} {\bibinfo
  {journal} {Phys. Chem. Chem. Phys.}\ }\textbf {\bibinfo {volume} {17}},\
  \bibinfo {pages} {14355} (\bibinfo {year} {2015})}\BibitemShut {NoStop}%
\bibitem [{\citenamefont {Spura}, \citenamefont {Elgabarty},\ and\
  \citenamefont {K\"uhne}(2015)}]{Spura_CorrectionOntheflycoupled_2015}%
  \BibitemOpen
  \bibfield  {author} {\bibinfo {author} {\bibfnamefont {T.}~\bibnamefont
  {Spura}}, \bibinfo {author} {\bibfnamefont {H.}~\bibnamefont {Elgabarty}}, \
  and\ \bibinfo {author} {\bibfnamefont {T.~D.}\ \bibnamefont {K\"uhne}},\
  }\href {\doibase 10.1039/C5CP90118A} {\bibfield  {journal} {\bibinfo
  {journal} {Phys. Chem. Chem. Phys.}\ }\textbf {\bibinfo {volume} {17}},\
  \bibinfo {pages} {19673} (\bibinfo {year} {2015})}\BibitemShut {NoStop}%
\bibitem [{\citenamefont {Schran}\ \emph {et~al.}(2017)\citenamefont {Schran},
  \citenamefont {Uhl}, \citenamefont {Behler},\ and\ \citenamefont
  {Marx}}]{Schran_Highdimensionalneuralnetwork_2017}%
  \BibitemOpen
  \bibfield  {author} {\bibinfo {author} {\bibfnamefont {C.}~\bibnamefont
  {Schran}}, \bibinfo {author} {\bibfnamefont {F.}~\bibnamefont {Uhl}},
  \bibinfo {author} {\bibfnamefont {J.}~\bibnamefont {Behler}}, \ and\ \bibinfo
  {author} {\bibfnamefont {D.}~\bibnamefont {Marx}},\ }\href {\doibase
  10.1063/1.4996819} {\bibfield  {journal} {\bibinfo  {journal} {J. Chem.
  Phys.}\ }\textbf {\bibinfo {volume} {148}},\ \bibinfo {pages} {102310}
  (\bibinfo {year} {2017})}\BibitemShut {NoStop}%
\bibitem [{\citenamefont {Schran}, \citenamefont {Brieuc},\ and\ \citenamefont
  {Marx}(2018)}]{Schran_ConvergedColoredNoise_2018}%
  \BibitemOpen
  \bibfield  {author} {\bibinfo {author} {\bibfnamefont {C.}~\bibnamefont
  {Schran}}, \bibinfo {author} {\bibfnamefont {F.}~\bibnamefont {Brieuc}}, \
  and\ \bibinfo {author} {\bibfnamefont {D.}~\bibnamefont {Marx}},\ }\href
  {\doibase 10.1021/acs.jctc.8b00705} {\bibfield  {journal} {\bibinfo
  {journal} {J. Chem. Theory Comput.}\ }\textbf {\bibinfo {volume} {14}},\
  \bibinfo {pages} {5068} (\bibinfo {year} {2018})}\BibitemShut {NoStop}%
\bibitem [{\citenamefont {Dagrada}\ \emph {et~al.}(2014)\citenamefont
  {Dagrada}, \citenamefont {Casula}, \citenamefont {Saitta}, \citenamefont
  {Sorella},\ and\ \citenamefont {Mauri}}]{Dagrada_QuantumMonteCarlo_2014}%
  \BibitemOpen
  \bibfield  {author} {\bibinfo {author} {\bibfnamefont {M.}~\bibnamefont
  {Dagrada}}, \bibinfo {author} {\bibfnamefont {M.}~\bibnamefont {Casula}},
  \bibinfo {author} {\bibfnamefont {A.~M.}\ \bibnamefont {Saitta}}, \bibinfo
  {author} {\bibfnamefont {S.}~\bibnamefont {Sorella}}, \ and\ \bibinfo
  {author} {\bibfnamefont {F.}~\bibnamefont {Mauri}},\ }\href {\doibase
  10.1021/ct401077x} {\bibfield  {journal} {\bibinfo  {journal} {J. Chem.
  Theory Comput.}\ }\textbf {\bibinfo {volume} {10}},\ \bibinfo {pages} {1980}
  (\bibinfo {year} {2014})}\BibitemShut {NoStop}%
\bibitem [{\citenamefont {Mouhat}\ \emph {et~al.}(2017)\citenamefont {Mouhat},
  \citenamefont {Sorella}, \citenamefont {Vuilleumier}, \citenamefont
  {Saitta},\ and\ \citenamefont
  {Casula}}]{Mouhat_FullyQuantumDescription_2017}%
  \BibitemOpen
  \bibfield  {author} {\bibinfo {author} {\bibfnamefont {F.}~\bibnamefont
  {Mouhat}}, \bibinfo {author} {\bibfnamefont {S.}~\bibnamefont {Sorella}},
  \bibinfo {author} {\bibfnamefont {R.}~\bibnamefont {Vuilleumier}}, \bibinfo
  {author} {\bibfnamefont {A.~M.}\ \bibnamefont {Saitta}}, \ and\ \bibinfo
  {author} {\bibfnamefont {M.}~\bibnamefont {Casula}},\ }\href {\doibase
  10.1021/acs.jctc.7b00017} {\bibfield  {journal} {\bibinfo  {journal} {J.
  Chem. Theory Comput.}\ }\textbf {\bibinfo {volume} {13}},\ \bibinfo {pages}
  {2400} (\bibinfo {year} {2017})}\BibitemShut {NoStop}%
\bibitem [{\citenamefont {McLachlan}\ and\ \citenamefont
  {Atela}(1992)}]{McLachlan_accuracysymplecticintegrators_1992}%
  \BibitemOpen
  \bibfield  {author} {\bibinfo {author} {\bibfnamefont {R.~I.}\ \bibnamefont
  {McLachlan}}\ and\ \bibinfo {author} {\bibfnamefont {P.}~\bibnamefont
  {Atela}},\ }\href {\doibase 10.1088/0951-7715/5/2/011} {\bibfield  {journal}
  {\bibinfo  {journal} {Nonlinearity}\ }\textbf {\bibinfo {volume} {5}},\
  \bibinfo {pages} {541} (\bibinfo {year} {1992})}\BibitemShut {NoStop}%
\bibitem [{\citenamefont {Brewer}, \citenamefont {Hulme},\ and\ \citenamefont
  {Manolopoulos}(1997)}]{Brewer_Manolopoulos_15dof_1997}%
  \BibitemOpen
  \bibfield  {author} {\bibinfo {author} {\bibfnamefont {M.~L.}\ \bibnamefont
  {Brewer}}, \bibinfo {author} {\bibfnamefont {J.~S.}\ \bibnamefont {Hulme}}, \
  and\ \bibinfo {author} {\bibfnamefont {D.~E.}\ \bibnamefont {Manolopoulos}},\
  }\href@noop {} {\bibfield  {journal} {\bibinfo  {journal} {J. Chem. Phys.}\
  }\textbf {\bibinfo {volume} {106}},\ \bibinfo {pages} {4832} (\bibinfo {year}
  {1997})}\BibitemShut {NoStop}%
\bibitem [{\citenamefont {Feynman}\ and\ \citenamefont
  {Hibbs}(1965)}]{feynman_pathintegral_1965}%
  \BibitemOpen
  \bibfield  {author} {\bibinfo {author} {\bibfnamefont {R.~P.}\ \bibnamefont
  {Feynman}}\ and\ \bibinfo {author} {\bibfnamefont {A.~R.}\ \bibnamefont
  {Hibbs}},\ }\href@noop {} {\emph {\bibinfo {title} {{Quantum mechanics and
  path integrals}}}}\ (\bibinfo  {publisher} {McGraw-Hill},\ \bibinfo {year}
  {1965})\BibitemShut {NoStop}%
\bibitem [{\citenamefont
  {Van~Vleck}(1928)}]{VanVleck_CorrespondencePrincipleStatistical_1928}%
  \BibitemOpen
  \bibfield  {author} {\bibinfo {author} {\bibfnamefont {J.~H.}\ \bibnamefont
  {Van~Vleck}},\ }\href@noop {} {\bibfield  {journal} {\bibinfo  {journal}
  {Proc. Natl. Acad. Sci.}\ }\textbf {\bibinfo {volume} {14}},\ \bibinfo
  {pages} {178} (\bibinfo {year} {1928})}\BibitemShut {NoStop}%
\bibitem [{\citenamefont {Miller}(1970)}]{Miller_Atom-Diatom_1970}%
  \BibitemOpen
  \bibfield  {author} {\bibinfo {author} {\bibfnamefont {W.~H.}\ \bibnamefont
  {Miller}},\ }\href {\doibase 10.1063/1.1674275} {\bibfield  {journal}
  {\bibinfo  {journal} {J. Chem. Phys.}\ }\textbf {\bibinfo {volume} {53}},\
  \bibinfo {pages} {1949} (\bibinfo {year} {1970})}\BibitemShut {NoStop}%
\bibitem [{\citenamefont
  {Heller}(1981{\natexlab{a}})}]{Heller_FrozenGaussian_1981}%
  \BibitemOpen
  \bibfield  {author} {\bibinfo {author} {\bibfnamefont {E.~J.}\ \bibnamefont
  {Heller}},\ }\href {\doibase 10.1063/1.442382} {\bibfield  {journal}
  {\bibinfo  {journal} {J. Chem. Phys.}\ }\textbf {\bibinfo {volume} {75}},\
  \bibinfo {pages} {2923} (\bibinfo {year} {1981}{\natexlab{a}})}\BibitemShut
  {NoStop}%
\bibitem [{\citenamefont
  {Heller}(1981{\natexlab{b}})}]{Heller_SCspectroscopy_1981}%
  \BibitemOpen
  \bibfield  {author} {\bibinfo {author} {\bibfnamefont {E.~J.}\ \bibnamefont
  {Heller}},\ }\href@noop {} {\bibfield  {journal} {\bibinfo  {journal} {Acc.
  Chem. Res.}\ }\textbf {\bibinfo {volume} {14}},\ \bibinfo {pages} {368}
  (\bibinfo {year} {1981}{\natexlab{b}})}\BibitemShut {NoStop}%
\bibitem [{\citenamefont {Heller}(1991)}]{Heller_Cellulardynamics_1991}%
  \BibitemOpen
  \bibfield  {author} {\bibinfo {author} {\bibfnamefont {E.~J.}\ \bibnamefont
  {Heller}},\ }\href {\doibase 10.1063/1.459848} {\bibfield  {journal}
  {\bibinfo  {journal} {J. Chem. Phys.}\ }\textbf {\bibinfo {volume} {94}},\
  \bibinfo {pages} {2723} (\bibinfo {year} {1991})}\BibitemShut {NoStop}%
\bibitem [{\citenamefont {Herman}\ and\ \citenamefont
  {Kluk}(1984)}]{Herman_Kluk_SCnonspreading_1984}%
  \BibitemOpen
  \bibfield  {author} {\bibinfo {author} {\bibfnamefont {M.~F.}\ \bibnamefont
  {Herman}}\ and\ \bibinfo {author} {\bibfnamefont {E.}~\bibnamefont {Kluk}},\
  }\href {\doibase 10.1016/0301-0104(84)80039-7} {\bibfield  {journal}
  {\bibinfo  {journal} {Chem. Phys.}\ }\textbf {\bibinfo {volume} {91}},\
  \bibinfo {pages} {27} (\bibinfo {year} {1984})}\BibitemShut {NoStop}%
\bibitem [{\citenamefont {Kay}(1994{\natexlab{a}})}]{Kay_Numerical_1994}%
  \BibitemOpen
  \bibfield  {author} {\bibinfo {author} {\bibfnamefont {K.~G.}\ \bibnamefont
  {Kay}},\ }\href@noop {} {\bibfield  {journal} {\bibinfo  {journal} {J. Chem.
  Phys.}\ }\textbf {\bibinfo {volume} {100}},\ \bibinfo {pages} {4432}
  (\bibinfo {year} {1994}{\natexlab{a}})}\BibitemShut {NoStop}%
\bibitem [{\citenamefont
  {Kay}(1994{\natexlab{b}})}]{Kay_Integralexpression_1994}%
  \BibitemOpen
  \bibfield  {author} {\bibinfo {author} {\bibfnamefont {K.~G.}\ \bibnamefont
  {Kay}},\ }\href@noop {} {\bibfield  {journal} {\bibinfo  {journal} {J. Chem.
  Phys.}\ }\textbf {\bibinfo {volume} {100}},\ \bibinfo {pages} {4377}
  (\bibinfo {year} {1994}{\natexlab{b}})}\BibitemShut {NoStop}%
\bibitem [{\citenamefont {Kay}(1994{\natexlab{c}})}]{Kay_Multidim_1994}%
  \BibitemOpen
  \bibfield  {author} {\bibinfo {author} {\bibfnamefont {K.~G.}\ \bibnamefont
  {Kay}},\ }\href@noop {} {\bibfield  {journal} {\bibinfo  {journal} {J. Chem.
  Phys.}\ }\textbf {\bibinfo {volume} {101}},\ \bibinfo {pages} {2250}
  (\bibinfo {year} {1994}{\natexlab{c}})}\BibitemShut {NoStop}%
\bibitem [{\citenamefont {Kay}(2006)}]{Kay_SCcorrections_2006}%
  \BibitemOpen
  \bibfield  {author} {\bibinfo {author} {\bibfnamefont {K.~G.}\ \bibnamefont
  {Kay}},\ }\href {\doibase 10.1016/j.chemphys.2005.06.019} {\bibfield
  {journal} {\bibinfo  {journal} {Chem. Phys.}\ }\textbf {\bibinfo {volume}
  {322}},\ \bibinfo {pages} {3} (\bibinfo {year} {2006})}\BibitemShut {NoStop}%
\bibitem [{\citenamefont {Church}, \citenamefont {Antipov},\ and\ \citenamefont
  {Ananth}(2017)}]{Nandini_Church_Mixedqcl_2017}%
  \BibitemOpen
  \bibfield  {author} {\bibinfo {author} {\bibfnamefont {M.}~\bibnamefont
  {Church}}, \bibinfo {author} {\bibfnamefont {S.~V.}\ \bibnamefont {Antipov}},
  \ and\ \bibinfo {author} {\bibfnamefont {N.}~\bibnamefont {Ananth}},\
  }\href@noop {} {\bibfield  {journal} {\bibinfo  {journal} {J. Chem. Phys.}\
  }\textbf {\bibinfo {volume} {146}},\ \bibinfo {pages} {234104} (\bibinfo
  {year} {2017})}\BibitemShut {NoStop}%
\bibitem [{\citenamefont {Antipov}, \citenamefont {Ye},\ and\ \citenamefont
  {Ananth}(2015)}]{Antipov_Nandini_Mixedqcl_2015}%
  \BibitemOpen
  \bibfield  {author} {\bibinfo {author} {\bibfnamefont {S.~V.}\ \bibnamefont
  {Antipov}}, \bibinfo {author} {\bibfnamefont {Z.}~\bibnamefont {Ye}}, \ and\
  \bibinfo {author} {\bibfnamefont {N.}~\bibnamefont {Ananth}},\ }\href@noop {}
  {\bibfield  {journal} {\bibinfo  {journal} {J. Chem. Phys.}\ }\textbf
  {\bibinfo {volume} {142}},\ \bibinfo {pages} {184102} (\bibinfo {year}
  {2015})}\BibitemShut {NoStop}%
\bibitem [{\citenamefont {Bonfanti}\ \emph {et~al.}(2018)\citenamefont
  {Bonfanti}, \citenamefont {Petersen}, \citenamefont {Eisenbrandt},
  \citenamefont {Burghardt},\ and\ \citenamefont
  {Pollak}}]{Bonfanti_ComputationS1S0_2018}%
  \BibitemOpen
  \bibfield  {author} {\bibinfo {author} {\bibfnamefont {M.}~\bibnamefont
  {Bonfanti}}, \bibinfo {author} {\bibfnamefont {J.}~\bibnamefont {Petersen}},
  \bibinfo {author} {\bibfnamefont {P.}~\bibnamefont {Eisenbrandt}}, \bibinfo
  {author} {\bibfnamefont {I.}~\bibnamefont {Burghardt}}, \ and\ \bibinfo
  {author} {\bibfnamefont {E.}~\bibnamefont {Pollak}},\ }\href {\doibase
  10.1021/acs.jctc.8b00355} {\bibfield  {journal} {\bibinfo  {journal} {J.
  Chem. Theory Comput.}\ }\textbf {\bibinfo {volume} {14}},\ \bibinfo {pages}
  {5310} (\bibinfo {year} {2018})}\BibitemShut {NoStop}%
\bibitem [{\citenamefont {Kaledin}\ and\ \citenamefont
  {Miller}(2003{\natexlab{a}})}]{Kaledin_Miller_Timeaveraging_2003}%
  \BibitemOpen
  \bibfield  {author} {\bibinfo {author} {\bibfnamefont {A.~L.}\ \bibnamefont
  {Kaledin}}\ and\ \bibinfo {author} {\bibfnamefont {W.~H.}\ \bibnamefont
  {Miller}},\ }\href {\doibase 10.1063/1.1562158} {\bibfield  {journal}
  {\bibinfo  {journal} {J. Chem. Phys.}\ }\textbf {\bibinfo {volume} {118}},\
  \bibinfo {pages} {7174} (\bibinfo {year} {2003}{\natexlab{a}})}\BibitemShut
  {NoStop}%
\bibitem [{\citenamefont {Kaledin}\ and\ \citenamefont
  {Miller}(2003{\natexlab{b}})}]{Kaledin_Miller_TAmolecules_2003}%
  \BibitemOpen
  \bibfield  {author} {\bibinfo {author} {\bibfnamefont {A.~L.}\ \bibnamefont
  {Kaledin}}\ and\ \bibinfo {author} {\bibfnamefont {W.~H.}\ \bibnamefont
  {Miller}},\ }\href {\doibase 10.1063/1.1589477} {\bibfield  {journal}
  {\bibinfo  {journal} {J. Chem. Phys.}\ }\textbf {\bibinfo {volume} {119}},\
  \bibinfo {pages} {3078} (\bibinfo {year} {2003}{\natexlab{b}})}\BibitemShut
  {NoStop}%
\bibitem [{\citenamefont {Tamascelli}\ \emph {et~al.}(2014)\citenamefont
  {Tamascelli}, \citenamefont {Dambrosio}, \citenamefont {Conte},\ and\
  \citenamefont {Ceotto}}]{Tamascelli_Ceotto_GPU_2014}%
  \BibitemOpen
  \bibfield  {author} {\bibinfo {author} {\bibfnamefont {D.}~\bibnamefont
  {Tamascelli}}, \bibinfo {author} {\bibfnamefont {F.~S.}\ \bibnamefont
  {Dambrosio}}, \bibinfo {author} {\bibfnamefont {R.}~\bibnamefont {Conte}}, \
  and\ \bibinfo {author} {\bibfnamefont {M.}~\bibnamefont {Ceotto}},\
  }\href@noop {} {\bibfield  {journal} {\bibinfo  {journal} {J. Chem. Phys.}\
  }\textbf {\bibinfo {volume} {140}},\ \bibinfo {pages} {174109} (\bibinfo
  {year} {2014})}\BibitemShut {NoStop}%
\bibitem [{\citenamefont {{Di Liberto}}\ and\ \citenamefont
  {Ceotto}(2016)}]{DiLiberto_Ceotto_Prefactors_2016}%
  \BibitemOpen
  \bibfield  {author} {\bibinfo {author} {\bibfnamefont {G.}~\bibnamefont {{Di
  Liberto}}}\ and\ \bibinfo {author} {\bibfnamefont {M.}~\bibnamefont
  {Ceotto}},\ }\href@noop {} {\bibfield  {journal} {\bibinfo  {journal} {J.
  Chem. Phys.}\ }\textbf {\bibinfo {volume} {145}},\ \bibinfo {pages} {144107}
  (\bibinfo {year} {2016})}\BibitemShut {NoStop}%
\bibitem [{\citenamefont {Zhuang}\ \emph {et~al.}(2012)\citenamefont {Zhuang},
  \citenamefont {Siebert}, \citenamefont {Hase}, \citenamefont {Kay},\ and\
  \citenamefont {Ceotto}}]{Zhuang_Ceotto_Hessianapprox_2012}%
  \BibitemOpen
  \bibfield  {author} {\bibinfo {author} {\bibfnamefont {Y.}~\bibnamefont
  {Zhuang}}, \bibinfo {author} {\bibfnamefont {M.~R.}\ \bibnamefont {Siebert}},
  \bibinfo {author} {\bibfnamefont {W.~L.}\ \bibnamefont {Hase}}, \bibinfo
  {author} {\bibfnamefont {K.~G.}\ \bibnamefont {Kay}}, \ and\ \bibinfo
  {author} {\bibfnamefont {M.}~\bibnamefont {Ceotto}},\ }\href@noop {}
  {\bibfield  {journal} {\bibinfo  {journal} {J. Chem. Theory Comput.}\
  }\textbf {\bibinfo {volume} {9}},\ \bibinfo {pages} {54} (\bibinfo {year}
  {2012})}\BibitemShut {NoStop}%
\bibitem [{\citenamefont {Ceotto}, \citenamefont {Zhuang},\ and\ \citenamefont
  {Hase}(2013)}]{Ceotto_Hase_AcceleratedSC_2013}%
  \BibitemOpen
  \bibfield  {author} {\bibinfo {author} {\bibfnamefont {M.}~\bibnamefont
  {Ceotto}}, \bibinfo {author} {\bibfnamefont {Y.}~\bibnamefont {Zhuang}}, \
  and\ \bibinfo {author} {\bibfnamefont {W.~L.}\ \bibnamefont {Hase}},\
  }\href@noop {} {\bibfield  {journal} {\bibinfo  {journal} {J. Chem. Phys.}\
  }\textbf {\bibinfo {volume} {138}},\ \bibinfo {pages} {054116} (\bibinfo
  {year} {2013})}\BibitemShut {NoStop}%
\bibitem [{\citenamefont {Buchholz}\ \emph {et~al.}(2018)\citenamefont
  {Buchholz}, \citenamefont {Fallacara}, \citenamefont {Gottwald},
  \citenamefont {Ceotto}, \citenamefont {Grossmann},\ and\ \citenamefont
  {Ivanov}}]{buchholz_ivanov_2018_ZPEL}%
  \BibitemOpen
  \bibfield  {author} {\bibinfo {author} {\bibfnamefont {M.}~\bibnamefont
  {Buchholz}}, \bibinfo {author} {\bibfnamefont {E.}~\bibnamefont {Fallacara}},
  \bibinfo {author} {\bibfnamefont {F.}~\bibnamefont {Gottwald}}, \bibinfo
  {author} {\bibfnamefont {M.}~\bibnamefont {Ceotto}}, \bibinfo {author}
  {\bibfnamefont {F.}~\bibnamefont {Grossmann}}, \ and\ \bibinfo {author}
  {\bibfnamefont {S.~D.}\ \bibnamefont {Ivanov}},\ }\href {\doibase
  https://doi.org/10.1016/j.chemphys.2018.06.008} {\bibfield  {journal}
  {\bibinfo  {journal} {Chem. Phys.}\ }\textbf {\bibinfo {volume} {515}},\
  \bibinfo {pages} {231 } (\bibinfo {year} {2018})},\ \bibinfo {note}
  {{Ultrafast Photoinduced Processes in Polyatomic Molecules: Electronic
  Structure, Dynamics and Spectroscopy (Dedicated to Wolfgang Domcke on the
  occasion of his 70th birthday)}}\BibitemShut {NoStop}%
\bibitem [{\citenamefont {Conte}\ \emph {et~al.}(2019)\citenamefont {Conte},
  \citenamefont {Gabas}, \citenamefont {Botti}, \citenamefont {Zhuang},\ and\
  \citenamefont {Ceotto}}]{Conte_HessianDatabase_2019}%
  \BibitemOpen
  \bibfield  {author} {\bibinfo {author} {\bibfnamefont {R.}~\bibnamefont
  {Conte}}, \bibinfo {author} {\bibfnamefont {F.}~\bibnamefont {Gabas}},
  \bibinfo {author} {\bibfnamefont {G.}~\bibnamefont {Botti}}, \bibinfo
  {author} {\bibfnamefont {Y.}~\bibnamefont {Zhuang}}, \ and\ \bibinfo {author}
  {\bibfnamefont {M.}~\bibnamefont {Ceotto}},\ }\href {\doibase
  10.1063/1.5109086} {\bibfield  {journal} {\bibinfo  {journal} {J. Chem.
  Phys.}\ }\textbf {\bibinfo {volume} {150}},\ \bibinfo {pages} {244118}
  (\bibinfo {year} {2019})}\BibitemShut {NoStop}%
\bibitem [{\citenamefont {Ceotto}\ \emph
  {et~al.}(2009{\natexlab{a}})\citenamefont {Ceotto}, \citenamefont {Atahan},
  \citenamefont {Shim}, \citenamefont {Tantardini},\ and\ \citenamefont
  {Aspuru-Guzik}}]{Ceotto_AspuruGuzik_PCCPFirstprinciples_2009}%
  \BibitemOpen
  \bibfield  {author} {\bibinfo {author} {\bibfnamefont {M.}~\bibnamefont
  {Ceotto}}, \bibinfo {author} {\bibfnamefont {S.}~\bibnamefont {Atahan}},
  \bibinfo {author} {\bibfnamefont {S.}~\bibnamefont {Shim}}, \bibinfo {author}
  {\bibfnamefont {G.~F.}\ \bibnamefont {Tantardini}}, \ and\ \bibinfo {author}
  {\bibfnamefont {A.}~\bibnamefont {Aspuru-Guzik}},\ }\href {\doibase
  10.1039/B820785B} {\bibfield  {journal} {\bibinfo  {journal} {Phys. Chem.
  Chem. Phys.}\ }\textbf {\bibinfo {volume} {11}},\ \bibinfo {pages} {3861}
  (\bibinfo {year} {2009}{\natexlab{a}})}\BibitemShut {NoStop}%
\bibitem [{\citenamefont {Ceotto}\ \emph
  {et~al.}(2009{\natexlab{b}})\citenamefont {Ceotto}, \citenamefont {Atahan},
  \citenamefont {Tantardini},\ and\ \citenamefont
  {Aspuru-Guzik}}]{Ceotto_AspuruGuzik_Multiplecoherent_2009}%
  \BibitemOpen
  \bibfield  {author} {\bibinfo {author} {\bibfnamefont {M.}~\bibnamefont
  {Ceotto}}, \bibinfo {author} {\bibfnamefont {S.}~\bibnamefont {Atahan}},
  \bibinfo {author} {\bibfnamefont {G.~F.}\ \bibnamefont {Tantardini}}, \ and\
  \bibinfo {author} {\bibfnamefont {A.}~\bibnamefont {Aspuru-Guzik}},\ }\href
  {\doibase 10.1063/1.3155062} {\bibfield  {journal} {\bibinfo  {journal} {J.
  Chem. Phys.}\ }\textbf {\bibinfo {volume} {130}},\ \bibinfo {pages} {234113}
  (\bibinfo {year} {2009}{\natexlab{b}})}\BibitemShut {NoStop}%
\bibitem [{\citenamefont {Ceotto}, \citenamefont {Tantardini},\ and\
  \citenamefont
  {Aspuru-Guzik}(2011)}]{Ceotto_AspuruGuzik_Curseofdimensionality_2011}%
  \BibitemOpen
  \bibfield  {author} {\bibinfo {author} {\bibfnamefont {M.}~\bibnamefont
  {Ceotto}}, \bibinfo {author} {\bibfnamefont {G.~F.}\ \bibnamefont
  {Tantardini}}, \ and\ \bibinfo {author} {\bibfnamefont {A.}~\bibnamefont
  {Aspuru-Guzik}},\ }\href {\doibase 10.1063/1.3664731} {\bibfield  {journal}
  {\bibinfo  {journal} {J. Chem. Phys.}\ }\textbf {\bibinfo {volume} {135}},\
  \bibinfo {pages} {214108} (\bibinfo {year} {2011})}\BibitemShut {NoStop}%
\bibitem [{\citenamefont {Ceotto}\ \emph {et~al.}(2011)\citenamefont {Ceotto},
  \citenamefont {Valleau}, \citenamefont {Tantardini},\ and\ \citenamefont
  {Aspuru-Guzik}}]{Ceotto_AspuruGuzik_Firstprinciples_2011}%
  \BibitemOpen
  \bibfield  {author} {\bibinfo {author} {\bibfnamefont {M.}~\bibnamefont
  {Ceotto}}, \bibinfo {author} {\bibfnamefont {S.}~\bibnamefont {Valleau}},
  \bibinfo {author} {\bibfnamefont {G.~F.}\ \bibnamefont {Tantardini}}, \ and\
  \bibinfo {author} {\bibfnamefont {A.}~\bibnamefont {Aspuru-Guzik}},\ }\href
  {\doibase 10.1063/1.3599469} {\bibfield  {journal} {\bibinfo  {journal} {J.
  Chem. Phys.}\ }\textbf {\bibinfo {volume} {134}},\ \bibinfo {pages} {234103}
  (\bibinfo {year} {2011})}\BibitemShut {NoStop}%
\bibitem [{\citenamefont {Ceotto}, \citenamefont {{Dell'Angelo}},\ and\
  \citenamefont {Tantardini}(2010)}]{Ceotto_Tantardini_Copper100_2010}%
  \BibitemOpen
  \bibfield  {author} {\bibinfo {author} {\bibfnamefont {M.}~\bibnamefont
  {Ceotto}}, \bibinfo {author} {\bibfnamefont {D.}~\bibnamefont
  {{Dell'Angelo}}}, \ and\ \bibinfo {author} {\bibfnamefont {G.~F.}\
  \bibnamefont {Tantardini}},\ }\href@noop {} {\bibfield  {journal} {\bibinfo
  {journal} {J. Chem. Phys.}\ }\textbf {\bibinfo {volume} {133}},\ \bibinfo
  {pages} {054701} (\bibinfo {year} {2010})}\BibitemShut {NoStop}%
\bibitem [{\citenamefont {Conte}, \citenamefont {Aspuru-Guzik},\ and\
  \citenamefont {Ceotto}(2013)}]{Conte_Ceotto_NH3_2013}%
  \BibitemOpen
  \bibfield  {author} {\bibinfo {author} {\bibfnamefont {R.}~\bibnamefont
  {Conte}}, \bibinfo {author} {\bibfnamefont {A.}~\bibnamefont {Aspuru-Guzik}},
  \ and\ \bibinfo {author} {\bibfnamefont {M.}~\bibnamefont {Ceotto}},\ }\href
  {\doibase 10.1021/jz401603f} {\bibfield  {journal} {\bibinfo  {journal} {J.
  Phys. Chem. Lett.}\ }\textbf {\bibinfo {volume} {4}},\ \bibinfo {pages}
  {3407} (\bibinfo {year} {2013})}\BibitemShut {NoStop}%
\bibitem [{\citenamefont {Gabas}, \citenamefont {Conte},\ and\ \citenamefont
  {Ceotto}(2017)}]{Gabas_Ceotto_Glycine_2017}%
  \BibitemOpen
  \bibfield  {author} {\bibinfo {author} {\bibfnamefont {F.}~\bibnamefont
  {Gabas}}, \bibinfo {author} {\bibfnamefont {R.}~\bibnamefont {Conte}}, \ and\
  \bibinfo {author} {\bibfnamefont {M.}~\bibnamefont {Ceotto}},\ }\href@noop {}
  {\bibfield  {journal} {\bibinfo  {journal} {J. Chem. Theory Comput.}\
  }\textbf {\bibinfo {volume} {13}},\ \bibinfo {pages} {2378} (\bibinfo {year}
  {2017})}\BibitemShut {NoStop}%
\bibitem [{\citenamefont {Buchholz}, \citenamefont {Grossmann},\ and\
  \citenamefont {Ceotto}(2016)}]{Buchholz_Ceotto_MixedSC_2016}%
  \BibitemOpen
  \bibfield  {author} {\bibinfo {author} {\bibfnamefont {M.}~\bibnamefont
  {Buchholz}}, \bibinfo {author} {\bibfnamefont {F.}~\bibnamefont {Grossmann}},
  \ and\ \bibinfo {author} {\bibfnamefont {M.}~\bibnamefont {Ceotto}},\
  }\href@noop {} {\bibfield  {journal} {\bibinfo  {journal} {J. Chem. Phys.}\
  }\textbf {\bibinfo {volume} {144}},\ \bibinfo {pages} {094102} (\bibinfo
  {year} {2016})}\BibitemShut {NoStop}%
\bibitem [{\citenamefont {Buchholz}, \citenamefont {Grossmann},\ and\
  \citenamefont {Ceotto}(2017)}]{Buchholz_Ceotto_applicationMixed_2017}%
  \BibitemOpen
  \bibfield  {author} {\bibinfo {author} {\bibfnamefont {M.}~\bibnamefont
  {Buchholz}}, \bibinfo {author} {\bibfnamefont {F.}~\bibnamefont {Grossmann}},
  \ and\ \bibinfo {author} {\bibfnamefont {M.}~\bibnamefont {Ceotto}},\
  }\href@noop {} {\bibfield  {journal} {\bibinfo  {journal} {J. Chem. Phys.}\
  }\textbf {\bibinfo {volume} {147}},\ \bibinfo {pages} {164110} (\bibinfo
  {year} {2017})}\BibitemShut {NoStop}%
\bibitem [{\citenamefont {Buchholz}, \citenamefont {Grossmann},\ and\
  \citenamefont {Ceotto}(2018)}]{Ceotto_Buchholz_SAM_2018}%
  \BibitemOpen
  \bibfield  {author} {\bibinfo {author} {\bibfnamefont {M.}~\bibnamefont
  {Buchholz}}, \bibinfo {author} {\bibfnamefont {F.}~\bibnamefont {Grossmann}},
  \ and\ \bibinfo {author} {\bibfnamefont {M.}~\bibnamefont {Ceotto}},\
  }\href@noop {} {\bibfield  {journal} {\bibinfo  {journal} {J. Chem. Phys.}\
  }\textbf {\bibinfo {volume} {148}},\ \bibinfo {pages} {114107} (\bibinfo
  {year} {2018})}\BibitemShut {NoStop}%
\bibitem [{\citenamefont {Wehrle}, \citenamefont {{\v{S}}ulc},\ and\
  \citenamefont
  {Van{\'\i}{\v{c}}ek}(2014)}]{Wehrle_Vanicek_Oligothiophenes_2014}%
  \BibitemOpen
  \bibfield  {author} {\bibinfo {author} {\bibfnamefont {M.}~\bibnamefont
  {Wehrle}}, \bibinfo {author} {\bibfnamefont {M.}~\bibnamefont {{\v{S}}ulc}},
  \ and\ \bibinfo {author} {\bibfnamefont {J.}~\bibnamefont
  {Van{\'\i}{\v{c}}ek}},\ }\href {\doibase 10.1063/1.4884718} {\bibfield
  {journal} {\bibinfo  {journal} {J. Chem. Phys.}\ }\textbf {\bibinfo {volume}
  {140}},\ \bibinfo {pages} {244114} (\bibinfo {year} {2014})}\BibitemShut
  {NoStop}%
\bibitem [{\citenamefont {Wehrle}, \citenamefont {Oberli},\ and\ \citenamefont
  {{Van{\'i}\v{c}ek}}(2015)}]{Wehrle_Vanicek_NH3_2015}%
  \BibitemOpen
  \bibfield  {author} {\bibinfo {author} {\bibfnamefont {M.}~\bibnamefont
  {Wehrle}}, \bibinfo {author} {\bibfnamefont {S.}~\bibnamefont {Oberli}}, \
  and\ \bibinfo {author} {\bibfnamefont {J.}~\bibnamefont
  {{Van{\'i}\v{c}ek}}},\ }\href {\doibase 10.1021/acs.jpca.5b03907} {\bibfield
  {journal} {\bibinfo  {journal} {J. Phys. Chem. A}\ }\textbf {\bibinfo
  {volume} {119}},\ \bibinfo {pages} {5685} (\bibinfo {year}
  {2015})}\BibitemShut {NoStop}%
\bibitem [{\citenamefont {Begu{\v s}i{\'c}}, \citenamefont {Cordova},\ and\
  \citenamefont {Van{\'i}{\v
  c}ek}(2019)}]{Begusic_SingleHessianthawedGaussian_2019}%
  \BibitemOpen
  \bibfield  {author} {\bibinfo {author} {\bibfnamefont {T.}~\bibnamefont
  {Begu{\v s}i{\'c}}}, \bibinfo {author} {\bibfnamefont {M.}~\bibnamefont
  {Cordova}}, \ and\ \bibinfo {author} {\bibfnamefont {J.}~\bibnamefont
  {Van{\'i}{\v c}ek}},\ }\href {\doibase 10.1063/1.5090122} {\bibfield
  {journal} {\bibinfo  {journal} {J. Chem. Phys.}\ }\textbf {\bibinfo {volume}
  {150}},\ \bibinfo {pages} {154117} (\bibinfo {year} {2019})}\BibitemShut
  {NoStop}%
\bibitem [{\citenamefont
  {Eckart}(1935)}]{Eckart_StudiesConcerningRotating_1935}%
  \BibitemOpen
  \bibfield  {author} {\bibinfo {author} {\bibfnamefont {C.}~\bibnamefont
  {Eckart}},\ }\href {\doibase 10.1103/PhysRev.47.552} {\bibfield  {journal}
  {\bibinfo  {journal} {Phys. Rev.}\ }\textbf {\bibinfo {volume} {47}},\
  \bibinfo {pages} {552} (\bibinfo {year} {1935})}\BibitemShut {NoStop}%
\bibitem [{\citenamefont {Wilson}, \citenamefont {Decius},\ and\ \citenamefont
  {Cross}(1980)}]{wilson1980molecular}%
  \BibitemOpen
  \bibfield  {author} {\bibinfo {author} {\bibfnamefont {E.}~\bibnamefont
  {Wilson}}, \bibinfo {author} {\bibfnamefont {J.}~\bibnamefont {Decius}}, \
  and\ \bibinfo {author} {\bibfnamefont {P.}~\bibnamefont {Cross}},\
  }\href@noop {} {\emph {\bibinfo {title} {Molecular {{Vibrations}}: {{The
  Theory}} of {{Infrared}} and {{Raman Vibrational Spectra}}}}},\ Dover Books
  on Chemistry Series\ (\bibinfo  {publisher} {{Dover Publications}},\ \bibinfo
  {year} {1980})\BibitemShut {NoStop}%
\bibitem [{\citenamefont {Miller}, \citenamefont {Handy},\ and\ \citenamefont
  {Adams}(1980)}]{Miller_ReactionpathHamiltonian_1980}%
  \BibitemOpen
  \bibfield  {author} {\bibinfo {author} {\bibfnamefont {W.~H.}\ \bibnamefont
  {Miller}}, \bibinfo {author} {\bibfnamefont {N.~C.}\ \bibnamefont {Handy}}, \
  and\ \bibinfo {author} {\bibfnamefont {J.~E.}\ \bibnamefont {Adams}},\ }\href
  {\doibase 10.1063/1.438959} {\bibfield  {journal} {\bibinfo  {journal} {J.
  Chem. Phys.}\ }\textbf {\bibinfo {volume} {72}},\ \bibinfo {pages} {99}
  (\bibinfo {year} {1980})}\BibitemShut {NoStop}%
\bibitem [{\citenamefont {Jellinek}\ and\ \citenamefont
  {Li}(1989)}]{Jellinek_SeparationEnergyOverall_1989}%
  \BibitemOpen
  \bibfield  {author} {\bibinfo {author} {\bibfnamefont {J.}~\bibnamefont
  {Jellinek}}\ and\ \bibinfo {author} {\bibfnamefont {D.~H.}\ \bibnamefont
  {Li}},\ }\href {\doibase 10.1103/PhysRevLett.62.241} {\bibfield  {journal}
  {\bibinfo  {journal} {Phys. Rev. Lett.}\ }\textbf {\bibinfo {volume} {62}},\
  \bibinfo {pages} {241} (\bibinfo {year} {1989})}\BibitemShut {NoStop}%
\bibitem [{\citenamefont
  {Watson}(1968)}]{Watson_Simplificationmolecularvibrationrotation_1968}%
  \BibitemOpen
  \bibfield  {author} {\bibinfo {author} {\bibfnamefont {J.~K.}\ \bibnamefont
  {Watson}},\ }\href {\doibase 10.1080/00268976800101381} {\bibfield  {journal}
  {\bibinfo  {journal} {Mol. Phys.}\ }\textbf {\bibinfo {volume} {15}},\
  \bibinfo {pages} {479} (\bibinfo {year} {1968})}\BibitemShut {NoStop}%
\bibitem [{\citenamefont {Avila}\ and\ \citenamefont {{Carrington
  Jr}}(2011)}]{Avila_Carrington_C2H4_2011}%
  \BibitemOpen
  \bibfield  {author} {\bibinfo {author} {\bibfnamefont {G.}~\bibnamefont
  {Avila}}\ and\ \bibinfo {author} {\bibfnamefont {T.}~\bibnamefont
  {{Carrington Jr}}},\ }\href@noop {} {\bibfield  {journal} {\bibinfo
  {journal} {J. Chem. Phys.}\ }\textbf {\bibinfo {volume} {135}},\ \bibinfo
  {pages} {064101} (\bibinfo {year} {2011})}\BibitemShut {NoStop}%
\bibitem [{\citenamefont {Kumar~P.}\ and\ \citenamefont
  {Marx}(2006)}]{KumarP_Understandinghydrogenscrambling_2006}%
  \BibitemOpen
  \bibfield  {author} {\bibinfo {author} {\bibfnamefont {P.}~\bibnamefont
  {Kumar~P.}}\ and\ \bibinfo {author} {\bibfnamefont {D.}~\bibnamefont
  {Marx}},\ }\href {\doibase 10.1039/B513089C} {\bibfield  {journal} {\bibinfo
  {journal} {Phys. Chem. Chem. Phys.}\ }\textbf {\bibinfo {volume} {8}},\
  \bibinfo {pages} {573} (\bibinfo {year} {2006})}\BibitemShut {NoStop}%
\bibitem [{\citenamefont {Harland}\ and\ \citenamefont
  {Roy}(2003)}]{Harland_Roy_SCIVRconstrained_2003}%
  \BibitemOpen
  \bibfield  {author} {\bibinfo {author} {\bibfnamefont {B.~B.}\ \bibnamefont
  {Harland}}\ and\ \bibinfo {author} {\bibfnamefont {P.-N.}\ \bibnamefont
  {Roy}},\ }\href@noop {} {\bibfield  {journal} {\bibinfo  {journal} {J. Chem.
  Phys.}\ }\textbf {\bibinfo {volume} {118}},\ \bibinfo {pages} {4791}
  (\bibinfo {year} {2003})}\BibitemShut {NoStop}%
\bibitem [{\citenamefont {Issack}\ and\ \citenamefont
  {Roy}(2005)}]{Issack_Geometricconstraintssemiclassical_2005}%
  \BibitemOpen
  \bibfield  {author} {\bibinfo {author} {\bibfnamefont {B.~B.}\ \bibnamefont
  {Issack}}\ and\ \bibinfo {author} {\bibfnamefont {P.-N.}\ \bibnamefont
  {Roy}},\ }\href {\doibase 10.1063/1.2004947} {\bibfield  {journal} {\bibinfo
  {journal} {J. Chem. Phys.}\ }\textbf {\bibinfo {volume} {123}},\ \bibinfo
  {pages} {084103} (\bibinfo {year} {2005})}\BibitemShut {NoStop}%
\bibitem [{\citenamefont {Issack}\ and\ \citenamefont
  {Roy}(2007{\natexlab{a}})}]{Issack_Geometricconstraintssemiclassical_2007}%
  \BibitemOpen
  \bibfield  {author} {\bibinfo {author} {\bibfnamefont {B.~B.}\ \bibnamefont
  {Issack}}\ and\ \bibinfo {author} {\bibfnamefont {P.-N.}\ \bibnamefont
  {Roy}},\ }\href {\doibase 10.1063/1.2423019} {\bibfield  {journal} {\bibinfo
  {journal} {J. Chem. Phys.}\ }\textbf {\bibinfo {volume} {126}},\ \bibinfo
  {pages} {024111} (\bibinfo {year} {2007}{\natexlab{a}})}\BibitemShut
  {NoStop}%
\bibitem [{\citenamefont {Issack}\ and\ \citenamefont
  {Roy}(2007{\natexlab{b}})}]{Issack_Semiclassicalinitialvalue_2007}%
  \BibitemOpen
  \bibfield  {author} {\bibinfo {author} {\bibfnamefont {B.~B.}\ \bibnamefont
  {Issack}}\ and\ \bibinfo {author} {\bibfnamefont {P.-N.}\ \bibnamefont
  {Roy}},\ }\href {\doibase 10.1063/1.2786456} {\bibfield  {journal} {\bibinfo
  {journal} {J. Chem. Phys.}\ }\textbf {\bibinfo {volume} {127}},\ \bibinfo
  {pages} {144306} (\bibinfo {year} {2007}{\natexlab{b}})}\BibitemShut
  {NoStop}%
\bibitem [{\citenamefont {Wong}\ \emph {et~al.}(2011)\citenamefont {Wong},
  \citenamefont {Benoit}, \citenamefont {Lewerenz}, \citenamefont {Brown},\
  and\ \citenamefont {Roy}}]{Wong_Roy_Formaldehyde_2011}%
  \BibitemOpen
  \bibfield  {author} {\bibinfo {author} {\bibfnamefont {S.~Y.~Y.}\
  \bibnamefont {Wong}}, \bibinfo {author} {\bibfnamefont {D.~M.}\ \bibnamefont
  {Benoit}}, \bibinfo {author} {\bibfnamefont {M.}~\bibnamefont {Lewerenz}},
  \bibinfo {author} {\bibfnamefont {A.}~\bibnamefont {Brown}}, \ and\ \bibinfo
  {author} {\bibfnamefont {P.-N.}\ \bibnamefont {Roy}},\ }\href {\doibase
  10.1063/1.3553179} {\bibfield  {journal} {\bibinfo  {journal} {J. Chem.
  Phys.}\ }\textbf {\bibinfo {volume} {134}},\ \bibinfo {pages} {094110}
  (\bibinfo {year} {2011})}\BibitemShut {NoStop}%
\bibitem [{\citenamefont {Ceotto}, \citenamefont {{Di Liberto}},\ and\
  \citenamefont {Conte}(2017)}]{ceotto_conte_DCSCIVR_2017}%
  \BibitemOpen
  \bibfield  {author} {\bibinfo {author} {\bibfnamefont {M.}~\bibnamefont
  {Ceotto}}, \bibinfo {author} {\bibfnamefont {G.}~\bibnamefont {{Di
  Liberto}}}, \ and\ \bibinfo {author} {\bibfnamefont {R.}~\bibnamefont
  {Conte}},\ }\href@noop {} {\bibfield  {journal} {\bibinfo  {journal} {Phys.
  Rev. Lett.}\ }\textbf {\bibinfo {volume} {119}},\ \bibinfo {pages} {010401}
  (\bibinfo {year} {2017})}\BibitemShut {NoStop}%
\bibitem [{\citenamefont {Gabas}, \citenamefont {Di~Liberto},\ and\
  \citenamefont {Ceotto}(2019)}]{Gabas_nucleobases2019}%
  \BibitemOpen
  \bibfield  {author} {\bibinfo {author} {\bibfnamefont {F.}~\bibnamefont
  {Gabas}}, \bibinfo {author} {\bibfnamefont {G.}~\bibnamefont {Di~Liberto}}, \
  and\ \bibinfo {author} {\bibfnamefont {M.}~\bibnamefont {Ceotto}},\ }\href
  {\doibase 10.1063/1.5100503} {\bibfield  {journal} {\bibinfo  {journal} {J.
  Chem. Phys.}\ }\textbf {\bibinfo {volume} {150}},\ \bibinfo {pages} {224107}
  (\bibinfo {year} {2019})}\BibitemShut {NoStop}%
\bibitem [{\citenamefont {Micciarelli}\ \emph {et~al.}(2018)\citenamefont
  {Micciarelli}, \citenamefont {Conte}, \citenamefont {Suarez},\ and\
  \citenamefont
  {Ceotto}}]{Micciarelli_Anharmonicvibrationaleigenfunctions_2018}%
  \BibitemOpen
  \bibfield  {author} {\bibinfo {author} {\bibfnamefont {M.}~\bibnamefont
  {Micciarelli}}, \bibinfo {author} {\bibfnamefont {R.}~\bibnamefont {Conte}},
  \bibinfo {author} {\bibfnamefont {J.}~\bibnamefont {Suarez}}, \ and\ \bibinfo
  {author} {\bibfnamefont {M.}~\bibnamefont {Ceotto}},\ }\href {\doibase
  10.1063/1.5041911} {\bibfield  {journal} {\bibinfo  {journal} {J. Chem.
  Phys.}\ }\textbf {\bibinfo {volume} {149}},\ \bibinfo {pages} {064115}
  (\bibinfo {year} {2018})}\BibitemShut {NoStop}%
\bibitem [{\citenamefont {Ma}\ \emph {et~al.}(2018)\citenamefont {Ma},
  \citenamefont {Di~Liberto}, \citenamefont {Conte}, \citenamefont {Hase},\
  and\ \citenamefont {Ceotto}}]{Ma_quantummechanicalinsight_2018}%
  \BibitemOpen
  \bibfield  {author} {\bibinfo {author} {\bibfnamefont {X.}~\bibnamefont
  {Ma}}, \bibinfo {author} {\bibfnamefont {G.}~\bibnamefont {Di~Liberto}},
  \bibinfo {author} {\bibfnamefont {R.}~\bibnamefont {Conte}}, \bibinfo
  {author} {\bibfnamefont {W.~L.}\ \bibnamefont {Hase}}, \ and\ \bibinfo
  {author} {\bibfnamefont {M.}~\bibnamefont {Ceotto}},\ }\href {\doibase
  10.1063/1.5054399} {\bibfield  {journal} {\bibinfo  {journal} {J. Chem.
  Phys.}\ }\textbf {\bibinfo {volume} {149}},\ \bibinfo {pages} {164113}
  (\bibinfo {year} {2018})}\BibitemShut {NoStop}%
\bibitem [{\citenamefont {{De Leon}}\ and\ \citenamefont
  {Heller}(1983)}]{DeLeon_Heller_SCeigenfunctions_1983}%
  \BibitemOpen
  \bibfield  {author} {\bibinfo {author} {\bibfnamefont {N.}~\bibnamefont {{De
  Leon}}}\ and\ \bibinfo {author} {\bibfnamefont {E.~J.}\ \bibnamefont
  {Heller}},\ }\href@noop {} {\bibfield  {journal} {\bibinfo  {journal} {J.
  Chem. Phys.}\ }\textbf {\bibinfo {volume} {78}},\ \bibinfo {pages} {4005}
  (\bibinfo {year} {1983})}\BibitemShut {NoStop}%
\bibitem [{\citenamefont
  {{Longuet-Higgins}}(1963)}]{Longuet-Higgins_symmetrygroupsnonrigid_1963}%
  \BibitemOpen
  \bibfield  {author} {\bibinfo {author} {\bibfnamefont {H.~C.}\ \bibnamefont
  {{Longuet-Higgins}}},\ }\href {\doibase 10.1080/00268976300100501} {\bibfield
   {journal} {\bibinfo  {journal} {Mol. Phys.}\ }\textbf {\bibinfo {volume}
  {6}},\ \bibinfo {pages} {445} (\bibinfo {year} {1963})}\BibitemShut {NoStop}%
\bibitem [{\citenamefont {Micciarelli}\ \emph {et~al.}(2019)\citenamefont
  {Micciarelli}, \citenamefont {Gabas}, \citenamefont {Conte},\ and\
  \citenamefont {Ceotto}}]{Micciarelli_effectivesemiclassicalapproach_2019}%
  \BibitemOpen
  \bibfield  {author} {\bibinfo {author} {\bibfnamefont {M.}~\bibnamefont
  {Micciarelli}}, \bibinfo {author} {\bibfnamefont {F.}~\bibnamefont {Gabas}},
  \bibinfo {author} {\bibfnamefont {R.}~\bibnamefont {Conte}}, \ and\ \bibinfo
  {author} {\bibfnamefont {M.}~\bibnamefont {Ceotto}},\ }\href {\doibase
  10.1063/1.5096968} {\bibfield  {journal} {\bibinfo  {journal} {J. Chem.
  Phys.}\ }\textbf {\bibinfo {volume} {150}},\ \bibinfo {pages} {184113}
  (\bibinfo {year} {2019})}\BibitemShut {NoStop}%
\bibitem [{\citenamefont {Peslherbe}\ and\ \citenamefont
  {Hase}(1994)}]{Peslherbe_Analysisextensionmodel_1994}%
  \BibitemOpen
  \bibfield  {author} {\bibinfo {author} {\bibfnamefont {G.~H.}\ \bibnamefont
  {Peslherbe}}\ and\ \bibinfo {author} {\bibfnamefont {W.~L.}\ \bibnamefont
  {Hase}},\ }\href {\doibase 10.1063/1.466648} {\bibfield  {journal} {\bibinfo
  {journal} {J. Chem. Phys.}\ }\textbf {\bibinfo {volume} {100}},\ \bibinfo
  {pages} {1179} (\bibinfo {year} {1994})}\BibitemShut {NoStop}%
\end{thebibliography}

%

\end{document}